\def\statusstring{Submitted to IEEE Transactions on Information Theory, 
                  November 20, 2010; revised on September 22, 2012;
                  date of current version, October 9, 2012.}

\documentclass[journal, twocolumn]{IEEEtran}

\def\baselinestretch{0.99}

\usepackage{amssymb} 
\usepackage{amsmath} 
\usepackage{amscd}
\usepackage{color}

\usepackage{theorem} 

\usepackage{latexsym} 
\usepackage{epsfig} 
\usepackage{psfrag} 
\usepackage{bm, bbm} 
\usepackage{xspace} 
\usepackage{graphicx} 
\usepackage{cite} 
\usepackage{url} 
\usepackage{subfigure,afterpage, wrapfig, pifont, multirow, rotating}

\IEEEtriggeratref{73}

\renewcommand{\mathbf}[1]{{\bm{#1}}}

\newcommand{\ignore}[1]{}

\newcommand{\confer}{\emph{cf.}}
\newcommand{\eg}{\emph{e.g.}}
\newcommand{\etal}{\emph{et al.}}
\newcommand{\ie}{\emph{i.e.}}
\newcommand{\etc}{\emph{etc.}}
\newcommand{\versus}{\emph{vs.}}

\renewcommand{\leq}{\leqslant}
\renewcommand{\geq}{\geqslant}

\newcommand{\R}{\mathbb{R}}
\newcommand{\Rp}{\mathbb{R}_{\geq 0}}
\newcommand{\Rpp}{\mathbb{R}_{>0}}
\newcommand{\Z}{\mathbb{Z}}
\newcommand{\Zp}{\Z_{\geq 0}}
\newcommand{\Zpp}{\Z_{>0}}

\newcommand{\Expec}{\mathsf{E}}

\DeclareMathOperator*{\argmax}{arg\,max}

\newcommand{\matr}[1]{\mathbf{#1}}
\newcommand{\vect}[1]{\mathbf{#1}}
\newcommand{\code}[1]{\mathcal{#1}}

\newcommand{\set}[1]{\mathcal{#1}}

\newcommand{\graph}[1]{\mathsf{#1}}

\DeclareMathOperator*{\argmin}{arg\,min}

\newcommand{\GF}[1]{\mathbb{F}_{#1}}

\newcommand{\defeq}{\triangleq}
\newcommand{\onestareq}{\overset{\text{(a)}}{=}}
\newcommand{\twostarseq}{\overset{\text{(b)}}{=}}
\newcommand{\threestarseq}{\overset{\text{(c)}}{=}}
\newcommand{\fourstarseq}{\overset{\text{(d)}}{=}}
\newcommand{\fivestarseq}{\overset{\text{(e)}}{=}}
\newcommand{\sixstarseq}{\overset{\text{(f)}}{=}}
\newcommand{\sevenstarseq}{\overset{\text{(g)}}{=}}

\newcommand{\onestar}{\text{(a)}}
\newcommand{\twostars}{\text{(b)}}
\newcommand{\threestars}{\text{(c)}}
\newcommand{\fourstars}{\text{(d)}}
\newcommand{\fivestars}{\text{(e)}}
\newcommand{\sixstars}{\text{(f)}}
\newcommand{\sevenstars}{\text{(g)}}

\newcommand{\codeC}{\code{C}}
\newcommand{\codeCchannel}{\code{C}_{\mathrm{ch}}}
\newcommand{\codeCedge}{\code{C}}
\newcommand{\codeCedgehalf}{\code{C}_{\mathrm{half}}}

\newcommand{\matrH}{\matr{H}}

\newcommand{\matrtheta}{\boldsymbol{\theta}}

\newcommand{\bel}{\beta}

\newcommand{\bels}{\bel^{*}}

\newcommand{\hbel}{\hat \bel}

\newcommand{\vbel}{\boldsymbol{\bel}}
\newcommand{\vbels}{\boldsymbol{\bel}^{*}}

\newcommand{\hvbel}{\boldsymbol{\hat \bel}}

\newcommand{\setA}{\set{A}}

\newcommand{\setAfull}{\set{A}_{\mathrm{full}}}
\newcommand{\setAhalf}{\set{A}_{\mathrm{half}}}
\newcommand{\setC}{\set{C}}

\newcommand{\setE}{\set{E}}

\newcommand{\setEfull}{\set{E}_{\mathrm{full}}}
\newcommand{\setEhalf}{\set{E}_{\mathrm{half}}}

\newcommand{\setF}{\set{F}}
\newcommand{\csetF}{\set{\tilde F}}

\newcommand{\setG}{\set{G}}

\newcommand{\setI}{\set{I}}
\newcommand{\setJ}{\set{J}}

\newcommand{\setR}{\set{R}}
\newcommand{\setS}{\set{S}}

\newcommand{\setX}{\set{X}}
\newcommand{\setY}{\set{Y}}

\newcommand{\rdiff}[1]{\frac{\mathrm{d}}{\mathrm{d}#1}}

\newcommand{\hcN}{\hat{\cgraph{N}}}

\newcommand{\veta}{\boldsymbol{\eta}}

\newcommand{\vomega}{\boldsymbol{\omega}}

\newcommand{\BGCD}{\mathrm{BGCD}}
\newcommand{\SGCD}{\mathrm{SGCD}}

\newcommand{\vA}{\vect{A}}
\newcommand{\va}{\vect{a}}
\newcommand{\cva}{\vect{\cover{a}}}

\newcommand{\vafull}{\va_{\mathrm{full}}}
\newcommand{\vahalf}{\va_{\mathrm{half}}}

\newcommand{\vC}{\vect{C}}
\newcommand{\vc}{\vect{c}}

\newcommand{\vcfull}{\vc_{\mathrm{full}}}
\newcommand{\vchalf}{\vc_{\mathrm{half}}}

\newcommand{\cf}{\tilde f}

\newcommand{\cc}{\cover{c}}

\newcommand{\cvc}{\vect{\cover{c}}}

\newcommand{\hcvc}{\vect{\hat{\cover{c}}}}

\newcommand{\vp}{\vect{p}}
\newcommand{\vq}{\vect{q}}

\newcommand{\vX}{\vect{X}}

\newcommand{\vx}{\vect{x}}

\newcommand{\tvx}{\vect{\tilde x}}

\newcommand{\BMAPD}{\mathrm{BMAPD}}
\newcommand{\SMAPD}{\mathrm{SMAPD}}
\newcommand{\hvxSMAPD}{\vect{\hat x}^{\mathrm{SMAPD}}}
\newcommand{\hxSMAPD}{\hat x^{\mathrm{SMAPD}}}

\newcommand{\hvxBMAPD}{\vect{\hat x}^{\mathrm{BMAPD}}}
\newcommand{\hxBMAPD}{\hat x^{\mathrm{BMAPD}}}

\newcommand{\vY}{\vect{Y}}

\newcommand{\vy}{\vect{y}}

\newcommand{\dleft}{d_{\mathrm{L}}}
\newcommand{\dright}{d_{\mathrm{R}}}

\newcommand{\wH}{w_{\mathrm{H}}}

\newcommand{\fp}[1]{\mathcal{#1}}

\newcommand{\fc}[1]{\mathcal{#1}}

\newcommand{\tr}{\mathsf{T}}

\newcommand{\lmpB}{\mathcal{B}}

\newcommand{\lmpBdown}[1]{\mathcal{B}'_{#1}}

\newcommand{\graphN}{\graph{N}}

\newcommand{\convhull}{\operatorname{conv}}

\newcommand{\conichull}{\operatorname{conic}}

\newcommand{\FBethe}{F_{\mathrm{B}}}

\newcommand{\UBethe}{U_{\mathrm{B}}}
\newcommand{\HBethe}{H_{\mathrm{B}}}
\newcommand{\UBethesub}[1]{U_{\mathrm{B},#1}}
\newcommand{\HBethesub}[1]{H_{\mathrm{B},#1}}

\newcommand{\ZBethe}{Z_{\mathrm{B}}}
\newcommand{\ZBetheM}[1]{Z_{\mathrm{B}, #1}}

\newcommand{\cover}[1]{\tilde{#1}}
\newcommand{\cset}[1]{\cover{\set{#1}}}
\newcommand{\cgraph}[1]{\cover{\graph{#1}}}

\newcommand{\card}[1]{\left\lvert #1 \right\rvert}
\newcommand{\cardbig}[1]{\big\lvert #1 \big\rvert}

\newcommand{\varphiM}{\boldsymbol{\varphi}_{M}}

\newcommand{\vPsiBME}{\boldsymbol{\Psi}_{\mathrm{BME}}}

\newcommand{\vpsi}{\boldsymbol{\psi}}

\newcommand{\meanvect}{\operatorname{\mathbf{mean}}}

\newcommand{\tvC}{\vect{\tilde{C}}}
\newcommand{\tvc}{\vect{\tilde{c}}}

\newcommand{\vaoldsample}[1]{\va_{\sample{#1}}}

\newcommand{\vcoldsample}[1]{\vc_{\sample{#1}}}
\newcommand{\vCsample}[1]{\tvC_{#1}}
\newcommand{\vcsample}[1]{\tvc_{#1}}
\newcommand{\vcsampletot}{\tvc}
\newcommand{\vqcsample}{\vect{q}^{(\vcsampletot)}}
\newcommand{\qcsample}{q^{(\vcsampletot)}}

\newcommand{\sample}[1]{\textrm{sample}(#1)}

\newcommand{\vxsampletot}{\tvx}

\newcommand{\Qxoldsample}[1]{Q_{\vx_{\textrm{sample}}}} 

\newcommand{\Qxsample}[1]{Q_{\vxsampletot}}
\newcommand{\setQsample}[1]{\set{Q}_{#1}}

\def\squarebox#1{\hbox to #1{\hfill\vbox to #1{\vfill}}}

\newcommand{\qedblack}{\hspace*{\fill}%
  $\blacksquare$\smallskip}
\newcommand{\qedwhite}{\hspace*{\fill}%
  $\square$\smallskip}

\newcommand{\defend}{\qedwhite} 
\newcommand{\exampleend}{\qedwhite} 
\newcommand{\remarkend}{\qedwhite} 
\newcommand{\assumptionend}{\qedwhite}

\newcommand{\FGibbs}{F_{\mathrm{G}}}
\newcommand{\UGibbs}{U_{\mathrm{G}}}
\newcommand{\HGibbs}{H_{\mathrm{G}}}
\newcommand{\ZGibbs}{Z_{\mathrm{G}}}
\newcommand{\ZGibbsext}[1]{Z_{\mathrm{G}}(#1)}

\newcommand{\setPi}{\Pi}

\newcommand{\del}{\partial}

\newtheorem{Lemma}{Lemma}
\newtheorem{Theorem}[Lemma]{Theorem}

\newtheorem{Definition}[Lemma]{Definition}
\newtheorem{Remark}[Lemma]{Remark}
\newtheorem{Example}[Lemma]{Example}

\newtheorem{Assumption}[Lemma]{Assumption}

\newenvironment{Proof}%
  {\noindent \emph{Proof:}}{\qedblack}

\newcommand{\cirrank}{\operatorname{circ}}
\newcommand{\ncomponents}{\#}

\newcommand{\perm}{\operatorname{perm}}

\title{Counting in Graph Covers: 
       A Combinatorial Characterization of the
       Bethe Entropy Function}

\author{Pascal O.~Vontobel
        \thanks{\statusstring \ Some of the material in this paper 
          was previously presented
          at the 46th Annual Allerton Conference on Communications, Control,
          and Computing, Monticello, IL, USA, Sep.~23--26, 2008, 
          at the 2009 Information Theory Workshop, Taormina, Italy,
          Oct.~10--13, 2009,
          and at the 48th Annual Allerton Conference on Communications,
          Control, and Computing, Monticello, IL, USA, Sep.~29--Oct.~1,
          2010.}%
        \thanks{P.~O.~Vontobel is with Hewlett-Packard Laboratories, 1501 Page
          Mill Road, Palo Alto, CA 94304, USA
          (e-mail: pascal.vontobel@ieee.org).}%
      }

\begin{document}

\markboth{Submitted to IEEE Transactions on Information Theory}%
         {Pascal O.~Vontobel}

\maketitle

\begin{abstract}
  We present a \emph{combinatorial} characterization of the Bethe entropy
  function of a factor graph, such a characterization being in contrast to the
  original, \emph{analytical}, definition of this function. We achieve this
  combinatorial characterization by counting valid configurations in finite
  graph covers of the factor graph.

  Analogously, we give a \emph{combinatorial} characterization of the Bethe
  partition function, whose original definition was also of an
  \emph{analytical} nature. As we point out, our approach has similarities to
  the replica method, but also stark differences.

  The above findings are a natural backdrop for introducing a decoder for
  graph-based codes that we will call symbolwise graph-cover decoding, a
  decoder that extends our earlier work on blockwise graph-cover
  decoding. Both graph-cover decoders are theoretical tools that help towards a
  better understanding of message-passing iterative decoding, namely blockwise
  graph-cover decoding links max-product (min-sum) algorithm decoding with
  linear programming decoding, and symbolwise graph-cover decoding links
  sum-product algorithm decoding with Bethe free energy function minimization
  at temperature one.

  In contrast to the Gibbs entropy function, which is a concave function, the
  Bethe entropy function is in general not concave everywhere. In particular,
  we show that every code picked from an ensemble of regular low-density
  parity-check codes with minimum Hamming distance growing (with high
  probability) linearly with the block length has a Bethe entropy function
  that is convex in certain regions of its domain.
\end{abstract}


\section{Introduction}
\label{sec:introduction:1}

\IEEEPARstart{W}{hat is the} meaning of the pseudo-marginal functions that are
computed by the sum-product algorithm, especially at a fixed point of the
sum-product algorithm? This question stood at the beginning of our
investigations. For factor graphs \emph{without} cycles the answer is clear,
and was stated succinctly already by Wiberg
\etal~\cite{Wiberg:Loeliger:Koetter:95, Wiberg:96}: the pseudo-marginal
functions at a fixed point of the sum-product algorithm (SPA) are the correct
marginal functions of the global function that is represented by the factor
graph. Note that here and hereafter we assume that SPA messages are updated
according to the so-called ``flooding message update schedule.'' For factor
graphs of finite size and without cycles this implies that the SPA reaches a
fixed point after a finite number of
iterations~\cite{Wiberg:Loeliger:Koetter:95, Wiberg:96}.

However, in the case of factor graphs \emph{with} cycles, the answer is a
priori not so clear, even for fixed points of the
SPA.\footnote{\label{footnote:assumptions:on:nfg:1} Here and in the following
  we assume that the local functions of the factor graph are
  non-negative. Moreover, we assume that we are only interested in SPA-based
  pseudo-marginal functions that are normalized, \ie, pseudo-marginal
  functions that sum to one. Therefore, without loss of generality, we may
  assume that at every iteration the SPA messages are normalized, \ie, that
  they sum to one. With this, SPA fixed points are well defined, also for
  factor graphs with cycles.} Of course, one can express the SPA-based
pseudo-marginal functions as marginal functions of the global functions of
computation-tree factor graphs, the latter being unwrapped versions of the
factor graph under consideration~\cite{Wiberg:Loeliger:Koetter:95,
  Wiberg:96}. However, the analysis of these objects has so far proven to be
rather difficult, a main reason for this being that the computation trees,
along with the global functions represented by them, change with the iteration
number.

\subsection{A Combinatorial Characterization of the Bethe Entropy Function in
  Terms of Finite Graph Covers}

Towards making progress on the above-mentioned question, this paper studies the
Bethe free energy function of a factor graph, a function that was introduced
by Yedidia, Freeman, and Weiss~\cite{Yedidia:Freeman:Weiss:05:1} and whose
importance stems from a very well-known theorem
in~\cite{Yedidia:Freeman:Weiss:05:1} which states that fixed points of the SPA
correspond to stationary points of the Bethe free energy
function. Consequently, it is clearly desirable to obtain a better
understanding of this function by characterizing it from different
perspectives.

Recall that the Bethe free energy function $\FBethe$ is defined to be
\begin{align*}
  \FBethe(\vbel)
    &\defeq
       \UBethe(\vbel)
       -
       T
       \cdot
       \HBethe(\vbel),
\end{align*}
where $\vbel$ is a (locally consistent) pseudo-marginal vector, $\UBethe$ is
the Bethe average energy function, $\HBethe$ is the Bethe entropy function,
and $T \! \geq \! 0$ is the temperature. (All the mathematical terms appearing
in this introduction will be suitably defined in later sections.) Both
$\UBethe$ and $\HBethe$ contribute significantly towards the shape of
$\FBethe$. However, the curvature of $\FBethe$ is exclusively determined by
the curvature of $\HBethe$; this is a consequence of the fact that $\UBethe$
is a linear function of its argument. Therefore, characterizing the function
$\FBethe$ is nearly tantamount to characterizing the function $\HBethe$.

In this paper we offer a combinatorial characterization of the Bethe entropy
function $\HBethe$ in terms of finite graph covers of the factor graph under
consideration. Recall that in earlier work~\cite{Koetter:Vontobel:03:1,
  Vontobel:Koetter:05:1:subm} we showed that every valid configuration in some
finite graph cover maps down to some pseudo-marginal vector with rational
coordinates, and vice-versa, every pseudo-marginal vector with rational
coordinates has at least one pre-image in some finite graph cover. (Actually,
the papers~\cite{Koetter:Vontobel:03:1, Vontobel:Koetter:05:1:subm} focused on
a special case of graphical models, namely graphical models that represent
binary low-density parity-check codes, and consequently dealt with
pseudo-codewords. However, the results therein are easily generalized to the
more general setup considered here.) The present paper discusses the following
extension of this result. Namely, letting $\vbel$ be a pseudo-marginal vector
with only rational coordinates and letting $\bar C_M(\vbel)$ be the average
number of pre-images of $\vbel$ among all the $M$-covers, we show that $\bar
C_M(\vbel)$ grows, when $M$ goes to infinity, like
\begin{align*}
  \bar C_M(\vbel)
    &= \exp 
         \big( 
           M \cdot \HBethe(\vbel)
           +
           o(M)
         \big).
\end{align*}
This characterization of the Bethe entropy function has clearly a
``combinatorial flavor,'' which is in contrast to the ``analytical flavor'' of
the original definition of the Bethe entropy function
in~\cite{Yedidia:Freeman:Weiss:05:1} (see
Definition~\ref{def:Bethe:free:energy:1} in the present paper).

\subsection{A Combinatorial Characterization of the Bethe Partition Function
                         in Terms of Finite Graph Covers}

This paper offers also a combinatorial characterization of the Bethe partition
function $\ZBethe$ in terms of finite graph covers of the factor graph under
consideration. This is again in contrast to the original, analytical,
definition of $\ZBethe$ that defines $\ZBethe$ via the minimum of
$\FBethe$. Compare this with the Gibbs partition function $\ZGibbs$: its
definition is combinatorial in the sense that $\ZGibbs$ is defined as a sum of
certain terms. (Of course, the Gibbs partition function can also be
characterized analytically via the minimum of the Gibbs free energy function
$\FGibbs$.)

More precisely, recall that the Gibbs partition function (or total sum) of a
factor graph $\graphN$ is
\begin{align*}
  \ZGibbs(\graphN)
    &\defeq
       \sum_{\va \in \setA}
         g(\va)^{1/T},
\end{align*}
where the sum is over all configurations of $\graphN$ and where $g$ is the
global function of $\graphN$. We show that the Bethe partition function can be
written as follows
\begin{align}
  \ZBethe(\graphN)
    &= \limsup_{M \to \infty} \ 
         \ZBetheM{M}(\graphN),
           \label{eq:ZBethe:intro:1}
\end{align}
where
\begin{align}
  \ZBetheM{M}(\graphN)
    &\defeq
       \sqrt[M]{\Big\langle \!
                  \ZGibbs(\cgraph{N})
                \! \Big\rangle_{\cgraph{N} \in \cset{N}_{M}}}.
                             \label{eq:ZBethe:intro:2}
\end{align}
Here the expression under the root sign represents the average of
$\ZGibbs(\cgraph{N})$ over all $M$-covers $\cgraph{N}$ of $\graphN$. Clearly,
the expression for $\ZBethe(\graphN)$ given
by~\eqref{eq:ZBethe:intro:1}--\eqref{eq:ZBethe:intro:2} has a ``combinatorial
flavor.''

Interestingly, the expression in~\eqref{eq:ZBethe:intro:2} is based on only
two rather simple concepts (besides the standard mathematical concepts of
taking limits, taking roots, and computing averages): we only need to define
the concept of an $M$-cover of a factor graph and the concept of the Gibbs
partition function of a factor graph. In our opinion, these concepts are quite
a bit simpler than the ones needed for defining $\FBethe$ and then
$\ZBethe(\graphN)$ in terms of the minimum of $\FBethe$. (A technical note on
the side: in order for the minimum of $\FBethe$ to make sense, we need the
assumption that was stated in Footnote~\ref{footnote:assumptions:on:nfg:1},
namely that all local functions of $\graphN$ are assumed to be
non-negative. As we will see, this assumption is also crucial for showing the
equivalence of the left- and right-hand sides of~\eqref{eq:ZBethe:intro:1}.)

Note that~\eqref{eq:ZBethe:intro:2} contains a finite sum. For small factor
graphs~$\graphN$ and small $M$ this fact can be exploited, for example, to
perform some brute-force computations and come up with conjectures of the
relationship of $\ZBetheM{M}(\graphN)$ with respect to (w.r.t.)
$\ZGibbs(\graphN)$. Afterwards, one can try to analytically prove these
conjectures about $\ZBetheM{M}(\graphN)$ for any finite $M$, and thereby prove
a similar conjecture for $\ZBethe(\graphN)$. This line of reasoning is
especially interesting if the proofs can be extended to hold for any factor
graph within some class of factor graphs.

\begin{figure}
  \begin{flushright}
    \epsfig{file=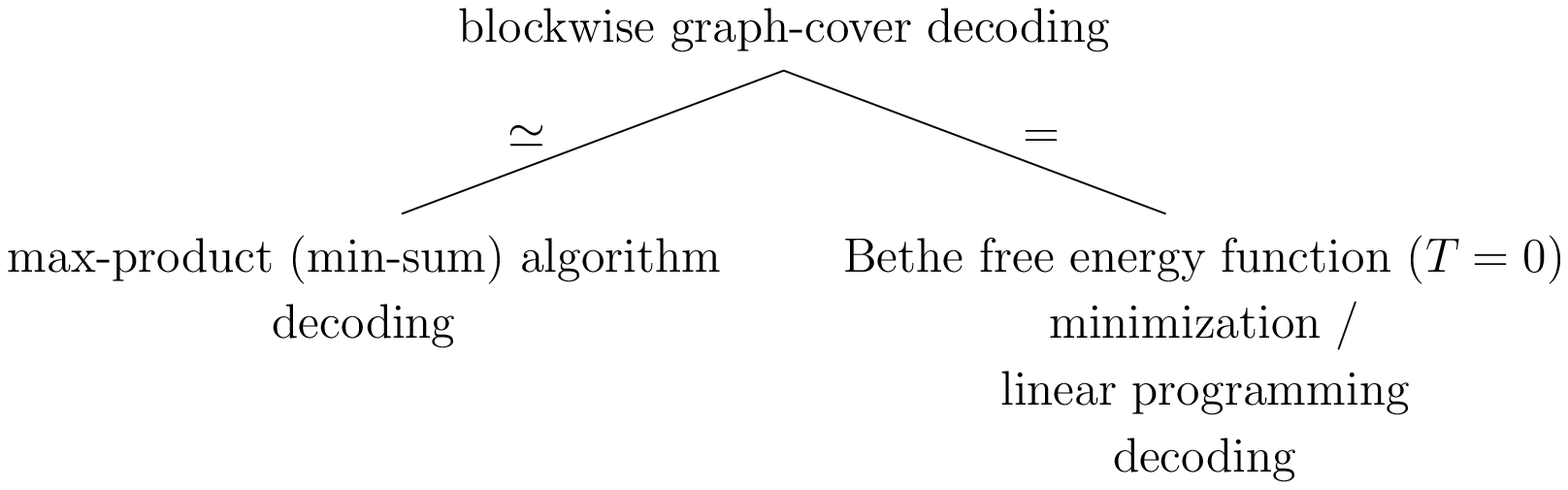, 
            scale=0.50}
    \hspace{0.30cm}\mbox{}
  \end{flushright}
  \caption{Blockwise graph-cover decoding forms a bridge between max-product
    (min-sum) algorithm decoding and linear programming decoding, the latter
    being equivalent to Bethe free energy minimization at temperature $T =
    0$. (See Section~\ref{sec:introduction:graph:cover:decoding:1} for more
    details.)}
  \label{fig:BGCD:overview:1}
\end{figure}

\begin{figure}
  \begin{flushright}
    \epsfig{file=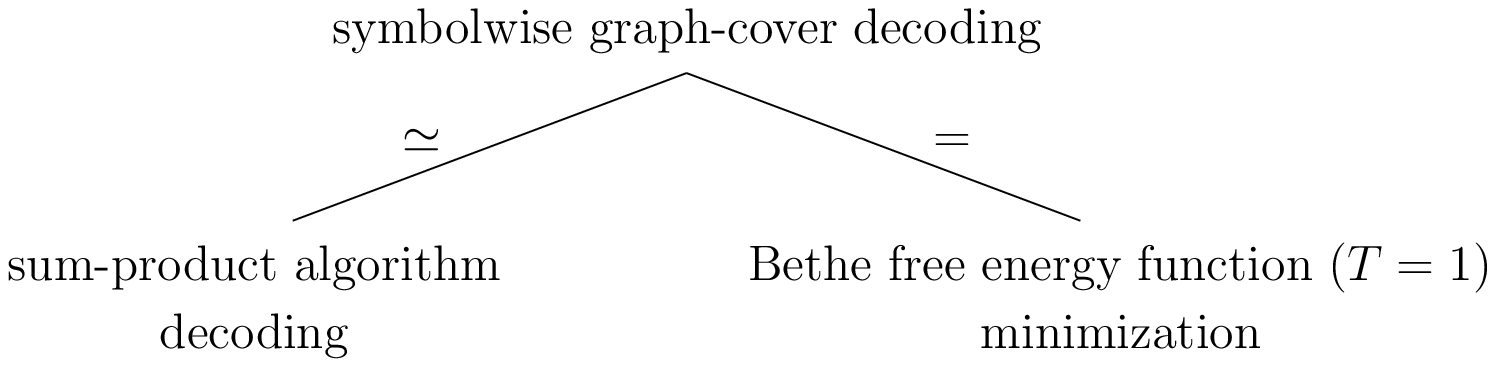, 
            scale=0.50}
    \hspace{0.30cm}\mbox{}
  \end{flushright}
  \caption{Symbolwise graph-cover decoding forms a bridge between sum-product
    algorithm decoding and Bethe free energy minimization at temperature $T =
    1$. (See Section~\ref{sec:introduction:graph:cover:decoding:1} for more
    details.)}
  \label{fig:SGCD:overview:1}
\end{figure}

\subsection{Graph-Cover Decoding}
\label{sec:introduction:graph:cover:decoding:1}

One of the main motivations of the papers~\cite{Koetter:Vontobel:03:1,
  Vontobel:Koetter:05:1:subm} to study finite graph covers of a factor graph
$\graphN$ was the fact that finite graph covers of $\graphN$ look locally the
same as $\graphN$. Consequently, any locally operating algorithm, like the
max-product algorithm or the sum-product algorithm, ``cannot distinguish'' if
they are operating on $\graphN$ or, implicitly, on any of its finite graph
covers. Clearly, for factor graphs \emph{with} cycles, this
``non-distinguishability'' observation implies fundamental limitations on the
conclusions that can be reached by locally operating algorithms because finite
graph covers of such factor graphs are ``non-trivial'' in the sense that they
contain valid configurations that ``cannot be explained'' by valid
configurations in the base factor graph. This is in sharp contrast to factor
graphs \emph{without cycles}: all $M$-covers of such factor graphs are
``trivial'' in the sense that they consist of $M$ independent copies of the
base factor graph, and so the set of valid configuration of any $M$-cover
equals the $M$-fold Cartesian product of the set of valid configurations of
the base factor graph with itself.

In fact, in the context of message-passing iterative decoding of graph-based
codes \emph{with} cycles, we argued in~\cite{Vontobel:Koetter:05:1:subm,
  Vontobel:Koetter:04:2} that these fundamental limitations of message-passing
iterative decoding, in particular of max-product (min-sum) algorithm decoding,
imply that these decoders behave less like blockwise maximum a-posteriori
decoding (which is equivalent to minimizing the Gibbs free energy function at
temperature $T = 0$), but much more like linear programming
decoding~\cite{Feldman:03:1, Feldman:Wainwright:Karger:05:1} (which is
equivalent to minimizing the Bethe free energy function at temperature $T =
0$). This was done with the help of a theoretical tool called blockwise
graph-cover decoding, a tool that in~\cite{Vontobel:Koetter:05:1:subm} was
simply called graph-cover decoding. Namely (see
Fig.~\ref{fig:BGCD:overview:1}),
\begin{itemize}

\item on the one hand we showed the equivalence of blockwise graph-cover
  decoding and linear programming decoding,

\item on the other hand we argued that blockwise graph-cover decoding is a
  good ``model'' for the behavior of the max-product (min-sum) algorithm
  decoding.

\end{itemize}
This latter connection, namely between blockwise graph-cover decoding and
max-product (min-sum) algorithm decoding, is in general only an approximate
one. However, in all cases where analytical tools are known that exactly
characterize the behavior of max-product algorithm decoding, the connection
between blockwise graph-cover decoding and the max-product (min-sum) algorithm
decoding is exact.

In this paper, we define symbolwise graph-cover decoding, which tries to
capture the essential limitations of sum-product algorithm decoding. It will
allow us to argue that for graph-based codes with cycles, sum-product
algorithm decoding behaves less like symbolwise maximum a-posteriori decoding
(which is equivalent to minimizing the Gibbs free energy function at
temperature $T = 1$), but much more like an algorithm that minimizes (at least
locally) the Bethe free energy function at temperature $T = 1$. This will be
done (see Fig.~\ref{fig:SGCD:overview:1}),
\begin{itemize}

\item on the one hand, by showing that symbolwise graph-cover decoding is
  equivalent to (globally) minimizing the Bethe free energy function at
  temperature $T = 1$,

\item and by arguing that symbolwise graph-cover decoding is a good ``model''
  for the behavior of sum-product algorithm decoding.

\end{itemize}
Similar to the above discussion about blockwise graph-cover decoding, this
latter connection, namely between symbolwise graph-cover decoding and
sum-product algorithm decoding, is in general an approximate one. However, in
many cases where analytical tools are known that exactly characterize the
behavior of sum-product algorithm decoding, the connection between symbolwise
graph-cover decoding and sum-product algorithm decoding is exact.

In any case, using the combinatorial characterization of the Bethe entropy
mentioned earlier in this introduction, one can state that a fixed point of
the sum-product algorithm corresponds to a certain pseudo-marginal vector of
the factor graph under consideration: it is, after taking a biasing
channel-output-dependent term properly into account, the pseudo-marginal
vector that has (locally) an extremal number of pre-images in all $M$-covers,
when $M$ goes to infinity.

\subsection{The Shape of the Bethe Entropy Function}

The paper concludes with a section on the concavity, or the lack thereof, of
the Bethe entropy function. Recall that the Gibbs entropy function is a
concave function, and therefore the Gibbs free energy function is a convex
function. However, the Bethe entropy function of factor graphs with cycles
does in general not exhibit this property. In fact, in this paper we show that
the factor graph associated with any code picked from Gallager's ensemble of
$(\dleft,\dright)$-regular low-density parity-check codes has a Bethe entropy
function that is convex in certain regions of its domain if the ensemble is
such that the minimum Hamming distance of its codes grows (with high
probability) linearly with the block length. This means that there is a
trade-off between two desirable objectives: on the one hand to pick a code
from a code ensemble with linearly growing minimum Hamming distance, on the
other hand to pick a code whose factor graph has a concave Bethe entropy
function, \ie, a convex Bethe free energy function.

\subsection{Related Work}
\label{sec:related:work:1}

Let us briefly discuss some work that is related to the content of this paper.
\begin{itemize}

\item Of course, what is called the Bethe approximation in the context of
  factor graphs has a long history in physics and goes back to ideas that were
  presented in a 1935 paper by Bethe~\cite{Bethe:35:1} (see also the 1936
  paper by Peierls~\cite{Peierls:36:1}). Bethe's approximation therein was
  mostly an assumption about the conditional independence between different
  sites in a crystalline alloy. Kurata, Kikuchi, and
  Watari~\cite{Kurata:Kikuchi:Watari:53:1} later on pointed out that this
  approximation is exact on what they called a ``Bethe lattice.'' (For further
  information on these and related topics in physics, we recommend, \eg,
  \cite{Eggarter:74:1, Thorpe:82:1, Baxter:82:1}.)

  Let us comment on the Bethe lattice of a lattice. Consider a lattice
  $\graph{L}$ and a factor graph $\graphN$. In factor-graph language, if
  $\graph{L}$ corresponds to $\graphN$, then the Bethe lattice~$\graph{\hat
    L}$ of $\graph{L}$ corresponds to the universal cover $\graph{\hat N}$ of
  $\graphN$, \ie, the limit of a computation tree of $\graphN$ with arbitrary
  root in the limit of infinitely many iterations. With the help of
  $\graph{\hat N}$ it is possible to give a combinatorial characterization of
  the Bethe partition function of $\graphN$ as some suitably normalized sum of
  the global function of $\graph{\hat N}$ over all its
  configurations. However, to make this rigorous, one has to formulate
  $\graph{\hat N}$ as a suitable limit of computation trees. This is not too
  difficult for very regular factor graphs or for factor graphs with suitable
  correlation decay properties. However, for general factor graphs the limit
  of the above-mentioned normalized sum is rather non-trivial. This difficulty
  is not quite surprising given, among other reasons, the fact that the SPA
  may asymptotically exhibit many different types of behaviors (fixed point,
  periodic, or even ``chaotic''), the fact that copies of factor-graph nodes
  have different multiplicities in finite-size computation trees (see, \eg,
  \cite{Frey:Koetter:Vardy:01:1}), or the fact that the fraction of leaf nodes
  among all nodes in a computation is non-vanishing in the limit of infinitely
  many iterations. Clearly, the expression in~\eqref{eq:ZBethe:intro:1} also
  contains a limit, however, in our opinion that limit is significantly
  simpler. Moreover, many effects that are responsible for the similarities
  and differences between the Bethe and the Gibbs partition function are
  already visible in finite graph covers with small cover degree.

\item Some computations that we will perform in the present paper are very
  similar to the computations that are necessary to derive the asymptotic
  growth rate of the average Hamming weight enumerator of proto-graph-based
  ensembles of (generalized) low-density parity-check (LDPC)
  codes~\cite{Fogal:McEliece:Thorpe:05:1, Divsalar:Jones:Dolinar:Thorpe:05:1,
    Fu:Anastasopoulos:07:1, Ravazzi:Fagnani:09:1, AbuSurra:Divsalar:Ryan:11:1,
    Flanagan:Paolini:Chiani:Fossorier:11:1, Divsalar:Dolecek:12:1} (see also
  the earlier work on uniform interleavers~\cite{Benedetto:Montorsi:96:1,
    Benedetto:Divsalar:Montorsi:Pollara:98:1}). However, besides some brief
  mention of the fundamental polytope in~\cite{Fogal:McEliece:Thorpe:05:1},
  these papers do not seem to elaborate on the connection of their results to
  the Bethe entropy function. (An exception is the very recent
  paper~\cite{Divsalar:Dolecek:12:1}.)

\item As already stated in the above introduction, the
  papers~\cite{Koetter:Vontobel:03:1, Vontobel:Koetter:05:1:subm} investigated
  some fundamental limitations that locally operating algorithms have compared
  to globally operating algorithms. It is worthwhile to point out that
  Angluin~\cite{Angluin:80:1}, in a paper that was published in 1980, used a
  very similar global-\versus-local argument to characterize networks of
  processors. Although on a philosophical level the starting point of her
  argument is very akin to the one in~\cite{Koetter:Vontobel:03:1,
    Vontobel:Koetter:05:1:subm}, her conclusions are quite different in nature
  (which is not so surprising given the differences between her setup and our
  setup).

  Much closer to the approach in~\cite{Koetter:Vontobel:03:1,
    Vontobel:Koetter:05:1:subm} is the relatively recent paper by Ruozzi
  \etal~\cite{Ruozzi:Thaler:Tatikonda:09:1} which showed that results on the
  limitations of locally operating algorithms on Gaussian graphical models (in
  particular results based on the concept of
  walk-summability~\cite{Malioutov:Johnson:Willsky:06:1}) can be re-derived by
  studying graph covers.

\item There are other papers where the Bethe approximation plays a central
  role in characterizing the suboptimal behavior of locally operating
  algorithms, in particular let us mention~\cite{Mooij:Kappen:05:1,
    Heskes:06:1, Chertkov:Chernyak:06:1, Walsh:Regalia:Johnson:06:1,
    Regalia:Walsh:07:1, Mooij:Kappen:07:1}.

\item Although some concepts in the present paper are evocative of concepts of
  the replica method (see, \eg, \cite{Mezard:Parisi:Virasoro:87:1}
  and~\cite[Chapter~8]{Mezard:Montanari:09:1},
  \cite[Appendix~I]{Tanaka:02:1}), there are also stark differences, as will
  be discussed in Section~\ref{sec:replica:method:1}. However, inspired by an
  earlier version of the present paper, Mori~\cite{Mori:11:1} recently showed
  an alternative (and simpler) approach to some computations that are done in
  the context of the replica method. See also the follow-up
  papers~\cite{Mori:Tanaka:12:1, Mori:Tanaka:12:2}.

\item Let $\matrtheta$ be a non-negative square matrix and let
  $\perm(\matrtheta)$ be the permanent of this matrix~\cite{Minc:78}. In the
  paper~\cite{Vontobel:11:3:subm}, many concepts of the present paper are
  specialized to a certain graphical model $\graphN(\matrtheta)$ for which
  $\ZGibbs\bigl( \graphN(\matrtheta) \bigr) = \perm(\matrtheta)$ holds. The
  reformulation of the Bethe partition function for such graphical models was
  subsequently used by Smarandache~\cite{Smarandache:11:1:subm} to give a
  proof for a conjecture about pseudo-codewords of LDPC codes.

  After the initial submission of the present paper and
  of~\cite{Vontobel:11:3:subm}, we became aware of a paper by Greenhill,
  Janson, and Ruci{\'n}ski~\cite{Greenhill:Janson:Rucinski:10:1} that, in the
  language of the present paper, introduces a graphical model
  $\graphN'(\matrtheta)$ for which $\ZGibbs\bigl( \graphN'(\matrtheta) \bigr)
  = \perm(\matrtheta)$ holds and for which they compute high-order
  approximations of $\ZBetheM{M}\bigl( \graphN'(\matrtheta) \bigr)$. However,
  note that $\graphN'(\matrtheta)$ is in general different from
  $\graphN(\matrtheta)$. A detailed discussion of connections between
  $\graphN(\matrtheta)$ and $\graphN'(\matrtheta)$ is given
  in~\cite[Section~VII.E]{Vontobel:11:3:subm}.

\item Based on the reformulation of the Bethe partition function in an earlier
  version of the present paper, Watanabe~\cite{Watanabe:11:1} stated a
  conjecture about the relationship of the number of independent sets of a
  graph and its Bethe approximation (along with other similar conjectures),
  and Ruozzi~\cite{Ruozzi:12:1} proved that the Gibbs partition function of a
  graphical model with log-supermodular function nodes is always lower bounded
  by its Bethe partition function, thereby proving a conjecture by Sudderth,
  Wainwright, and Willsky~\cite{Sudderth:Wainwright:Willsky:07:1}.

\item Graph covers were used in the recent
  paper~\cite{Parvaresh:Vontobel:12:1} to explain why the Bethe partition
  function is very close to the Gibbs partition function for certain graphical
  models that appear in the context of constrained coding. They were also used
  in~\cite{Vontobel:12:1} to prove properties of the Bethe approximation of
  the so-called pattern maximum likelihood distribution.

\end{itemize}

\subsection{Overview of the Paper}

This paper is structured as follows. We conclude this first section with a
subsection on notations and definitions. Then, in
Section~\ref{sec:normal:factor:graphs:1} we review the basics of normal factor
graphs, \ie, the type of factor graphs that we will use in this paper, and in
Section~\ref{sec:FGibbs:and:ZGibbs:1} we discuss the Gibbs free energy
function and related functions. The Gibbs free energy function is again the
topic of Section~\ref{sec:Gibbs:free:energy:arises:naturally:1}, where we
present a simple setup where this function arises naturally. Afterwards, in
Section~\ref{sec:local:marginal:polytope:and:Bethe:approximation:1} we move on
to introduce the Bethe approximation and the functions that come with it.

After reviewing the main facts about graph covers in
Section~\ref{sec:finite:graph:covers:1}, we come to the main part of this
paper, namely Section~\ref{sec:counting:in:finite:graph:covers:1} where we
present the promised combinatorial characterization of the Bethe entropy
function and the Bethe partition function.

In contrast to the previous sections that considered a general factor graph
setup, the next three sections focus on factor graphs that appear in coding
theory. Namely, Section~\ref{sec:normal:factor:graphs:for:channel:coding:1}
reviews some relevant concepts, Section~\ref{sec:graph:cover:decoding:1}
discusses blockwise and symbolwise graph-cover decoding, and
Section~\ref{sec:influence:minimum:Hamming:distance:1} investigates the
influence of the minimum Hamming distance of a code upon the Bethe entropy
function of its factor graph.

Finally, the paper is concluded in Section~\ref{sec:conclusions:1}. The longer
proofs of the lemmas and theorems in the main text are collected in the
appendices.

\subsection{Basic Notations and Definitions}
\label{sec:notation:1}

This subsection discusses the most important notations that will be used in
this paper. More notational definitions will be given in later sections.

We let $\Z$, $\Zp$, $\Zpp$, $\R$, $\Rp$, $\Rpp$, and $\GF{2}$ be,
respectively, the ring of integers, the set of non-negative integers, the set
of positive integers, the field of real numbers, the set of non-negative real
numbers, the set of positive real numbers, and the Galois field of size
two. Scalars are denoted by non-boldface characters, whereas vectors and
matrices by boldface characters. All logarithms in this paper will be natural
logarithms, and so entropies will be measured in nats. (The only exception are
figures where entropies are shown in bits.) As usually done in information
theory, we define $\log(0) \defeq -\infty$ and $0 \cdot \log(0) \defeq 0$.

Sets are denoted by calligraphic letters, and the size of a finite set $\setS$
is written like $\card{\setS}$. The convex hull and the conic
hull~\cite{Boyd:Vandenberghe:04:1} of some set $\setR \in \R^n$ are,
respectively, denoted by $\convhull(\setR)$ and $\conichull(\setR)$.

We use square brackets in two different ways. Namely, for any $L \in \Zpp$ we
define $[L] \defeq \{ 1, \ldots, L \}$, and for any statement $S$ we follow
Iverson's convention by defining $[S] \defeq 1$ if $S$ is true and $[S] \defeq
0$ otherwise.

Finally, for a finite set $\setS$, we define $\setPi_{\setS}$ to be the set of
vectors representing probability mass functions over $\setS$, \ie,
\begin{align*}
  \setPi_{\setS}
    &\defeq 
       \left\{
           \vect{p} = \big( p_s \big)_{s \in \setS}
         \ \middle| \ 
           p_s \geq 0 \text{ for all $s \in \setS$}, \ 
           \sum_{s \in \setS} p_s = 1
         \right\}.
\end{align*}

\section{Normal Factor Graphs}
\label{sec:normal:factor:graphs:1}

Factor graphs are a convenient way to represent multivariate
functions~\cite{Kschischang:Frey:Loeliger:01}. In this paper we use a variant
called normal factor graphs (NFGs)~\cite{Forney:01:1} (also called
Forney-style factor graphs~\cite{Loeliger:04:1}), where variables are
associated with edges.

The key aspects of an NFG are best explained with the help of an
example.

\begin{figure}
  \begin{center}
    \epsfig{file=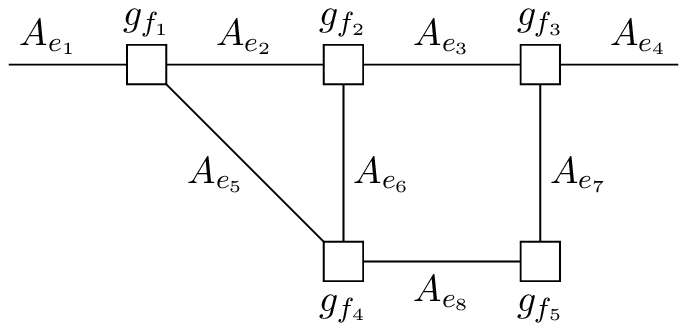, scale=1}
  \end{center}
  \caption{NFG $\graphN$ used in Example~\ref{example:simple:ffg:1}.}
  \label{fig:example:simple:ffg:1}
\end{figure}

\begin{Example}
  \label{example:simple:ffg:1}

  Consider the multivariate function
  \begin{align*}
    g(a_{e_1}, \ldots, a_{e_8})
      &\defeq
         g_{f_1}(a_{e_1}, a_{e_2}, a_{e_5})
         \cdot
         g_{f_2}(a_{e_2}, a_{e_3}, a_{e_6}) \\
      &\quad\ 
         \cdot
         g_{f_3}(a_{e_3}, a_{e_4}, a_{e_7})
         \cdot
         g_{f_4}(a_{e_5}, a_{e_6}, a_{e_8}) \\
      &\quad\ 
         \cdot
         g_{f_5}(a_{e_7}, a_{e_8}),
  \end{align*}
  where the so-called global function $g$ is the product of the so-called
  local functions $g_{f_1}$, $g_{f_2}$, $g_{f_3}$, $g_{f_4}$, and
  $g_{f_5}$. The decomposition of this global function as a product of local
  functions can be depicted with the help of an NFG $\graphN$ as shown in
  Fig.~\ref{fig:example:simple:ffg:1}. In particular, the NFG $\graphN$
  consists of
  \begin{itemize}
    
  \item the function nodes $f_1$, $f_2$, $f_3$, $f_4$, and $f_5$;

  \item the half-edges $e_1$ and $e_4$ (sometimes also called ``external
    edges'');

  \item the full-edges $e_2$, $e_3$, $e_5$, $e_6$, $e_7$, and $e_8$ (sometimes
    also called ``internal edges'').

  \end{itemize}
  In general,
  \begin{itemize}

  \item a function node $f$ represents the local function $g_f$;

  \item with an edge $e$ we associate the variable $A_e$ (note that a
    realization of the variable $A_e$ is denoted by $a_e$);

  \item an edge $e$ is incident on a function node $f$ if and only if $a_e$
    appears as an argument of the local function $g_f$.

  \end{itemize}
  Note that the NFG $\graphN$ contains three cycles, one involving the edges
  $e_2$, $e_5$, $e_6$, one involving the edges $e_3$, $e_6$, $e_7$, $e_8$, and
  one involving the edges $e_2$, $e_3$, $e_5$, $e_7$, $e_8$. As is well known
  from the literature on graphical models, and as we can also see from other
  parts of this paper, the existence/absence of cycles in an NFG has
  significant implications for its properties, in particular with respect to
  the behavior of locally operating algorithms like the max-product algorithm
  and the sum-product algorithm.  \exampleend
\end{Example}

We now present the general definition of an NFG that we will use in this
paper.

\begin{Definition}
  \label{def:normal:factor:graph:1}

  An NFG $\graphN(\setF, \setE, \setA, \setG)$ consists of the following
  objects.
  \begin{itemize}
   
  \item A graph $(\setF, \setE)$ with vertex set $\setF$ (also known as the
    function node set) and with edge set $\setE = \setEhalf \cup \setEfull$,
    where $\setEhalf$ and $\setEfull$ represent, respectively, the set of
    half-edges and the set of full-edges.

  \item A collection of alphabets $\setA \defeq \{ \setA_e \}_{e \in \setE}$,
    where the alphabet $\setA_e$ is associated with the edge $e \in \setE$. In
    the following, with a slight abuse of notation, $\setA$ will also stand
    for the set $\setA = \prod_{e \in \setE} \setA_e$, \ie, the Cartesian
    product of the alphabets $\{ \setA_e \}_{e \in \setE}$. Moreover, we also
    define the sets $\setAhalf \defeq \prod_{e \in \setEhalf} \setA_e$ and
    $\setAfull \defeq \prod_{e \in \setEfull} \setA_e$. Clearly, $\setA =
    \setAhalf \times \setAfull$.
  
  \item A collection of functions $\setG \defeq \{ g_f \}_{f \in \setF}$
    (called local functions), where the function $g_f$ is associated with the
    function node $f \in \setF$ and further specified below. \defend

  \end{itemize}
\end{Definition}

\begin{Definition}
  \label{def:normal:factor:graph:2}

  Given an NFG $\graphN(\setF,\setE, \setA, \setG)$, we make
  the following definitions.
  \begin{itemize}

  \item For every $f \in \setF$, we define $\setE_f$ to be the set of edges
    incident on $f$.

  \item A vector $\va$ in the set $\setA$, \ie,
    \begin{align*}
      \va 
        &\defeq
           (a_e)_{e \in \setE}
         \in \setA,
    \end{align*}
    will be called a configuration of the NFG. For a given
    vector $\va$, we also define for every $f \in \setF$ the sub-vector
    \begin{align*}
      \va_f
        &\defeq
           (a_{f,e})_{e \in \setE_f}
         \defeq
           (a_e)_{e \in \setE_f}.
    \end{align*}
    Note that we will also use the notation $\va_f = (a_{f,e})_{e \in
      \setE_f}$ when there is not necessarily an underlying configuration
    $\va$ of the whole NFG.

  \item For every $f \in \setF$, the local function $g_f$ is an arbitrary
    mapping
    \begin{align*}
      g_f: \
        \prod_{e \in \setE_f}
          \setA_e
            \to \R,
       \quad
       \va_f
         \mapsto
           g_f(\va_f).
    \end{align*}

  \item For every $f \in \setF$, we define the function node alphabet
    $\setA_f$ to be the set
    \begin{align*}
      \setA_f
        &\defeq
            \left\{
              \va_f \in \prod_{e \in \setE_f} \setA_e
            \
            \middle|
            \
              g_f(\va_f) \neq 0
            \right\}.
    \end{align*}
    This set is also known as the local constraint code of the function node
    $f$.

  \item The global function $g$ is defined to be the mapping
    \begin{align*}
      g: \
         \setA
           \to \R,
         \quad
         \va
           \mapsto
             \prod_f
               g_f(\va_f).
    \end{align*}
    Equivalently, in the case where we distinguish between half- and
    full-edges, $g$ represents the mapping
    \begin{align*}
      g: \
         \setAhalf \times \setAfull
           \to \R,
         \quad
         (\vahalf, \vafull)
           \mapsto
             \prod_f
               g_f(\va_f).
    \end{align*}

  \item A configuration $\va$ with $g(\va) \neq 0$ is called a valid
    configuration. The set of all valid configurations, \ie,
    \begin{align*}
      \codeCedge
        &\defeq
           \left\{
             \va \in \setA
           \, \middle| \!\!
             \begin{array}{rl}
               \text{$a_e \in \setA_e$,} & \!\!\!\!
               \text{$e \in \setE$} \\
               \text{$\va_f \in \setA_f$,} & \!\!\!\!
               \text{$f \in \setF$}
             \end{array} \!\!
           \right\},
    \end{align*}
    is called the global behavior of $\graphN$, the full behavior of
    $\graphN$, or the edge-based code realized by $\graphN$.

  \item The projection of $\codeCedge$ onto $\setEhalf$, \ie,
    \begin{align*}
      \codeCedgehalf
        &\defeq
           \big\{ 
             (c_e)_{e \in \setEhalf}
           \ \big| \ 
             \vc \in \codeCedge
           \big\} \\
        &= \left\{
             \vahalf \in \setAhalf
           \, \middle| \!\!
             \begin{array}{c}
               \text{there exists an $\vafull \in \setAfull$} \\ 
               \text{such that $(\vahalf,\vafull) \in \codeCedge$}
             \end{array} \!\!
           \right\},
    \end{align*}
    is called the half-edge-based code realized by $\graphN$.

  \end{itemize}
\end{Definition}

A comment concerning the above definition: in the following, when confusion
can arise what NFG an object is referring to, we will use more
precise notations like $g_{\graphN}$, $\codeCedge(\graphN)$, \etc, instead of
$g$, $\codeCedge$, \etc.

\begin{figure}
  \begin{center}
    \epsfig{file=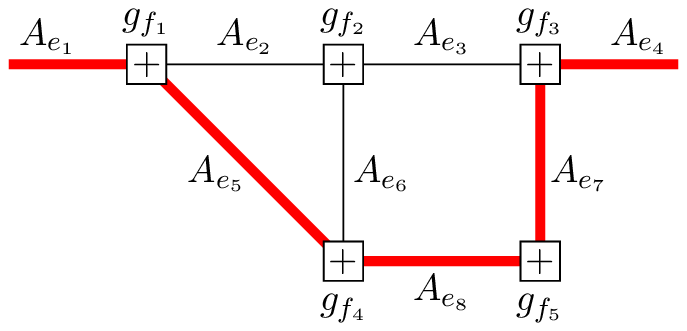, scale=1}
  \end{center}
  \caption{Configuration $\va$ on the NFG $\graphN$ from
    Fig.~\ref{fig:example:simple:ffg:1}: for every $e \in \setE$, if $a_e = 0$
    then the edge $e$ is thin and in black, whereas if $a_e = 1$ then the edge
    $e$ is thick and in red. (See Example~\ref{example:simple:ffg:1:cont:0}
    for more details.)}
  \label{fig:example:simple:ffg:1:mod:2}
\end{figure}

\begin{Example}
  \label{example:simple:ffg:1:cont:0}

  Consider again the NFG $\graphN$ that is discussed in
  Example~\ref{example:simple:ffg:1} and depicted in
  Fig.~\ref{fig:example:simple:ffg:1}. It is shown again in
  Fig.~\ref{fig:example:simple:ffg:1:mod:2}. Assume that its details are as
  follows:
  \begin{itemize}

  \item The variable alphabets are $\setA_e = \{ 0, 1 \}$, $e \in \setE$.

  \item The local functions are
    \begin{align*}
      g_f(\va_f)
        &= \begin{cases}
             1 & \text{if $\sum_{e \in \setE_j} a_{f,e} = 0 \ (\mathrm{mod} \ 2)$} \\
             0 & \text{otherwise}
           \end{cases},
           \quad f \in \setF.
     \end{align*}
  \end{itemize}
  Therefore the local constraint codes are
  \begin{align*}
    \setA_f
      &= \left\{
           \va_f \in \{ 0, 1 \}^{|\setE_f|}
         \
         \middle|
         \
           \sum_{e \in \setE_f}
             a_{f,e}
             = 0 \ (\mathrm{mod} \ 2)
         \right\}, \ f \in \setF,
  \end{align*}
  \ie, single parity-check codes of length $|\setE_f|$.

  The configuration shown in Fig.~\ref{fig:example:simple:ffg:1:mod:2}
  corresponds to the variable assignment
  \begin{align*}
    \va
      &= (a_{e_1}, a_{e_2}, a_{e_3}, a_{e_4}, a_{e_5}, a_{e_6}, a_{e_7}, a_{e_8}) \\
      &= (1,0,0,1,1,0,1,1).
  \end{align*}
  The configuration $\va$ has the following sub-vectors
  \begin{alignat*}{2}
    \va_1
      &\!=\! (a_{e_1}, a_{e_2}, a_{e_5})
      &  =  & (1,0,1), \ 
    \va_2
       \!=\! (a_{e_2}, a_{e_3}, a_{e_6})
       \!=\! (0,0,0), \\
    \va_3
      &\!=\! (a_{e_3}, a_{e_4}, a_{e_7})
      &  =  & (0,1,1), \ 
    \va_4
       \!=\! (a_{e_5}, a_{e_6}, a_{e_8})
       \!=\! (1,0,1), \\
    \va_5
      &\!=\!  (a_{e_7}, a_{e_8})
      &  =  & (1,1).
  \end{alignat*}
  Because $\va_f \in \setA_f$ for all $f \in \setF$, the configuration $\va$
  is a valid configuration. One can easily check that the global function
  value of $\va$ is $g(\va) = 1$. (In fact, for this NFG the global function
  value of all valid configurations is $1$.)

  The set of all valid configurations of $\graphN$ turns out to be
  \begin{align}
    \codeC
      &= \left\{
           \begin{array}{c}
             (0,0,0,0,0,0,0,0), \
             (0,1,0,0,1,1,0,0), \\
             (0,0,1,0,0,1,1,1), \
             (0,1,1,0,1,0,1,1), \\
             (1,0,0,1,1,0,1,1), \
             (1,1,0,1,0,1,1,1), \\
             (1,0,1,1,1,1,0,0), \
             (1,1,1,1,0,0,0,0)\phantom{,}
           \end{array}
         \right\},
           \label{eq:example:simple:ffg:1:valid:configurations:1}
  \end{align}
  and its projection unto $\setEhalf = \{ e_1, e_4 \}$ is
  \begin{align*}
    \codeCedgehalf
      &= \big\{
           (0,0), \ (1,1)
         \big\}.
  \end{align*}

  \exampleend

\end{Example}

Although the definition of NFGs requires the global function to be such that
all variables are arguments of at most two local functions, this does not
really impose a major restriction on the expressive power of NFGs. Namely,
this requirement can easily be circumvented by replacing a global function by
a suitably modified global function that contains additional variables and
additional local functions. (We refer to~\cite{Forney:01:1, Loeliger:04:1} for
further details.)

In the following, when there is no ambiguity, we will use the short-hands
$\sum_{\vc}$\,, $\sum_{\va}$\,, $\sum_{\va_f}$\,, and $\sum_{a_e}$ for,
respectively, $\sum_{\vc \in \setC}$\,, $\sum_{\va \in \setA}$\,,
$\sum_{\va_f}$\,, and $\sum_{a_e \in \setA_e}$.

\begin{Assumption}
  For the rest of the paper, we assume that for all $f \in \setF$, the
  co-domain of the local function $g_f$ is $\Rp$, \ie, the set of non-negative
  real numbers. Consequently, for every $\va \in \setA$ it holds that $g(\va)
  \in \Rp$.  \assumptionend
\end{Assumption}

The above assumption is not a significant restriction since many interesting
problems can be cast in terms of an NFG that satisfies this assumption. As
will be evident from the upcoming sections, the main reason for imposing the
above assumption is the fact that the definitions of the Gibbs average energy
function (and therefore the Gibbs free energy function) and the Bethe average
energy function (and therefore the Bethe free energy function) contain
expressions that involve the logarithm of the global and the local functions.

However, we stress that the above assumption is \emph{not} necessary for
defining the Gibbs entropy function and the Bethe entropy function. Therefore
the upcoming results on these functions hold for the more general setup of
Definition~\ref{def:normal:factor:graph:2}.

\section{The Gibbs Free Energy Function and \\
                 the Gibbs Partition Function}
\label{sec:FGibbs:and:ZGibbs:1}

This section reviews the concept of the Gibbs free energy function of an NFG,
along with some related functions. The temperature $T$ appears in them as a
parameter.

\begin{Assumption}
  \label{assumption:temperature:1}

  Throughout this paper, the temperature $T$ will be some fixed non-negative
  real number, \ie, $T \in \Rp$. (If a definition requires $T \in \Rpp$, then
  this will be pointed out.) \assumptionend
\end{Assumption}

The Gibbs free energy function is defined such that its minimal value, along
with the location of its minimal value, encode important information about the
global function that is represented by the NFG. In particular,
for $T \in \Rpp$ the minimal value equals the temperature $T$ times the
negative logarithm of the partition function, where the partition function is
the sum of the $(1/T)$-th power of the global function over all
configurations.

\begin{Definition}
  \label{def:F:Gibbs:1}

  Consider an NFG $\graphN(\setF,\setE, \setA, \setG)$. For any temperature $T
  \in \Rp$, the Gibbs free energy function associated with $\graphN$ is
  defined to be (see, \eg, \cite{Yedidia:Freeman:Weiss:05:1}) 
  \begin{align*}
    \FGibbs: \ 
      &\setPi_{\codeCedge}
         \to 
         \R, \quad
       \vp
         \mapsto 
           \UGibbs(\vp)
           -
           T
           \cdot
           \HGibbs(\vp),
  \end{align*}
  where
  \begin{alignat*}{2}
    \UGibbs: \ 
       &
       \setPi_{\codeCedge}
         \to 
         \R, \quad
       \vp
         &
         \, \mapsto\, 
         & 
         -
         \sum_{\vc}
           p_{\vc}
           \cdot
           \log
             \big(
               g(\vc)
             \big), \\
    \HGibbs: \ 
       &
       \setPi_{\codeCedge}
         \to 
         \R, \quad
       \vp
         &
         \, \mapsto\,  
         &
         -
         \sum_{\vc}
           p_{\vc}
           \cdot
           \log
             \big(
               p_{\vc}
             \big).
  \end{alignat*}
  Here, $\UGibbs$ is called the Gibbs average energy function and $\HGibbs$ is
  called the Gibbs entropy function.
  
  Moreover, for $T \in \Rpp$, the (Gibbs) partition function associated with
  $\graphN$ is defined to be (see, \eg, \cite{Yedidia:Freeman:Weiss:05:1}) 
  \begin{align}
    \ZGibbs
      &\defeq
         \sum_{\va \in \setA}
           g(\va)^{1/T}
       = \sum_{\vc \in \codeCedge}
           g(\vc)^{1/T}.
             \label{eq:Z:Gibbs:1}
  \end{align}
  \vskip-0.40cm
  \defend
\end{Definition}

Note that ``function'' in ``partition function'' refers to the fact that the
expression in~\eqref{eq:Z:Gibbs:1} typically is a function of some parameters
like the temperature $T$. (A better word for ``partition function'' would
possibly be ``partition sum'' or ``state sum,'' which would more closely
follow the German ``Zustandssumme'' whose first letter is used to denote the
partition function.)

\begin{Lemma}
  \label{lemma:F:Gibbs:minimum:1}

  Consider an NFG $\graphN(\setF,\setE, \setA, \setG)$. For $T \in \Rp$, the
  function $\FGibbs(\vp)$ is convex in $\vp$. Moreover, for $T \in \Rpp$, the
  function $\FGibbs(\vp)$ is minimized by $\vp = \vp^{*}$, where
  \begin{align}
    \vp^{*}_{\vc}
      &\defeq
         \frac{g(\vc)^{1/T}}
              {\ZGibbs},
           \quad \vc \in \codeCedge.
             \label{eq:F:Gibbs:minimum:location:1}
  \end{align}
  At its minimum, $\FGibbs$ takes on the value
  \begin{align}
    \FGibbs(\vp^{*})
      &= -
         T
         \cdot
         \log
           (
             \ZGibbs
           ),
             \label{eq:Z:Gibbs:2}
  \end{align}
  which is also known as the Helmholtz free energy.
\end{Lemma}

\begin{Proof}
  For $T = 0$ we have $\FGibbs(\vp) = \UGibbs(\vp)$, $\vp \in
  \setPi_{\codeC}$. The convexity of $\FGibbs$ then follows from the convexity
  of $\UGibbs$, which follows from the fact that $\UGibbs$ is a linear
  function of its argument.

  For $T \in \Rpp$, the Gibbs free energy function $\FGibbs$ can be expressed
  in terms of a relative entropy functional, namely
  \begin{align*}
    \FGibbs(\vp)
      &= T
         \cdot
         D
           \left(
               (p_{\vc})_{\vc \in \codeCedge}
             \ \middle\Vert \ 
               \left( \frac{g(\vc)^{1/T}}{\ZGibbs} \right)_{\!\! \vc \in \codeCedge}
           \right)
         -
         T
         \cdot
         \log(\ZGibbs).
  \end{align*}
  The statements in the lemma then follow easily from standard properties of
  the relative entropy functional (see, \eg, \cite{Cover:Thomas:91}).
\end{Proof}

Note that, with appropriate care, results involving the Gibbs free energy
function and related functions at temperature $T = 0$ can be recovered from
studying the case $T \in \Rpp$ and taking the limit $T \downarrow 0$. However,
as mentioned in Assumption~\ref{assumption:temperature:1}, in this paper the
temperature $T$ is a fixed parameter and we will not consider such limits.

Let us briefly discuss a variant of the above Gibbs free energy
function. Namely, for some $\vchalf \in \codeCedgehalf$, we will say that $\vp
\in \setPi_{\codeCedge}$ is compatible with $\vchalf$ if
\begin{align*}
  \vp_{\vc'} \
    &\begin{cases}
       \ \geq 0
         & \text{(for all $\vc' \in \codeCedge$ with $\vchalf' = \vchalf$)}, \\
       \ = 0
       & \text{(for all $\vc' \in \codeCedge$ with $\vchalf' \neq \vchalf$)}
     \end{cases}.
\end{align*}
With this definition, as an alternative to the minimization problem
in Lemma~\ref{lemma:F:Gibbs:minimum:1}, we can consider a minimization problem
where we minimize over all $\vp$ that are compatible with some given
$\vchalf$. Technically, we can accomplish this by defining a modified Gibbs
free energy function $\FGibbs'$ that equals the Gibbs free energy function
$\FGibbs$ for $\vp$'s which are compatible with this $\vchalf$, and that is
infinite otherwise.

\begin{Definition}
  \label{def:F:Gibbs:2}

  Consider an NFG $\graphN(\setF,\setE, \setA, \setG)$. For any temperature $T
  \in \Rp$, the modified Gibbs free energy function associated with $\graphN$
  is defined to be
  \begin{alignat*}{2}
    \FGibbs': \
      &\codeCedgehalf \times \setPi_{\setC}
      & \ \to \
      & \R  \cup \{ +\infty \} \\ 
      &(\vchalf, \vp)
      & \ \mapsto \ 
      &\begin{cases}
         \FGibbs(\vp) & \text{($\vp$ is compatible with $\vchalf$)} \\
         +\infty      & \text{(otherwise)}
       \end{cases}
  \end{alignat*}
  Moreover, for $T \in \Rpp$, the modified (Gibbs) partition function
  associated with $\graphN$ is defined to be
  \begin{align*}
    \ZGibbs': \ 
      &\codeCedgehalf
       \to
       \R, \quad
    \vchalf
       \mapsto\sum_{\vc' \in \codeC: \, \vchalf' = \vchalf}
         g(\vc')^{1/T}.
  \end{align*}
\end{Definition}

\begin{Lemma}
  \label{lemma:F:Gibbs:minimum:2}

  Consider an NFG $\graphN(\setF,\setE, \setA, \setG)$ and fix some $\vchalf
  \in \codeCedgehalf$. For $T \in \Rp$, the function $\FGibbs'(\vchalf, \vp)$
  is convex in $\vp$. Moreover, for $T \in \Rpp$, the function
  $\FGibbs'(\vchalf, \vp)$ is minimized by $\vp = \vp^{*}$, where
  \begin{align*}
    \vp^{*}_{\vc'}
      &\defeq
         \begin{cases}
           \frac{g(\vc')^{1/T}}
                {\ZGibbs'(\vchalf)},
             & \text{($\vc' \in \setC$ with $\vchalf' = \vchalf$)}, \\
           0 & \text{(otherwise)}
         \end{cases}.
  \end{align*}
  At its minimum, $\FGibbs'(\vchalf, \, \cdot \,)$ takes on the value
  \begin{align*}
    \FGibbs'(\vchalf, \vp^{*})
      &= -
         T
         \cdot
         \log
           \big(
             \ZGibbs'(\vchalf)
           \big).
  \end{align*}
\end{Lemma}

\begin{Proof}
  Similar to the proof of Lemma~\ref{lemma:F:Gibbs:minimum:1}.
\end{Proof}

Clearly, if the NFG $\graphN$ does not contain any half-edges, then the
modified Gibbs partition function is essentially a scalar. Note that this
modified Gibbs partition function (at temperature $T = 1$) equals what
Forney~\cite{Forney:11:1} calls the partition function of an NFG and what
Al-Bashabsheh and Mao~\cite{AlBashabsheh:Mao:11:1} call the exterior function
of an NFG.

\section{Why the Gibbs Free Energy Function Arises Rather Naturally}
\label{sec:Gibbs:free:energy:arises:naturally:1}

Of course, besides the function Gibbs free energy function $\FGibbs$, there
are many ways to formulate a function $F: \, \setPi_{\codeC} \to \R$ such that
\begin{itemize}

\item the minimum of $F(\vp)$ is achieved at $\vp = \vp^{*}$, where $\vp^{*}$
  equals the expression in~\eqref{eq:F:Gibbs:minimum:location:1},

\item and such that the minimum value $F(\vp^{*})$ equals the expression
  in~\eqref{eq:Z:Gibbs:2}.

\end{itemize}
The goal of this section is to discuss a setup where $\FGibbs$ arises rather
naturally as a function that has these properties. This section also reviews
some important concepts from the method of types (see, \eg,
\cite{Csiszar:Korner:81, Cover:Thomas:91}), many of which inspired the
concepts that were briefly mentioned in Section~\ref{sec:introduction:1} and
that will be introduced more thoroughly in
Sections~\ref{sec:finite:graph:covers:1}
and~\ref{sec:counting:in:finite:graph:covers:1}. Note that we do not make any
novelty claims for the observations discussed in the present section.

Throughout this section we consider an NFG $\graphN(\setF,\setE, \setA,
\setG)$. For simplicity we consider only the case where $\setA_e = \{ 0, 1 \}
\subset \R$ for all $e \in \setE$. Moreover, throughout this discussion, the
temperature will be fixed to $T = 1$. (Both assumptions are not critical, and
with a suitable formalism, the results can easily be generalized.)

We start by defining a probability mass function $P_{\vC}(\vc)$ on
$\codeCedge$ that is induced by the global function on $\graphN$, namely
\begin{align*}
  P_{\vC}(\vc)
    &\defeq
       \frac{g(\vc)}
            {\ZGibbs},
            \quad \vc \in \codeCedge,
\end{align*}
where $\ZGibbs$ is defined in~\eqref{eq:Z:Gibbs:1}. 
Now, assume that for every $e \in \setE$ and $a_e \in \setA_e$ we want to
compute the marginal
\begin{align*}
  P_{A_e}(a_e)
    &\defeq
       \sum_{\vc: \, c_e = a_e}
         P_{\vC}(\vc).
\end{align*}
(This computational problem comes up, for example, as part of symbolwise
maximum a-posterior decoding, \confer~Section~\ref{sec:SMAPD:1}.) To that end,
define the vector $\veta \defeq (\eta_{e,1})_{e \in \setE}$ with components
$\eta_{e,1} \defeq P_{A_e}(1)$. Clearly, once we have computed the vector
$\veta$, we have all the desired marginals because $P_{A_e}(0) = 1 -
P_{A_e}(1) = 1 - \eta_{e,1}$, $e \in \setE$. It can easily be verified that
$\veta$ satisfies
\begin{align}
  \veta
    &= \sum_{\vc \in \codeCedge}
         P_{\vC}(\vc)
         \cdot
         \vc
     = \Expec[\vC].
         \label{eq:veta:expression:1}
\end{align}

In practice, the sum in~\eqref{eq:veta:expression:1} is very often intractable
because the set $\codeCedge$ is very large. However,
\eqref{eq:veta:expression:1} shows that $\veta$ is some expectation value and
so we can try to approximate it by stochastic averaging. Therefore, let
\begin{align*}
  \vcoldsample{1}, \ \vcoldsample{2}, \ \ldots, \ \vcoldsample{M}
\end{align*}
be $M$ i.i.d.\ sample vectors distributed according to $P_{\vC}$. Then
\begin{align}
  \veta
    &\approx
       \frac{1}{M}
       \sum_{m \in [M]}
         \vaoldsample{m}.
           \label{eq:veta:approximation:1}
\end{align}
(This expression is akin to the expression
in~\cite[Section~13.2]{Mezard:Montanari:09:1} on ``decoding by sampling.'')
This approach can work well for certain NFGs. However, in general it is,
unfortunately, difficult to efficiently obtain enough i.i.d.\ samples so that
$\veta$ can be estimated with sufficient accuracy. The difficulty of
generating i.i.d.\ samples happens for example in the case of NFGs that
represent good codes, see the discussion
in~\cite[Section~13.2]{Mezard:Montanari:09:1}. Therefore, the expression
in~\eqref{eq:veta:approximation:1} does not offer a shortcut for the main
computational step in symbolwise maximum a-posteriori decoding. (Of course,
this observation is not really surprising given the well-known computational
complexity of that decoder.)

Nevertheless, conceptually the expression in~\eqref{eq:veta:approximation:1}
is very useful as it suggests the following considerations that will lead to a
function that fulfills the promises that were stated at the beginning of this
section. Namely, let
\begin{align*}
  \vCsample{1}, \ \vCsample{2}, \ \ldots, \ \vCsample{M}
\end{align*}
be $M$ i.i.d.\ random vectors with distribution $P_{\vC}$. Then
\begin{align}
  \veta
    &= \Expec[\vC]
     \onestareq
       \frac{1}{M}
       \sum_{m \in [M]}
         \Expec[\vCsample{m}]
     = \Expec
         \left[
           \frac{1}{M}
           \sum_{m \in [M]}
             \vCsample{m}
         \right] \nonumber \\
    &= \sum_{\vcsample{1} \in \codeCedge}
       \cdots
       \sum_{\vcsample{M} \in \codeCedge}
         \left(
           \prod_{m \in [M]}
             P_{\vC}(\vcsample{m})
         \right)
         \frac{1}{M}
         \sum_{m \in [M]}
           \vcsample{m},
             \label{eq:veta:reformulation:1}
\end{align}
where step~$\onestar$ follows trivially from the definition of $\vCsample{m}$,
$m \in [M]$, and was inspired by~\eqref{eq:veta:approximation:1}. As simple as
it is, this step is actually the only ``non-trivial'' step in the whole
discussion here. The rest will simply be a ``mechanical'' application of the
method of types (see, \eg, \cite{Csiszar:Korner:81, Cover:Thomas:91}) towards
simplifying the expression in~\eqref{eq:veta:reformulation:1}.

Therefore, let us recall the relevant definitions from the method of types.

\begin{Definition}
  \label{def:method:of:types:1}

  Consider an NFG $\graphN(\setF,\setE, \setA, \setG)$ and fix some integer $M
  \in \Zpp$.

  \begin{itemize}

  \item \textbf{(Mapping)} Define the mapping
    \begin{align*}
      \varphiM: \
        & \codeCedge^M
          \to
          \setPi_{\codeCedge}, \ \ 
          \vcsampletot \defeq ( \vcsample{m} )_{m \in [M]}
          \mapsto 
          \vqcsample,     
    \end{align*}
    where
    \begin{align*}
      \qcsample_{\vc}
        &\defeq
           \frac{1}{M}
           \cdot
           (\text{number of appearances of $\vc$ in $\vcsampletot$}) \\
        &= \frac{1}{M}
           \sum_{m \in [M]}
             \big[ \vcsample{m} \! = \! \vc \big],
               \quad \vc \in \codeCedge.
    \end{align*}
    (In the above expression we have used Iverson's convention that was
    defined in Section~\ref{sec:notation:1}.)

  \item \textbf{(Type)} Let $\vcsampletot$ be a sequence over $\codeCedge$ of
    length $M$, \ie, $\vcsampletot \defeq ( \vcsample{m} )_{m \in [M]} \in
    \codeCedge^M$. Then the vector
    \begin{align*}
      \vqcsample
        &\defeq \varphiM(\vcsampletot)
    \end{align*}
    is called the type of $\vcsampletot$, or the empirical probability
    distribution of $\vcsampletot$.

  \item \textbf{(Set of all possible types)} The set $\setQsample{M} \subseteq
    \setPi_{\codeCedge}$ is defined to be the set of all possible types that
    are based on sequences over $\codeCedge$ of length $M$, \ie,
    \begin{align*}
      \setQsample{M}
        &\defeq
           \varphiM\big( \codeCedge^M \big).
    \end{align*}
  
  \item \textbf{(Type class)} For any $\vq \in \setQsample{M}$, the type class
    of $\vq$ is defined to be the set of all vectors in $\codeCedge^M$ with
    type $\vq$. Equivalently, the type class of $\vq$ is defined to the
    pre-image of $\vq$ under the mapping $\varphiM$, \ie,
    \begin{align*}
      \set{T}_M(\vq)
        &\defeq
           \varphiM^{-1}(\vq)
         = \Big\{
               \vcsampletot \in \codeCedge^M
             \Bigm| 
               \vqcsample = \vq
           \Big\}.
    \end{align*}

  \item \textbf{(Mean vector)} Assume $\setA_e = \{ 0, 1 \} \subset \R$ for
    all $e \in \setE$. For any type $\vq \in \setQsample{M}$, the mean vector
    associated with $\vq$ is defined to be
    \begin{align*}
      \meanvect(\vq)
        &\defeq
           \sum_{\vc \in \codeCedge}
             q_{\vc}
             \cdot
             \vc.
    \end{align*}
    \vskip-0.25cm
    \defend
  
  \end{itemize}
\end{Definition}

The following lemma contains some well-known properties of the objects that
were introduced in the above definition.

\begin{Lemma}
  \label{lemma:method:of:types:properties:1}
  
  Consider an NFG $\graphN(\setF,\setE, \setA, \setG)$ and fix some integer $M
  \in \Zpp$.
  
  \begin{itemize}

  \item The size of the set $\setQsample{M}$ is upper bounded as follows
    \begin{align*}
      \card{\setQsample{M}}
        &\leq (M+1)^{\card{\codeCedge}}.
    \end{align*}
    Because $\card{\codeCedge}$ is a fixed number for a given NFG $\graphN$,
    this upper bound is a polynomial in $M$.

  \item Let $\vcsampletot \defeq ( \vcsample{m} )_{m \in [M]} \in \codeCedge^M$,
    \ie, $\vcsampletot$ is a sequence over $\codeCedge$ of length $M$. Then
    \begin{align*}
      \prod_{m \in [M]}
        P_{\vC}(\vcsample{m})
        &= \frac{1}{\ZGibbs^M}
           \exp
             \left(
               -
               M
               \cdot
               \UGibbs\left( \vqcsample \right)
             \right),
    \end{align*}
    where $\UGibbs$ and $\ZGibbs$ are, respectively, the Gibbs average energy
    function and the Gibbs partition function associated with $\graphN$, see
    Definition~\ref{def:F:Gibbs:1}.

  \item For any type $\vq \in \setQsample{M}$, let $C_M(\vq)$ be the size of
    the type class $\set{T}_M(\vq)$ of $\vq$. Then
    \begin{align*}
      C_M(\vq)
        &= \exp
             \big(
               M \cdot \HGibbs(\vq)
               +
               o(M)
             \big),
    \end{align*}
    where $\HGibbs$ is the Gibbs entropy function associated with $\graphN$,
    see Definition~\ref{def:F:Gibbs:1}. (Because $C_M(\vq)$ is counting
    certain objects, this characterization of the Gibbs entropy function has
    clearly a ``combinatorial flavor,'' which is in contrast to the
    ``analytical flavor'' of the Gibbs entropy function in
    Definition~\ref{def:F:Gibbs:1}.)

  \item Let $\vcsampletot \defeq ( \vcsample{m} )_{m \in [M]} \in \codeCedge^M$,
    \ie, $\vcsampletot$ is a sequence over $\codeCedge$ of length $M$. Then
    \begin{align*}
      \frac{1}{M}
      \sum_{m \in [M]}
        \vcsample{m}
        &= \meanvect\left( \vqcsample \right).
    \end{align*}

  \end{itemize}
\end{Lemma}

\begin{Proof}
  See, \eg, \cite{Csiszar:Korner:81, Cover:Thomas:91}.
\end{Proof}

With this, the $M$-fold summation $\sum_{\vcsample{1}} \cdots
\sum_{\vcsample{M}}$ in~\eqref{eq:veta:reformulation:1} can be replaced by the
double summation $\sum_{\vq \in \setQsample{M}} \sum_{\vcsampletot \in
  \set{T}_M(\vq)}$, and we obtain
\begin{align}
  \veta
    &= \sum_{\vq \in \setQsample{M}} \
         \sum_{\vcsampletot \in \set{T}_M(\vq)} \
           \underbrace{
             \left(
               \prod_{m \in [M]}
                 P_{\vC}(\vcsample{m})
             \right)
           }_{\onestareq \ \frac{1}{\ZGibbs^M} \cdot \exp(-M \cdot \UGibbs(\vq))}
           \cdot
           \underbrace{
             \frac{1}{M}
             \sum_{m \in [M]}
               \vcsample{m}
           }_{\twostarseq \ \meanvect(\vq)} \nonumber \\
    &= \sum_{\vq \in \setQsample{M}} \ \ 
         \sum_{\vcsampletot \in \set{T}_M(\vq)}
           \frac{1}{\ZGibbs^M}
           \cdot
           \exp\big( -M \cdot \UGibbs(\vq) \big)
           \cdot
           \meanvect(\vq) \nonumber \\
    &\threestarseq
       \sum_{\vq \in \setQsample{M}}
         \frac{1}{\ZGibbs^M}
         \cdot 
         \exp\big( -M \cdot \UGibbs(\vq) \big)
         \cdot
         \meanvect(\vq)
         \cdot
         \underbrace{
           \sum_{\vcsampletot \in \set{T}_M(\vq)}
             1
         }_{\fourstarseq \ C_M(\vq)} \nonumber \\
    &\fivestarseq
       \sum_{\vq \in \setQsample{M}}
         s_M(\vq)
         \cdot
         \meanvect(\vq),
           \label{eq:veta:reformulation:2}
\end{align}
where at steps~$\onestar$ and~$\twostars$ we have used
Lemma~\ref{lemma:method:of:types:properties:1}, where at step~$\threestars$ we
have used the fact that the terms appearing in the summation are independent
of $\vcsampletot$ given their type $\vq$, where at step~$\fourstars$ we have
used Lemma~\ref{lemma:method:of:types:properties:1}, and where at
step~$\fivestars$ we have used the abbreviation
\begin{align}
  s_M(\vq)
    &\defeq
       \frac{1}{\ZGibbs^M}
       \cdot 
       \exp\big( - M \cdot \UGibbs(\vq) \big)
       \cdot
       C_M(\vq).
         \label{eq:def:s:M:1}
\end{align}
Similarly, we obtain
\begin{align}
  1
    &= \sum_{\vcsample{1}}
       \cdots
       \sum_{\vcsample{M}}
         \prod_{m \in [M]}
           P_{\vC}(\vcsample{m}) \nonumber \\
    &= \sum_{\vq \in \setQsample{M}} \
         \sum_{\vcsampletot \in \set{T}_M(\vq)} \
           \underbrace{
             \prod_{m \in [M]}
               P_{\vC}(\vcsample{m})
           }_{= \ \frac{1}{\ZGibbs^M} \cdot \exp(-M \cdot \UGibbs(\vq))}
             \nonumber \\
    &= \sum_{\vq \in \setQsample{M}}
         \frac{1}{\ZGibbs^M}
         \cdot
         \exp\big( -M \cdot \UGibbs(\vq) \big)
         \cdot
         \underbrace{
           \sum_{\vcsampletot \in \set{T}_M(\vq)}
             1
         }_{= \ C_M(\vq)} \nonumber \\
    &= \sum_{\vq \in \setQsample{M}}
         s_M(\vq),
           \label{eq:sum:sM:1}
\end{align}
\ie, $s_M$ is a probability mass function on $\setQsample{M}$. Moreover, using
Lemma~\ref{lemma:method:of:types:properties:1}, it follows
from~\eqref{eq:def:s:M:1} that
\begin{align}
  s_M(\vq)
    &= \frac{1}{\ZGibbs^M}
       \cdot 
       \exp
         \Big(
           - M \cdot \big( \UGibbs(\vq) - \HGibbs(\vq) \big)
           + o(M)
         \Big) \nonumber \\
    &= \frac{1}{\ZGibbs^M}
       \cdot 
       \exp
         \big(
           - M \cdot \FGibbs(\vq)
           + o(M)
         \big),
           \quad \vq \in \setQsample{M}.
             \label{eq:s:M:reformulation:1}
\end{align}
Because~\eqref{eq:veta:reformulation:2} holds for any $M$, we might as well
take the limit $M \to \infty$. Then, because in the limit $M \to \infty$ the
probability mass function $s_M$ is concentrated more and more around
\begin{align*}
  \vq^{*}
    &\defeq
       \argmin_{\vq \in \setPi_{\codeCedge}} \, 
         \FGibbs(\vq),
\end{align*}
and because the size of $\setQsample{M}$ grows at most polynomially in $M$, it
follows that in the limit $M \to \infty$ the sum
in~\eqref{eq:veta:reformulation:2} can be simplified, and we obtain
\begin{align*}
  \veta
    &= \meanvect(\vq^{*}).
\end{align*}
Moreover, using~\eqref{eq:s:M:reformulation:1}, the expression
in~\eqref{eq:sum:sM:1} can be rewritten for finite $M$ to read
\begin{align*}
  \ZGibbs
    &= \sqrt[M]
       {\sum_{\vq \in \setQsample{M}}
          \exp
            \big(
              - M \cdot \FGibbs(\vq)
              + o(M)
            \big)
       }.
\end{align*}
Taking the limit $M \to \infty$, and using the fact that the size of
$\setQsample{M}$ grows at most polynomially in $M$, we can write
\begin{align*}
  \ZGibbs
    &= \exp
         \big(
           - \FGibbs(\vq^{*})
         \big)
     = \exp
         \left(
           -
           \argmin_{\vq \in \setPi_{\codeCedge}} \, 
             \FGibbs(\vq)
         \right),
\end{align*}
\ie,
\begin{align*}
  -
  \log(\ZGibbs)
    &= \argmin_{\vq \in \setPi_{\codeCedge}} \, 
         \FGibbs(\vq).
\end{align*}
We conclude this section with some remarks.
\begin{itemize}

\item Note again that the only ``non-trivial'' step in the above derivation
  was step~$\onestar$ in~\eqref{eq:veta:reformulation:1}.

\item Recall that we assumed that $T = 1$; however, the results in this
  section can easily be generalized to $T \in \Rpp$.

\item We could have started this section by defining
  \begin{align*}
    P_{\vA}(\va)
      &\defeq
         \frac{g(\va)}{\ZGibbs},
           \quad \va \in \setA, 
  \end{align*}
  and then we could have continued by replacing $\codeCedge$ by $\setA$,
  $\setPi_{\codeCedge}$ by $\setPi_{\setA}$, \etc\ (Clearly, $P_{\vA}(\va) =
  P_{\vC}(\va)$ if $\va \in \codeCedge$ and $P_{\vA}(\va) = 0$ if $\va \notin
  \codeCedge$.) Both approaches, the approach taken in this section and this
  alternative approach, have their advantages and disadvantages, but
  ultimately they yield equivalent results.

\end{itemize}

\section{The Local Marginal Polytope and \\
               the Bethe Approximation}
\label{sec:local:marginal:polytope:and:Bethe:approximation:1}

In many problems it is desirable to compute the Gibbs partition function of
some graphical model. However, the direct evaluation of~\eqref{eq:Z:Gibbs:1}
is usually intractable. Moreover, although the above reformulation of the
Gibbs partition function via the minimum value of some function,
see~\eqref{eq:Z:Gibbs:2}, is an elegant reformulation of the Gibbs partition
function computation problem, this does not yield any computational savings
yet.

Nevertheless, it suggests to look for a function that is tractable and whose
minimum is close to the minimum of the Gibbs free energy function. An ansatz
for such a function is the so-called Bethe free energy
function~\cite{Yedidia:Freeman:Weiss:05:1}. The Bethe free energy function is
interesting because a theorem by Yedidia, Freeman, and
Weiss~\cite{Yedidia:Freeman:Weiss:05:1} says that fixed points of the
sum-product algorithm correspond to stationary points of the Bethe free energy
function. (For further motivations for the Bethe approximation we refer to the
discussion in~\cite{Yedidia:Freeman:Weiss:05:1, Yedidia:01:1,
  Wainwright:Jordan:08:1}.)

Before we can state the definition of the Bethe free energy function, we need
the concept of the local marginal polytope.

\begin{Definition}
  \label{def:lmp:1}

  Let $\graphN(\setF,\setE, \setA, \setG)$ be an NFG and let
  \begin{align*}
    \vbel
      &\defeq
         \big(
           (\vbel_f)_{f \in \setF},
           (\vbel_e)_{e \in \setE}
         \big)
  \end{align*}
  be a collection of vectors based on the real vectors
  \begin{alignat*}{2}
    \vbel_f
      &\defeq
         (\bel_{f,\va_f})_{\va_f \in \setA_f},
           \quad && f \in \setF \\
    \vbel_e
      &\defeq
         (\bel_{e,a_e})_{a_e \in \setA_e},
           \quad && e \in \setE.
  \end{alignat*}
  Then, for $f \in \setF$, the $f$th local marginal polytope (or $f$th belief
  polytope) $\lmpB_f$ is defined to be the set
  \begin{align*}
    \lmpB_f
      &\defeq
         \setPi_{\setA_f},
  \end{align*}
  and for $e \in \setE$, the $e$th local marginal polytope (or $e$th belief
  polytope) $\lmpB_e$ is defined to be the set
  \begin{align*}
    \lmpB_e
      &\defeq
         \setPi_{\setA_e}.
  \end{align*}
  With this, the local marginal polytope (or belief polytope) $\lmpB$ is
  defined to be the set
  \begin{align*}
    \lmpB
      &= \left\{
           \ \vbel \ 
         \ 
         \middle|
         \ 
           \begin{array}{c}
             \vbel_f \in \lmpB_f \text{ for all $f \in \setF$} \\[0.05cm]
             \vbel_e \in \lmpB_e \text{ for all $e \in \setE$} \\[0.25cm]
             \sum\limits_{\va'_f \in \setA_f: \, a'_{f,e} = a_e} 
               \bel_{f, \va'_f} = \bel_{e, a_e} \\
             \text{for all $f \in \setF$,
                   $e \in \setE_f$,
                   $a_e \in \setA_e$}
           \end{array}
         \right\},
  \end{align*}
  where $\vbel \in \lmpB$ is called a pseudo-marginal vector, or more
  precisely, a locally consistent pseudo-marginal vector. The constraints that
  were listed last in the definition of $\lmpB$ will be called ``edge
  consistency constraints.''
  \defend
\end{Definition}

\begin{Definition}
  \label{def:Bethe:free:energy:1}

  For any temperature $T \in \Rp$, the Bethe free energy function associated
  with some NFG $\graphN(\setF,\setE, \setA, \setG)$ is
  defined to be the function (see~\cite{Yedidia:Freeman:Weiss:05:1})
  \begin{align*}
    \FBethe: \ 
      &\lmpB
         \to \R, \quad
       \vbel
       \mapsto
         \UBethe(\vbel)
         -
         T
         \cdot \HBethe(\vbel),
  \end{align*}
  where
  \begin{align*}
    \UBethe: \ 
      &\lmpB
         \to \R, 
       \quad
       \vbel
       \mapsto
         \sum_f
           \UBethesub{f}(\vbel_f), \\
    \HBethe: \ 
      &\lmpB
         \to \R,
       \quad
       \vbel
       \mapsto
         \sum_f
           \HBethesub{f}(\vbel_f)
         -
         \sum_{e \in \setEfull}
           \HBethesub{e}(\vbel_e),
  \end{align*}
  with
  \begin{alignat*}{2}
    \UBethesub{f}: \ 
      &\lmpB_f
         \to \R, \quad
       \vbel_f
       &
       \, \mapsto\,
       &
         -
         \sum_{\va_f}
           \bel_{f,\va_f}
           \cdot
           \log
             \big(
               g_f(\va_f)
             \big), \\
    \HBethesub{f}: \ 
      &\lmpB_f
         \to \R, \quad
       \vbel_f
       &
       \, \mapsto\,
       &
         -
         \sum_{\va_f}
           \bel_{f,\va_f}
           \cdot
           \log
             \big(
               \bel_{f,\va_f}
             \big), \\
    \HBethesub{e}: \ 
      &\lmpB_e
         \to \R, \quad
       \vbel_e
       &
       \, \mapsto\,
       &
         -
         \sum_{a_e}
           \bel_{e,a_e}
           \cdot
           \log
             \big(
               \bel_{e,a_e}
             \big).
  \end{alignat*}
  Here, $\UBethe$ is the Bethe average energy function and $\HBethe$ is the
  Bethe entropy function. 
  \defend
\end{Definition}

Note that in the above definition of $\HBethe(\vbel)$, the term
$\HBethesub{e}(\vbel_e)$ appears with coefficient $-1$ for full-edges $e \in
\setEfull$, whereas it appears with coefficient $0$ for half-edges $e \in
\setEhalf$. Therefore, the latter terms are omitted.\footnote{These
  coefficients are consistent with the coefficients
  in~\cite{Yedidia:Freeman:Weiss:05:1}. Namely, because half/full-edges in
  NFGs correspond to variable nodes of degree one/two in
  factor graphs~\cite{Loeliger:04:1}, and because the Bethe entropy function
  term corresponding to a degree-$d$ variable node in a factor graph appears
  with coefficient $-(d-1)$ in the Bethe entropy function definition, we see
  that for a full-edge the corresponding coefficient must be $-(2-1) = -1$,
  and that for a half-edge the corresponding coefficient must be $-(1-1) =
  0$.}

\begin{Definition}
  \label{def:Bethe:partition:function:1}

  For any $T \in \Rpp$, the Bethe partition function associated with some NFG
  $\graphN(\setF,\setE, \setA, \setG)$ is defined to be
  \begin{align*} 
    \ZBethe
      &\defeq
         \exp
           \left(
             -
             \frac{1}{T}
             \cdot
             \min_{\vbel \in \lmpB}
               \FBethe(\vbel)
           \right).
  \end{align*}
  \defend
\end{Definition}

Let us comment on a variety of issues with respect to the above definition of
the Bethe partition function.
\begin{itemize}

\item Note that here $\ZBethe$ is \emph{defined} such that a similar statement
  can be made as in Lemma~\ref{lemma:F:Gibbs:minimum:1}.

\item For NFGs \emph{without} cycles one can show that $\ZBethe = \ZGibbs$.

\item The Bethe free energy function of NFGs \emph{with}
  cycles can have non-global local extrema, which is in contrast to the Gibbs
  free energy function which is convex and therefore has no non-global local
  extrema.

\item Similar to Lemma~\ref{lemma:F:Gibbs:minimum:2}, we can consider a
  modified Bethe free energy function that equals the Bethe free energy
  function for $\vbel$'s that are compatible with some $\vahalf \in
  \setAhalf$, and that is infinite otherwise. Of course, the modified Bethe
  partition function will be a function of $\vahalf \in \setAhalf$.

\end{itemize}

In the next section, we will present an alternative characterization of the
local marginal polytope, namely in terms of so-called finite graph covers,
thereby generalizing some observations that were made
in~\cite{Koetter:Vontobel:03:1, Vontobel:Koetter:05:1:subm}. This will then
lead the way to Section~\ref{sec:counting:in:finite:graph:covers:1}, where we
will show that the Bethe entropy function, and consequently also the Bethe
partition function, cannot only be characterized \emph{analytically} (as was
done here in Definitions~\ref{def:Bethe:free:energy:1}
and~\ref{def:Bethe:partition:function:1}), but also
\emph{combinatorially}. This combinatorial approach is based on counting valid
configurations in finite graph covers of the underlying NFG.

\section{Finite Graph Covers}
\label{sec:finite:graph:covers:1}

This section reviews the concept of a finite graph cover of a graph; for more
details we refer the interested reader to~\cite{Stark:Terras:96:1}. We also
refer to~\cite{Koetter:Vontobel:03:1, Vontobel:Koetter:05:1:subm,
  Koetter:Li:Vontobel:Walker:07:1, Kelley:Sridhara:07:1,
  Axvig:Dreher:Morrison:Psota:Perez:Walker:09:1}, where finite graph covers
were used in the context of coding theory, especially for the analysis of
linear programming decoding and message-passing iterative decoding.

\begin{Definition}[see, \eg, \cite{Massey:77:1, Stark:Terras:96:1}]
  \label{def:graph:cover:1}

  A {\em cover} of a graph $\graph{G}$ with vertex set $\set{V}$ and edge set
  $\set{E}$ is a graph $\graph{G}$ with vertex set $\cset{V}$ and edge set
  $\cset{E}$, along with a surjection $\pi: \cset{V} \to \set{V}$ which is a
  graph homomorphism (\ie, $\pi$ takes adjacent vertices of $\graph{G}$ to
  adjacent vertices of $\graph{G}$) such that for each vertex $v \in \set{V}$
  and each $\cover{v} \in \pi^{-1}(v)$, the neighborhood $\del(\cover{v})$ of
  $\cover{v}$ is mapped bijectively to $\del(v)$. A cover is called an {\em
    $M$-cover}, where $M \in \Zpp$, if $\bigl| \pi^{-1}(v) \bigr| = M$ for
  every vertex $v$ in $\set{V}$.\footnote{The number $M$ is also known as the
    degree of the cover. (Not to be confused with the degree of a vertex.)}
  \defend
\end{Definition}

\begin{figure}
  \begin{center}
    \epsfig{file=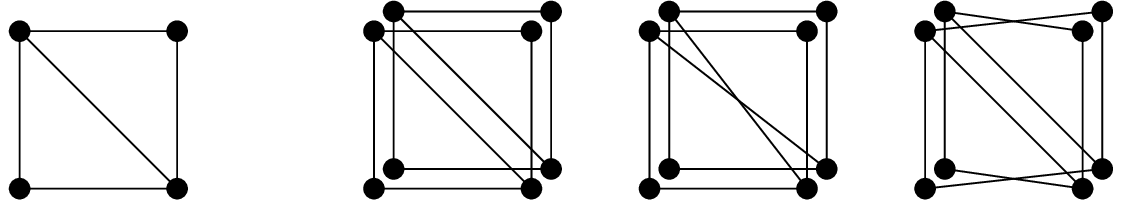, width=8.5cm} \\[1cm]
    \epsfig{file=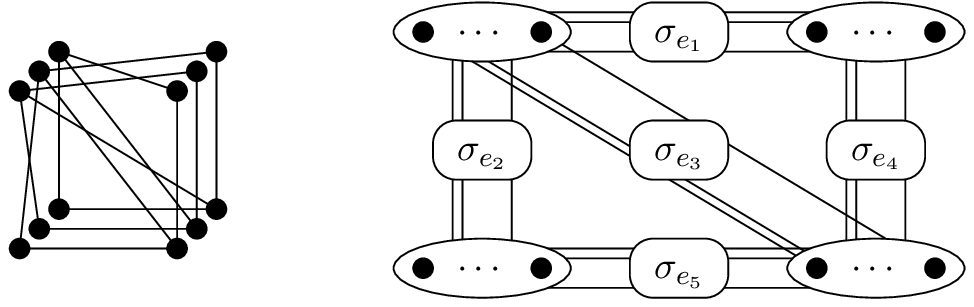, width=8.5cm}
  \end{center}
  \caption{Top left: base graph $\graph{G}$. Top right: sample of possible
    $2$-covers of $\graph{G}$. Bottom left: a possible $3$-cover of
    $\graph{G}$. Bottom right: a possible $M$-cover of $\graph{G}$. Here,
    $\sigma_{e_1}, \ldots, \sigma_{e_5}$ are arbitrary edge
    permutations.}
  \label{fig:graph:cover:samples:1}
\end{figure}

A consequence of this definition is that if $\cgraph{G}$ is an $M$-cover of
$\graph{G}$ then we can choose its vertex set $\cset{V}$ to be $\cset{V}
\defeq \set{V} \times [M]$: if $(v,m) \in \cset{V}$ then $\pi\big( (v,m) \big)
= v$ and if $\big( (v_1,m_1), (v_2,m_2) \big) \in \cset{E}$ then $\pi\big(
\big\{ (v_1,m_1), (v_2,m_2) \big\} \big) = \{ v_1, v_2 \}$. Another
consequence is that any $M_2$-cover of any $M_1$-cover of the base graph is an
$(M_2 \cdot M_1)$-cover of the base graph.

\begin{figure}
  \begin{center}
    \epsfig{file=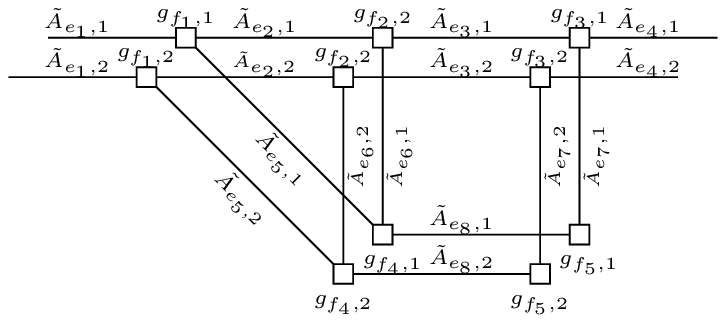, scale=1} \\[0.5cm]
    \epsfig{file=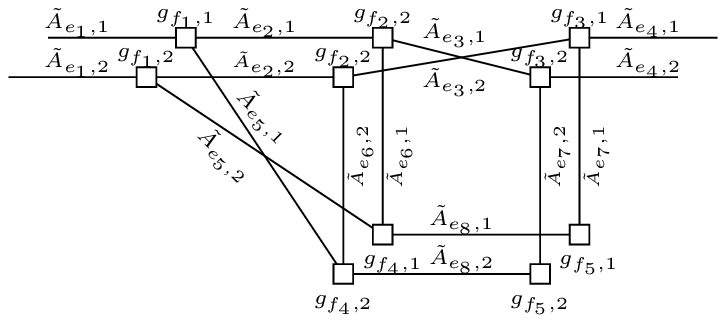, scale=1}
  \end{center}
  \caption{Two possible $2$-covers of the NFG $\graphN$ that is shown in
    Fig.~\ref{fig:example:simple:ffg:1}.}
  \label{fig:example:simple:ffg:1:cover:1}
\end{figure}

\begin{Example}[\!\!\cite{Vontobel:Koetter:05:1:subm}]
  \label{example:four:vertex:five:edges:graph:finite:graph:covers:1}

  Let $\graph{G}$ be a (base) graph with $4$ vertices and $5$ edges as shown
  in Fig.~\ref{fig:graph:cover:samples:1} (top
  left). Figs.~\ref{fig:graph:cover:samples:1} (top right),
  \ref{fig:graph:cover:samples:1} (bottom left),
  and~\ref{fig:graph:cover:samples:1} (bottom right) show, respectively,
  possible $2$-, $3$-, and $M$-covers of $\graph{G}$. Note that any $2$-cover
  of $\graph{G}$ must have $8 = 2 \cdot 4$ vertices and $10 = 2 \cdot 5$
  edges, that any $3$-cover of $\graph{G}$ must have $12 = 3 \cdot 4$ vertices
  and $15 = 3 \cdot 5$ edges, and that any $M$-cover must have $M \cdot 4$
  vertices and $M \cdot 5$ edges. As depicted in
  Fig.~\ref{fig:graph:cover:samples:1} (bottom right), any $M$-cover of
  $\graph{G}$ is entirely specified by $\card{\set{E}}$ edge permutations,
  where $\setE$ is the edge set of $\graph{G}$.  \exampleend
\end{Example}

\begin{figure*}
  \hskip1.5cm
  \epsfig{file=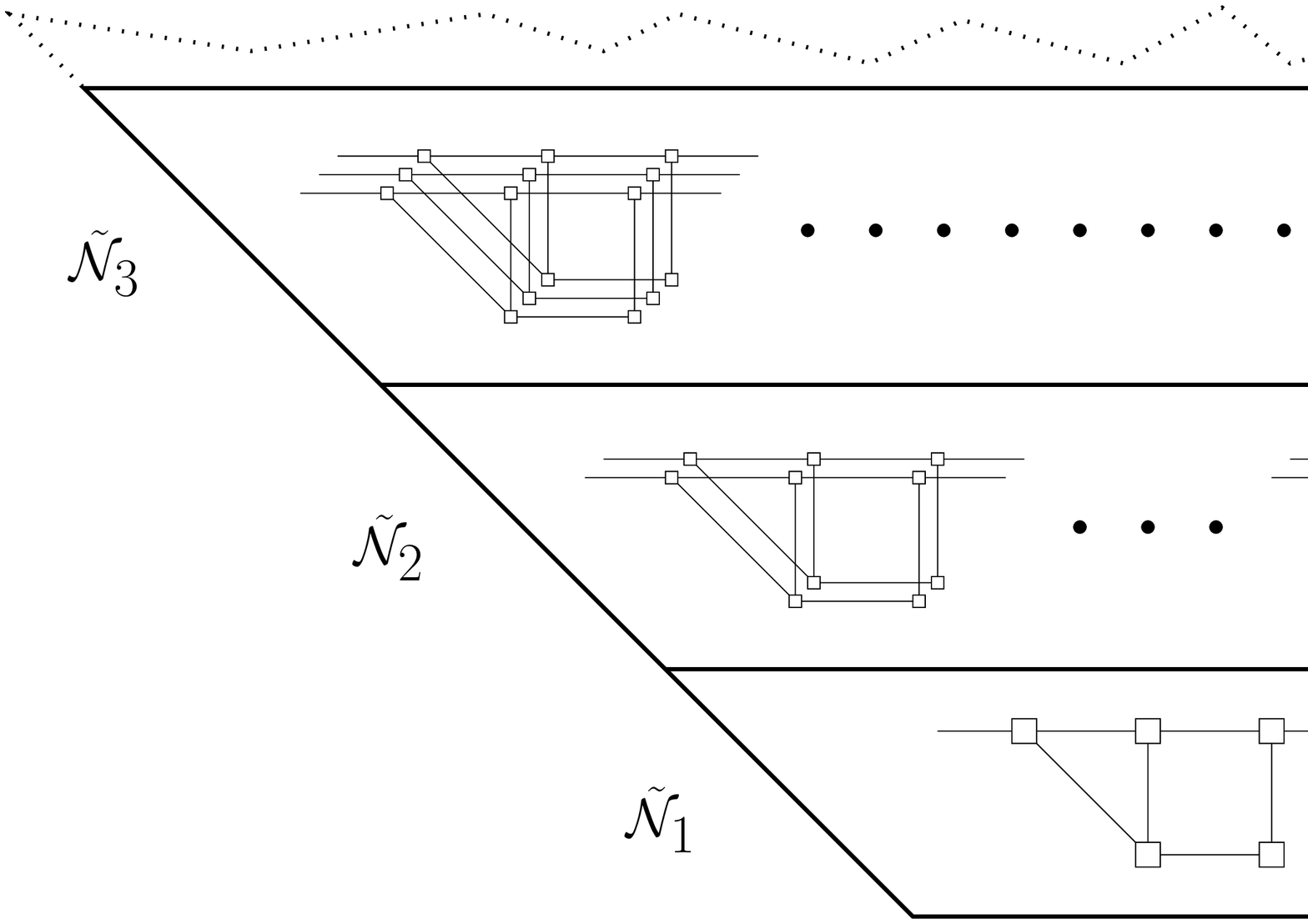, scale=0.4}
  \vskip0.5cm
  \caption{Hierarchy of all finite graph covers of the (base) NFG $\graphN$
    from Fig.~\ref{fig:example:simple:ffg:1}, where the $M$th level lists
    all graphs in $\cset{N}_M$, \ie, all $M$-covers. (For a base NFG with
    $\card{\setEfull}$ full-edges, there are $\cardbig{\cset{N}_M} =
    (M!)^{\card{\setEfull}}$ graph covers at the $M$th level.) The
    pseudo-marginal mappings $\varphiM$, $M \in \Zpp$, and their images are
    specified in Definitions~\ref{def:down:projection:1}
    and~\ref{def:down:projection:2}.}
  \label{fig:example:simple:ffg:1:cover:hierarchy:1}
\end{figure*}

As we can see from this example, an $M$-cover $\cgraph{G}$ of $\graph{G}$ may
consist of several connected components also if $\graph{G}$ consists of only
one connected component. In general, letting $\ncomponents\graph{G}$ and
$\ncomponents\cgraph{G}$ denote the number of connected components of
$\graph{G}$ and $\cgraph{G}$, respectively, one can easily verify that
\begin{align}
  \ncomponents\graph{G}
     \,\leq\, 
       \ncomponents\cgraph{G}
    &\,\leq\, 
       M
         \!\cdot\!
         \ncomponents\graph{G}.
           \label{eq:ncomponents:inequalities:1}
\end{align}

Because NFGs are graphs, we can also consider finite graph covers of NFGs, as
is done in the next example. (Note that we do \emph{not} apply edge
permutations to copies of half-edges.)

\begin{Example}
  \label{example:simple:ffg:1:cont:1}

  Consider again the NFG $\graphN$ that is discussed in
  Example~\ref{example:simple:ffg:1} and depicted in
  Fig.~\ref{fig:example:simple:ffg:1}. Two possible $2$-covers of this
  (base) NFG are shown in Fig.~\ref{fig:example:simple:ffg:1:cover:1}. The
  first graph cover is ``trivial'' in the sense that it consists of two
  disjoint copies of the NFG in Fig.~\ref{fig:example:simple:ffg:1}. The
  second graph cover is ``more interesting'' in the sense that the edge
  permutations are such that the two copies of the base NFG are
  intertwined. (Of course, both graph covers are equally valid.)

  Note that the $M$ copies of a function node $g_f$ are denoted by $\{ g_{f,m}
  \}_{m \in [M]}$, and that the $M$ copies of a variable label $A_e$ are
  denoted by $\{ \cover{A}_{e,m} \}_{m \in [M]}$. In that respect, we chose
  the variable labels to be such that if a full-edge $e$ connects function
  nodes $f_i$ and $f_j$, $i < j$, then the variable label $\cover{A}_{e,m}$,
  $m \in [M]$, will be associated with the edge that connects the function
  nodes $(f_i,m)$ and $(f_j,\sigma_e(m))$, where $\sigma_e: \, [M] \to [M]$
  describes the permutation that is applied to the $M$ copies of the edge
  $e$. Similarly, if a half-edge $e$ is connected to the function node $f$
  then the variable label $\cover{A}_{e,m}$, $m \in [M]$, will be associated
  with the half-edge that is connected to the function node $(f,m)$. That
  being said, it is important to note that the results that are presented in
  this paper are invariant to the chosen labeling convention.  \exampleend
\end{Example}

\begin{Definition}
  \label{def:set:of:graph:covers:1}

  Consider an NFG $\graphN(\setF,\setE, \setA, \setG)$. We define the set
  $\cset{N}_M$ to be the set of all $M$-covers $\cgraph{N}$ of $\graphN$.
  \defend
\end{Definition}

\noindent
Note that in this definition we consider only \emph{labeled} $M$-covers as
follows:
\begin{itemize}

\item All vertices of an $M$-cover have distinct labels, say $(f,m)$ or
  $g_{f,m}$ with $(f,m) \in \setF \times [M]$.

\item All edges of an $M$-cover have distinct labels, say $(e,m)$ or $A_{e,m}$
  with $(e,m) \in \setE \times [M]$.

\item We \emph{do not identify} $M$-covers whose graphs are isomorphic but
  whose \emph{vertex labels} are distinct.

\item However, we \emph{do identify} $M$-covers whose graphs (including vertex
  labels) are isomorphic but whose \emph{edge labels} are distinct.

\end{itemize}
(For reasons of simplicity, these vertex and edge labels are sometimes omitted
in drawings.) These conventions are reflected in the following lemma that
counts $M$-covers.

\begin{Lemma}
  \label{lemma:set:of:graph:covers:1}

  Consider an NFG $\graphN(\setF,\setE, \setA, \setG)$. Then
  \begin{align}
    \cardbig{\cset{N}_M}
      &= (M!)^{\card{\setEfull(\graphN)}} \, .
           \label{eq:number:M:covers:1}
  \end{align}
\end{Lemma}

\begin{Proof}
  An $M$-cover $\cgraph{N}$ of $\graphN$ can be obtained as follows.
  \begin{enumerate}

  \item For every $f \in \setF(\graphN)$, draw $M$ copies of $f$ (with
    distinct labels).

  \item For every $e = \{ f, f' \} \in \setEfull(\graphN)$, connect the $M$
    copies of $f$ and the $M$ copies of $f'$ by $M$ disjoint edges.

  \item For every $e = \{ f \} \in \setEhalf(\graphN)$, attach a half-edge to
    the $M$ copies of $f$.

  \end{enumerate}
  Because the second step can be done independently for every $e \in
  \setEfull(\graphN)$, because for every such edge there are $M!$ ways of
  connecting the $M$ copies of $f$ and the $M$ copies of $f'$ by $M$ disjoint
  edges, and because the obtained $M$-covers are all distinct, the result
  follows.
\end{Proof}

\begin{Example}
  \label{example:simple:ffg:1:cont:2}

  Consider again the NFG $\graphN$ that is discussed in
  Example~\ref{example:simple:ffg:1} and depicted in
  Fig.~\ref{fig:example:simple:ffg:1}. We can order all the finite graph
  covers of this base NFG according to the hierarchy shown in
  Fig.~\ref{fig:example:simple:ffg:1:cover:hierarchy:1}, where the $M$th
  level lists all graphs in $\cset{N}_M$, \ie, all $M$-covers. (Note that
  there is exactly one $1$-cover, namely the base NFG itself.)  The inverted
  pyramid alludes to the fact that the number of $M$-covers is growing with
  $M$, \ie, $\cardbig{\cset{N}_M}$ is growing with $M$,
  see~\eqref{eq:number:M:covers:1}.  \exampleend
\end{Example}

The following definition specifies a collection of mappings that will be
crucial for the rest of this section and for the next section. These mappings
are inspired by similar mappings that appear in the method of types,
see~Definition~\ref{def:method:of:types:1}.

\begin{Definition}
  \label{def:down:projection:1}

  Let $\graphN(\setF,\setE, \setA, \setG)$ be an NFG with local marginal
  polytope $\lmpB$ (see Definition~\ref{def:lmp:1}). For any $M \in \Zpp$ we
  define the pseudo-marginal mapping
  \begin{alignat*}{2}
    \varphiM: \ 
    &
    \Big\{
      \big(
        \cgraph{N}, \, 
        \cvc
      \big)
      \ \Big| \ 
        \cgraph{N} \in \cset{N}_{M}, \ 
        \cvc \in \codeCedge(\cgraph{N})
    \Big\}
      &\ \to \ & \lmpB, \\
    &
    \big(
      \cgraph{N}, \,
      \cvc
    \big)
      &\ \mapsto\ &
         \vbel.
  \end{alignat*}
  Here, for a given $\cgraph{N} \in \cset{N}_M$ and $\cvc \in
  \codeCedge(\cgraph{N})$, the components of $\vbel$ are defined as
  follows
  \begin{align}
    \bel_{f, \va_f}
      &\defeq
         \frac{1}{M}
         \sum_{m \in [M]}
           \big[ \cvc_{f,m} \! = \! \va_f \big],
             \quad f \in \setF, \, \va_f \in \setA_f,
             \label{eq:varphi:1} \\
    \bel_{e, a_e}
      &\defeq
         \frac{1}{M}
         \sum_{m \in [M]}
           \big[ \cc_{e,m} \! = \! a_e \big],
             \quad e \in \setE, \, a_e \in \setA_e.
             \label{eq:varphi:2}
  \end{align}
  (In the above expressions we have used Iverson's convention that was defined
  in Section~\ref{sec:notation:1}.)
  \defend
\end{Definition}

Note that one cannot mention a valid configuration $\cvc$ without mentioning
the $M$-cover $\cgraph{N}$ in which it lives. Therefore, although the
expressions in~\eqref{eq:varphi:1} and~\eqref{eq:varphi:2} do \emph{not}
involve the $M$-cover $\cgraph{N}$ explicitly, the domain of $\varphiM$ must
be over $( \text{graph},\text{valid configuration})$-pairs $(\cgraph{N},
\cvc)$, and not just over valid configurations $\cvc$.

\begin{figure}
  \begin{center}
    \epsfig{file=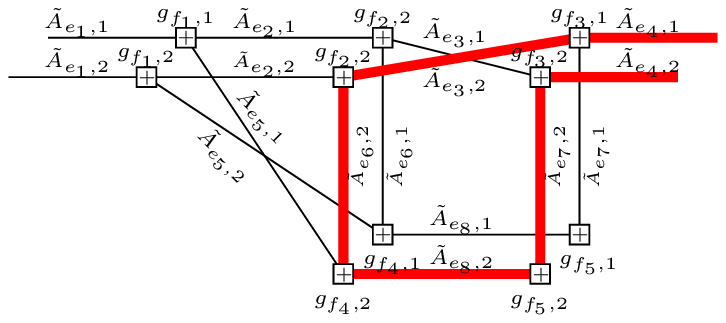, scale=1}
  \end{center}
  \caption{Valid configuration $\tvc$ on a possible $2$-cover of the NFG
    $\graphN$ in Fig.~\ref{fig:example:simple:ffg:1}. (This $2$-cover is
    identical to the second $2$-cover in
    Fig.~\ref{fig:example:simple:ffg:1:cover:1}.) For every $(e,m) \in \setE
    \times [M]$, if $\cover{c}_{e,m} = 0$ then the edge $(e,m)$ is thin and in
    black, whereas if $\cover{c}_{e,m} = 1$ then the edge $(e,m)$ is thick and
    in red. (See Example~\ref{example:simple:ffg:1:cont:3} for more details.)}
  \label{fig:example:simple:ffg:1:cover:1:configuration:1}
\end{figure}

\begin{Example}
  \label{example:simple:ffg:1:cont:3}

  Consider again the NFG $\graphN$ that is discussed in
  Example~\ref{example:simple:ffg:1:cont:0} and depicted in
  Fig.~\ref{fig:example:simple:ffg:1:mod:2}, which goes back to
  Example~\ref{example:simple:ffg:1} and
  Fig.~\ref{fig:example:simple:ffg:1}. A possible $2$-cover $\cgraph{N}$ of
  $\graphN$ is shown in
  Fig.~\ref{fig:example:simple:ffg:1:cover:1:configuration:1}. Applying the
  pseudo-marginal mapping to the valid configuration $\tvc \in
  \codeC(\cgraph{N})$ shown in
  Fig.~\ref{fig:example:simple:ffg:1:cover:1:configuration:1}, we obtain the
  pseudo-marginal vector $\vbel \defeq \varphiM(\tvc)$ with the following
  components. (We show only a selection of the obtained pseudo-marginals.)
  \begin{itemize}

  \item For $\beta_{f_1,\va_1}$ with $\va_1 = (a_{e_1}, a_{e_2}, a_{e_3})$:
    \begin{align*}
      \beta_{f_1,(000)}
        &\!=\! 1, \ 
      \beta_{f_1,(001)}
         \!=\! 0, \
      \beta_{f_1,(010)}
         \!=\! 0, \ 
      \beta_{f_1,(011)}
         \!=\! 0, \\
      \beta_{f_1,(100)}
        &\!=\! 0, \
      \beta_{f_1,(101)}
         \!=\! 0, \
      \beta_{f_1,(110)}
         \!=\! 0, \
      \beta_{f_1,(111)}
         \!=\! 0.
    \end{align*}

  \item For $\beta_{f_2,\va_2}$ with $\va_2 = (a_{e_2}, a_{e_3}, a_{e_6})$:
    \begin{alignat*}{5}
      \beta_{f_2,(000)}
        &{} \!= &{} \frac{1}{2}, \ 
      \beta_{f_2,(001)}
        &{} \!= &{} 0, \
      \beta_{f_2,(010)}
        &{} \!= &{} 0, \ 
      \beta_{f_2,(011)}
        &{} \!= &{} \frac{1}{2}, \\
      \beta_{f_2,(100)}
        &{} \!= &{} 0, \
      \beta_{f_2,(101)}
        &{} \!= &{} 0, \
      \beta_{f_2,(110)}
        &{} \!= &{} 0, \
      \beta_{f_2,(111)}
        &{} \!= &{} 0.
    \end{alignat*}

  \item For $\beta_{f_5,\va_5}$ with $\va_5 = (a_{e_7}, a_{e_8})$:
    \begin{align*}
      \beta_{f_5,(00)}
        &= \frac{1}{2}, \ 
      \beta_{f_5,(01)}
         = 0, \
      \beta_{f_5,(10)}
         = 0, \ 
      \beta_{f_5,(11)}
         = \frac{1}{2}.
    \end{align*}

  \item For $\beta_{e_4,a_4}$:
    \begin{align*}
      \beta_{e_4,0}
        &= 0, \  
      \beta_{e_4,1}
         = 1.
    \end{align*}

  \item For $\beta_{e_7,a_7}$:
    \begin{align*}
      \beta_{e_7,0}
        &= \frac{1}{2}, \ 
      \beta_{e_7,1}
         = \frac{1}{2}.
    \end{align*}
  \exampleend

  \end{itemize}

\end{Example}

The upcoming Theorem~\ref{theorem:down:projection:1} will show that the local
marginal polytope $\lmpB$ is a valid choice as a co-domain of the
pseudo-marginal mapping $\varphiM$. For this theorem, the following definition
is useful.

\begin{Definition}
  \label{def:down:projection:2}

  Consider an NFG $\graphN(\setF,\setE, \setA, \setG)$. For every $M \in
  \Zpp$, we define $\lmpBdown{M}$ to be the image of the pseudo-marginal
  mapping $\varphiM$, \ie,
  \begin{align*}
    \lmpBdown{M}
      &\defeq
         \operatorname{image}(\varphiM).
  \end{align*}
  Moreover, we define $\lmpBdown{}$ to be the union of all $\lmpBdown{M}$, \ie,
  \begin{align}
    \lmpBdown{}
      &\defeq
         \bigcup_{M \in \Zpp}
           \lmpBdown{M}. 
             \label{eq:def:lmp:B:prime:1}
  \end{align}
  \vskip-0.5cm
  \defend
\end{Definition}

In words, $\lmpBdown{}$ is the set where for every $\vbel \in \lmpBdown{}$ there is some
$M \in \Zpp$ such that there is an $M$-cover $\graphN \in \cset{N}_M$ with a
valid configuration $\cvc \in \codeCedge(\cgraph{N})$ in it such that $\cvc$
maps down to $\vbel$ under the pseudo-marginal mapping
$\varphiM$. Generalizing the language of~\cite{Kelley:Sridhara:07:1}, we will
call
\begin{align*}
  \begin{array}{lcl}
    \lmpBdown{M} & \!\!\!\!: & \text{set of all $M$-cover lift-realizable 
                                     ps.-marg.\ vectors,} \\
    \lmpBdown{}   & \!\!\!\!: & \text{set of all lift-realizable
                                     pseudo-marginal vectors.}
  \end{array}
\end{align*}

For any $M_1, M_2 \in \Zpp$ with $M_1$ dividing $M_2$, one can show the following
chain of set inclusions
\begin{align*}
  \lmpBdown{M_1}
    &\subseteq
       \lmpBdown{M_2}
     \subseteq
       \lmpBdown{}.
\end{align*}
The second set inclusion follows from~\eqref{eq:def:lmp:B:prime:1}; we leave
it as an exercise for the reader to verify the first set inclusion.

\begin{Theorem}
  \label{theorem:down:projection:1}

  Let $\graphN(\setF,\setE, \setA, \setG)$ be an NFG with local marginal
  polytope $\lmpB$. The set of all lift-realizable pseudo-marginal vectors 
  satisfies
  \begin{align*}
    \lmpBdown{}
      &= \lmpB \cap \mathbb{Q}^{\operatorname{dim}(\lmpB)},
  \end{align*}
  which implies that $\lmpBdown{}$ is dense in $\lmpB$. Moreover, all vertices of
  $\lmpB$ are in $\lmpBdown{}$.
\end{Theorem}

\begin{Proof}
  This is a more or less a straightforward extension of the characterization
  in~\cite{Koetter:Vontobel:03:1, Vontobel:Koetter:05:1:subm} of the
  fundamental polytope in terms of valid configurations in finite graph
  covers. We omit the details.
\end{Proof}

Let us conclude this section with a few comments.

\begin{Remark}
  \label{remark:lmpB:1}

  Let $\graphN(\setF,\setE, \setA, \setG)$ be an NFG with local marginal
  polytope $\lmpB$. For any $M \in \Zpp$, consider the set of a all $M$-cover
  lift-realizable pseudo-marginal vectors $\lmpBdown{M}$.

  \begin{itemize}
  
  \item It holds that
    \begin{align*}
      \card{\lmpBdown{M}}
        &\leq (M + 1)^{\operatorname{dim}(\lmpB)}.
    \end{align*}
    This follows from the observation that every component of $\vbel \in
    \lmpBdown{M}$ takes values in the set $\bigl\{ \frac{0}{M}, \frac{1}{M},
    \frac{2}{M}, \ldots, \frac{M}{M} \bigr\}$. Because
    $\operatorname{dim}(\lmpB)$ is a fixed number for a given NFG $\graphN$,
    it follows that the number of elements of $\lmpBdown{M}$ grows at most
    polynomially in $M$. This important fact will allow us to use the method
    of types in the next section. (Compare this observation also with a
    similar statement in Lemma~\ref{lemma:method:of:types:properties:1}.)
  
  \item Although the focus of this paper is mostly on the behavior of
    $\lmpBdown{M}$ when $M$ goes to infinity, the set $\lmpBdown{M}$ for $M = 1$, \ie,
    the set $\lmpBdown{1}$, is also of special interest. The reason for this is
    that $\convhull(\lmpBdown{1})$ contains all pseudo-marginal vectors that are
    globally realizable. Here, a pseudo-marginal vector $\vbel$ is called
    globally realizable~\cite{Wainwright:Jordan:08:1} when there is a $\vp \in
    \setPi_{\codeCedge}$ such that $\vbel$ contains the true marginals of
    $\vp$, \ie,
    \begin{alignat*}{2}
      \bel_{f,\va_f}
        &= \sum_{\vc: \, \vc_f = \va_f}
             \vp_{\vc},
               \quad && f \in \setF, \, \va_f \in \setA_f, \\
      \bel_{e,a_e}
        &= \sum_{\vc: \, c_e = a_e}
             \vp_{\vc},
               \quad && e \in \setE, \, a_e \in \setA_e.
    \end{alignat*}

  \item For the NFG $\graphN$ discussed in
    Examples~\ref{example:simple:ffg:1:cont:0}
    and~\ref{example:simple:ffg:1:cont:3}, one can verify that the local
    marginal polytope of $\graphN$ satisfies
    \begin{align}
      \lmpB
        &\supsetneq
           \convhull(\lmpBdown{1}),
             \label{eq:lmpB:vs:lmpBone:1}
    \end{align}
    \ie, $\lmpB$ is strictly larger than $\convhull(\lmpBdown{1})$. This can be
    shown as follows. Consider the valid configuration $\tvc$ of the $2$-cover
    shown in Fig.~\ref{fig:example:simple:ffg:1:cover:1:configuration:1}:
    its associated pseudo-marginal vector $\vbel$ does not lie in
    $\convhull(\lmpBdown{1})$. Indeed, because all variable alphabets are $\{ 0, 1
    \}$ and because $\bel_{e,0} = 1 - \bel_{e,1}$ for all $e \in \setE$, one
    can verify that the condition that the vector $\vbel$ is in
    $\convhull(\lmpBdown{1})$ is equivalent to the condition that the vector
    \begin{align*}
      (\bel_{e_1,1}, \bel_{e_2,1}, \ldots, \bel_{e_7,1})
        &=
        \left(
          \frac{0}{2}, \frac{0}{2}, \frac{1}{2}, \frac{2}{2}, 
          \frac{0}{2}, \frac{1}{2}, \frac{1}{2}, \frac{1}{2} 
        \right)
    \end{align*}
    is in the convex hull of the set $\codeC$ of valid configurations of
    $\graphN$ as listed
    in~\eqref{eq:example:simple:ffg:1:valid:configurations:1}. However, the
    latter is not the case. Therefore, $\convhull(\lmpBdown{2}) \supsetneq
    \convhull(\lmpBdown{1})$. Combining this with $\lmpB \supseteq
    \convhull(\lmpBdown{2})$, we find that the $\lmpB$
    satisfies~\eqref{eq:lmpB:vs:lmpBone:1}. In conclusion, valid
    configurations like $\tvc$ in
    Fig.~\ref{fig:example:simple:ffg:1:cover:1:configuration:1} are the
    reason why $\lmpB$ is strictly larger than $\convhull(\lmpBdown{1})$.
    \remarkend

  \end{itemize}
\end{Remark}

\section{Counting in Finite Graph Covers}
\label{sec:counting:in:finite:graph:covers:1}

The definition of the Bethe entropy function and the Bethe partition function
in Definitions~\ref{def:Bethe:free:energy:1}
and~\ref{def:Bethe:partition:function:1}, respectively, were entirely
analytical. In this subsection we will present a combinatorial
characterization of these functions in terms of counting certain valid
configurations in graph covers, a characterization that was first outlined
in~\cite{Vontobel:08:3, Vontobel:09:Talk:5}.

We start with the definition of a certain averaging operator. This definition
is motivated by the fact that many results in this section are based on
associating a real number to every $M$-cover of a base NFG and on computing
the average of this value over all $M$-covers.

\begin{Definition}
  \label{def:average:over:graph:covers:1}

  Let $\graphN(\setF,\setE, \setA, \setG)$ be an NFG. For any
  $M \in \Zpp$ and any function $\chi_M: \, \cset{N}_M \to \R$ we define the
  averaging operator to be
  \begin{align*}
    \Big\langle \!
      \chi_M(\cgraph{N})
    \! \Big\rangle_{\cgraph{N} \in \cset{N}_M}
      &\defeq
         \frac{1}{\cardbig{\cset{N}_M}}
         \sum_{\cgraph{N} \in \cset{N}_M}
           \chi_M(\cgraph{N}).
  \end{align*}
  \vskip-0.25cm
  \defend
\end{Definition}

\subsection{The Bethe Entropy Function}

The next definition introduces the function that will be key towards the
promised combinatorial characterization of the Bethe entropy function.

\begin{Definition}
  \label{def:counting:valid:configurations:compatible:with:beliefs:1}

  Let $\graphN(\setF,\setE, \setA, \setG)$ be an NFG and let $\lmpBdown{}$ be its
  set of all lift-realizable pseudo-marginal vectors. Then, for every $M \in
  \Zpp$ and every $\vbel \in \lmpBdown{}$ we define
  \begin{alignat}{2}
    \chi_{M,\vbel}: \ \ 
      &\cset{N}_M
       &&\ \to \ 
         \R, \nonumber \\
      &\cgraph{N}
       &&\ \mapsto \ 
         \card{
           \left\{
             \cvc \in \codeCedge(\cgraph{N})
           \ \middle| \ 
             \varphiM\big( \cgraph{N},\cvc \big) = \vbel
         \right\}}.
           \label{eq:average:number:of:valid:configurations:1}
  \end{alignat}
  \defend
\end{Definition}

Note that for an $M$-cover $\cgraph{N}$ of $\graphN$ the value of
$\chi_{M,\vbel}(\cgraph{N})$ represents the number of valid configurations in
$\cgraph{N}$ that map down to $\vbel$. Consequently,
\begin{align}
  \bar C_M(\vbel)
    &\defeq 
       \Big\langle
         \chi_{M,\vbel}(\cgraph{N})
       \Big\rangle_{\cgraph{N} \in \cset{N}_M}
         \label{eq:def:C:bar:1}
\end{align}
is the average number of valid configurations that map down to $\vbel$, where
the averaging is over all $M$-covers of $\graph{N}$. (Observe that this is the
same $\bar C_M(\vbel)$ as in Section~\ref{sec:introduction:1}.) Letting
$\varphiM^{-1}$ denote the inverse of the mapping $\varphiM$, the quantity
$\bar C_M(\vbel)$ can also be written in terms of the pre-image of $\vbel$
under the mapping $\varphiM$, \ie,
\begin{align*}
  \bar C_M(\vbel)
    &= \frac{\card{\varphiM^{-1}({\vbel})}}
            {\cardbig{\cset{N}_M}}.
\end{align*}

\begin{Lemma}
  \label{lemma:average:number:pre:images:in:graph:covers:1}

  Let $\graphN(\setF,\setE, \setA, \setG)$ be some NFG, and for every $M \in
  \Zpp$ let $\lmpBdown{M}$ be its set of all $M$-cover lift-realizable
  pseudo-marginal vectors. Then for every $\vbel \in \lmpBdown{M}$ we have
  \begin{align*}
    \bar C_M(\vbel)
      &= \left(
           \prod_{f \in \setF}\!
             {M \choose M \cdot \vbel_f}
         \right)
         \!\cdot\!
         \left(
           \prod_{e \in \setEfull}\!\!
             {M \choose M \cdot \vbel_e}
         \right)^{-1},
  \end{align*}
  where we have used the multinomial coefficients
  \begin{align}
    {M \choose M \cdot \vbel_f}
      &\defeq
         \frac{M!}
         {\prod_{\va_f} (M \bel_{f,\va_f})!},
           \label{eq:M:choose:bel:1} \\
    {M \choose M \cdot \vbel_e}
      &\defeq
         \frac{M!}
         {\prod_{\va_e} (M \bel_{e,a_e})!}.
           \label{eq:M:choose:bel:2}
  \end{align}
  (Note that the components of $M \cdot \vbel$ are non-negative integers and
  so these expressions are indeed well defined.)
\end{Lemma}

\begin{Proof}
  See
  Appendix~\ref{sec:proof:lemma:average:number:pre:images:in:graph:covers:1}.
\end{Proof}

Note that the multinomial coefficients that appear in the above expression for
$\bar C_M(\vbel)$ have different origins. Namely, the multinomial coefficients
in the numerator of $\bar C_M(\vbel)$ stem from counting locally valid
configurations at the function nodes (see the proof of
Lemma~\ref{lemma:average:number:pre:images:in:graph:covers:1} in
Appendix~\ref{sec:proof:lemma:average:number:pre:images:in:graph:covers:1} for
the definition of ``locally valid configurations''), whereas the multinomial
coefficients in the denominator of $\bar C_M(\vbel)$ stem from counting the
number of edge connections that lead to overall valid configurations, and from
the division by the total number of $M$-covers.

The next theorem states the first main result of this section (and also of
this paper). It connects the asymptotic behavior of $\bar C_M(\vbel)$ with the
Bethe entropy function value of $\vbel$. Therefore, this result gives the
promised combinatorial characterization of the Bethe entropy function value of
$\vbel$. (The second main result of this section will be the combinatorial
characterization of the Bethe partition function presented in
Theorem~\ref{theorem:degree:M:partition:function:1}.)

\begin{Theorem}
  \label{theorem:asymptotic:average:number:pre:images:in:graph:covers:1}

  Let $\graphN(\setF,\setE, \setA, \setG)$ be some NFG and let $\lmpBdown{}$ be its
  set of lift-realizable pseudo-marginal vectors. For any $\vbel \in \lmpBdown{}$
  we have
  \begin{align*}
    \limsup_{M \to \infty}
      \frac{1}{M}
        \log
          \big(
            \bar C_M(\vbel)
          \big)
          &= \HBethe(\vbel).
  \end{align*}
\end{Theorem}

\begin{Proof}
  There are infinitely many $M \in \Zpp$ such that $\vbel \in
  \lmpBdown{M}$. This can be seen as follows. Namely, by definition of
  $\lmpBdown{}$, there must be at least one $M^* \in \Zpp$ such that $\vbel \in
  \lmpBdown{M^*}$. However, because $\vbel \in \lmpBdown{M^*}$ implies that
  $\vbel \in \lmpBdown{M}$ holds for any $M \in \Zpp$ that is divisible by
  $M^*$, there are in fact infinitely many $M \in \Zpp$ such that $\vbel \in
  \lmpBdown{M}$. The theorem statement then follows by combining
  Lemma~\ref{lemma:average:number:pre:images:in:graph:covers:1}, the results
  \begin{align*}
    {M \choose M \cdot \vbel_f}
      &= \exp
           \left(
             -
             M
             \cdot
             \sum_{\va_f}
               \bel_{f,\va_f}
               \log
               \left(
                 \bel_{f,\va_f}
               \right)
             +
             o(M)
           \right), \\
    {M \choose M \cdot \vbel_e}
      &= \exp
           \left(
             -
             M
             \cdot
             \sum_{a_e}
               \bel_{e,a_e}
               \log
               \left(
                 \bel_{e,a_e}
               \right)
             +
             o(M)
           \right),
  \end{align*}
  for $\vbel \in \lmpBdown{M}$ (which are consequences of Stirling's approximation
  of the factorial function), and the definition of the Bethe entropy function
  in Definition~\ref{def:Bethe:free:energy:1}.
\end{Proof}

A straightforward consequence of
Theorem~\ref{theorem:asymptotic:average:number:pre:images:in:graph:covers:1}
is that for $\vbel \in \lmpBdown{M}$ we have
\begin{align}
  \bar C_M(\vbel)
    &= \exp\big( M \cdot \HBethe(\vbel) + o(M) \big).
         \label{eq:asymptotic:average:number:pre:images:in:graph:covers:1}
\end{align}
Therefore, the Bethe entropy function value of $\vbel$ has the meaning of
being the asymptotic growth rate of the average number of valid configurations
in $M$-covers that map down to $\vbel$, where the averaging is over all
$M$-covers of $\graph{N}$, and where asymptotic is in the sense that $M$ goes
to infinity.

At this point, we encourage the reader to compare the observations that were
made so far in this section with similar statements that were made in
Lemma~\ref{lemma:method:of:types:properties:1} with respect to
$C_M(\vq)$. There are many similarities, but also some key differences. One
key difference is the following:
\begin{itemize}

\item In the setup of the present section, we count the average number of
  certain valid configurations in $M$-covers of some NFG $\graphN$. Most
  importantly, every $M$-cover is obtained by suitably ``intertwining'' $M$
  independent and identical copies of $\graphN$.

\item In the setup of Section~\ref{sec:Gibbs:free:energy:arises:naturally:1},
  we count certain valid configurations in $\graphN^M$, which corresponds to
  counting certain valid configurations in the $M$-cover of $\graphN$ that
  consists of $M$ independent and identical copies of $\graphN$.\footnote{More
    precise would be ``\ldots which corresponds to counting certain valid
    configurations in one of the $M$-covers of $\graphN$ that consists of $M$
    independent and identical copies of $\graphN$.''}

\end{itemize}
In conclusion: when counting certain valid configurations in ``intertwined''
$M$-covers we get the Bethe entropy function, whereas when counting certain
valid configurations in ``non-intertwined'' $M$-covers we get the Gibbs
entropy function. (See also the comments at the end of the upcoming
Section~\ref{sec:degree:M:Bethe:partition:function:1}.)

\subsection{The Bethe Average Energy Function}

In this subsection we show how the global function of an $M$-cover of some
base NFG can be expressed in terms of the Bethe average energy
function of this base NFG.

\begin{Theorem}
  \label{theorem:asymptotic:average:function:value:pre:images:in:graph:covers:1}

  Let $\graphN(\setF,\setE, \setA, \setG)$ be some NFG and let $M \in \Zpp$.
  Then for any $M$-cover $\cgraph{N}$ of $\graphN$ and any $\cvc \in
  \codeCedge(\cgraph{N})$ we have
  \begin{align*}
    -
    \frac{1}{M}
    \log 
      g_{\cgraph{N}}(\cvc)
      &= \Big.
          \UBethe(\vbel)
         \Big|_{\vbel = \varphiM(\cgraph{N},\cvc)}.
  \end{align*}
  (Note that this expression does not involve a limit $M \to \infty$.)
\end{Theorem}

\begin{Proof}
  Let $\csetF$ be the set of function nodes of $\cgraph{N}$. Then
  \begin{align*}
    g_{\cgraph{N}}(\cvc)
      &= \prod_{\cf \in \csetF}
           g_{\cf}(\cvc_{\cf})
       = \prod_{f \in \setF}
           \prod_{\va_f}
             \big(
               g_{f}(\va_f)
             \big)^{M \cdot \bel_{f,\va_f}}.
  \end{align*}
  Taking the logarithm on both sides, multiplying both sides by $-1/M$, and
  using the definition of the Bethe average energy function (see
  Definition~\ref{def:Bethe:free:energy:1}), we obtain the expression stated
  in the theorem.
\end{Proof}

Recall the definition of globally realizable pseudo-marginal vectors from
Remark~\ref{remark:lmpB:1}. One can easily show that for every $\vbel \in
\convhull(\lmpBdown{1})$ it holds that
\begin{align*}
  \UBethe(\vbel)
    &= \UGibbs(\vp),
\end{align*}
where $\vp \in \setPi_{\codeC}$ is the distribution whose marginals are given
by $\vbel$. In this sense, the Bethe average energy function can be seen as a
``straightforward extension'' of the Gibbs average energy function from the
domain $\convhull(\lmpBdown{1})$ to the domain $\lmpB$.

\subsection{The Bethe Free Energy Function}

An immediate consequence of
Theorems~\ref{theorem:asymptotic:average:number:pre:images:in:graph:covers:1}
and~\ref{theorem:asymptotic:average:function:value:pre:images:in:graph:covers:1}
is that for any temperature $T \in \Rpp$, any $M \in \Zpp$, any $M$-cover
$\cgraph{N}$ of $\graphN$, and any $\cvc \in \codeCedge(\cgraph{N})$ it holds
that
\begin{align}
  g_{\cgraph{N}}(\cvc)^{1/T}
  \cdot
  \bar C_M(\vbel)
    &= \exp
         \left(
           -
           \frac{M}{T} \cdot \FBethe(\vbel)
           +
           o(M)
         \right),
         \label{eq:combinatorial:characterization:Bethe:free:energy:1}
\end{align}
where $\vbel \defeq \varphiM(\cgraph{N}, \cvc)$.

\subsection{The Degree-$M$ Bethe Partition Function}
\label{sec:degree:M:Bethe:partition:function:1}

The above developments motivate the following definition of a degree-$M$ Bethe
partition function, which, as we will show, has the property that in the limit
$M \to \infty$ it converges to the Bethe partition function. Note that in
contrast to the definition of the Bethe partition function in
Definition~\ref{def:Bethe:partition:function:1}, which was analytical, the
definition of the degree-$M$ Bethe partition function is combinatorial.

\begin{Definition}
  \label{def:degree:M:partition:function:1}
 
  Let $\graphN(\setF,\setE, \setA, \setG)$ be an NFG. For any
  temperature $T \in \Rpp$ and any $M \in \Zpp$, we define the degree-$M$
  Bethe partition function to be
  \begin{align*}
    \ZBetheM{M}(\graphN)
      &\defeq
         \sqrt[M]{\Big\langle \!
                    \ZGibbs(\cgraph{N})
                  \! \Big\rangle_{\cgraph{N} \in \cset{N}_{M}}}.
  \end{align*}
  (Note that the right-hand side of the above expression is based on the Gibbs
  partition function, see~\eqref{eq:Z:Gibbs:1}, and not on the Bethe partition
  function.)  \defend
\end{Definition}

From the above expression we see that the degree-$M$ Bethe partition function
is defined to be the $M$th root of the average Gibbs partition function,
where the averaging is done over all $M$-covers of $\graphN$.

With this we are in a position to formulate the second main result of this
section (and of this paper).

\begin{Theorem}
  \label{theorem:degree:M:partition:function:1}
 
  For any NFG $\graphN(\setF,\setE, \setA, \setG)$ and any
  temperature $T \in \Rpp$ it holds that
  \begin{align*}
    \limsup_{M \to \infty} \ 
      \ZBetheM{M}(\graphN)
      &= \ZBethe(\graphN).
  \end{align*}
\end{Theorem}

\begin{Proof}
  See Appendix~\ref{sec:proof:theorem:degree:M:partition:function:1}.
\end{Proof}

\begin{figure*}
  \begin{center}
    \begin{tabular}{|c|c|@{\hskip0.5cm}c@{\hskip0.5cm}|}
      \hline
      & NFG & valid configurations \\
      \hline
      & & \\
      \begin{sideways}\hspace{-0.70cm}$2$-cover $\cgraph{N}_{2,4}$\end{sideways} 
      & 
      \begin{minipage}[c]{0.20\linewidth}
        \begin{center}
          \begin{tabular}{c}
            \epsfig{file=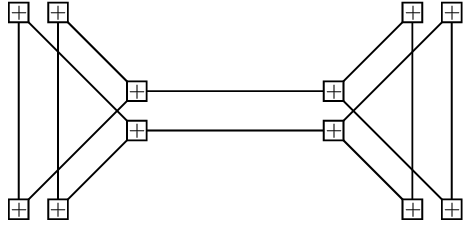, scale=0.50}
          \end{tabular}
        \end{center}
      \end{minipage}
      &
      \begin{minipage}[c]{0.60\linewidth}
        \begin{center}
          \begin{tabular}{cccc}
            \epsfig{file=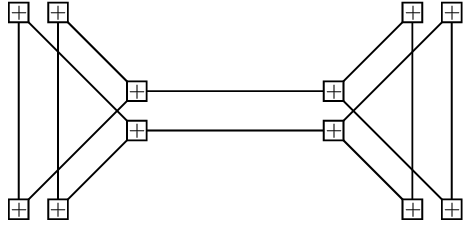, scale=0.50}
            &
            \epsfig{file=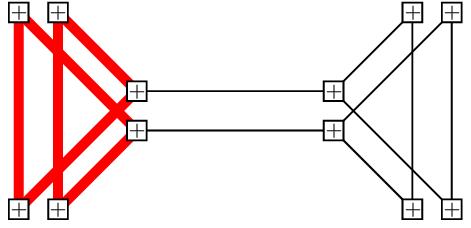, scale=0.50}
            &
            \epsfig{file=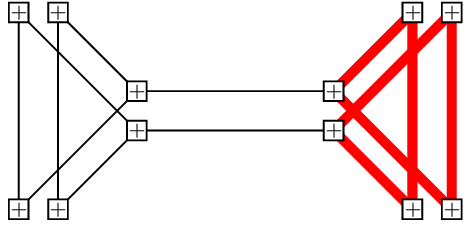, scale=0.50}
            &
            \epsfig{file=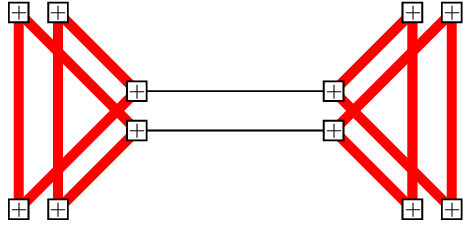, scale=0.50} \\[0.25cm]
            \epsfig{file=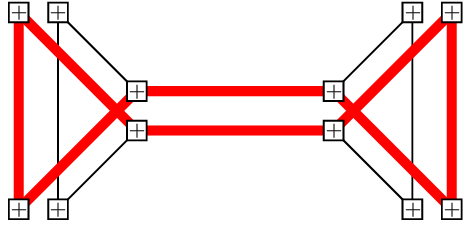, scale=0.50}
            &
            \epsfig{file=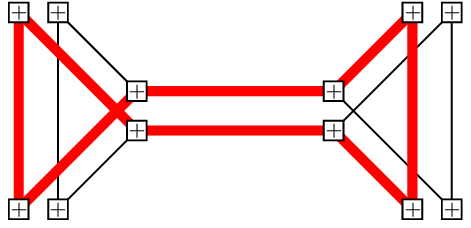, scale=0.50}
            &
            \epsfig{file=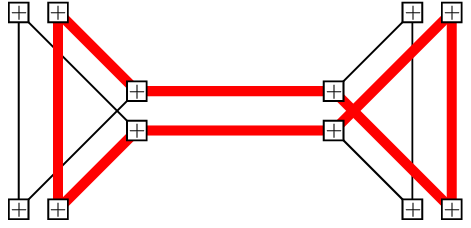, scale=0.50}
            &
            \epsfig{file=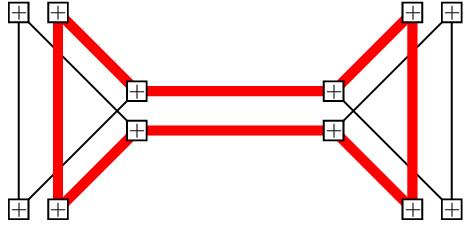, scale=0.50}
          \end{tabular}
        \end{center}
      \end{minipage} \\
      & & \\
      \hline
      & & \\
      \begin{sideways}\hspace{-0.70cm}$2$-cover $\cgraph{N}_{2,3}$\end{sideways} 
      &
      \begin{minipage}[c]{0.20\linewidth}
        \begin{center}
          \begin{tabular}{c}
            \epsfig{file=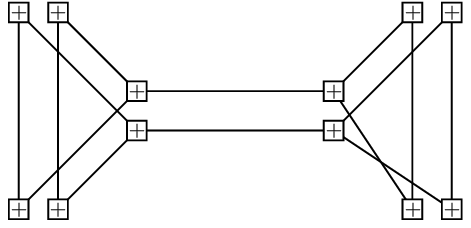, scale=0.50}
          \end{tabular}
        \end{center}
      \end{minipage}
      &
      \begin{minipage}[c]{0.60\linewidth}
        \begin{center}
          \begin{tabular}{cccc}
            \epsfig{file=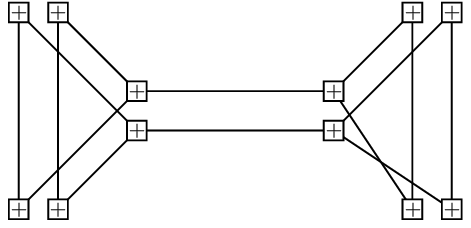, scale=0.50}
            &
            \epsfig{file=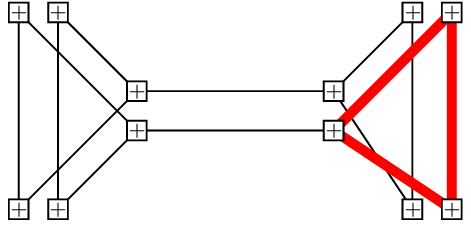, scale=0.50}
            &
            \epsfig{file=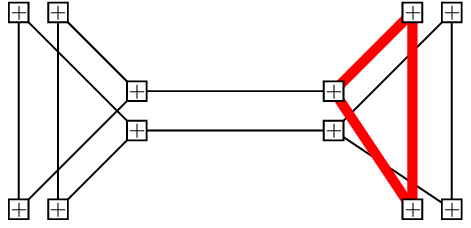, scale=0.50}
            &
            \epsfig{file=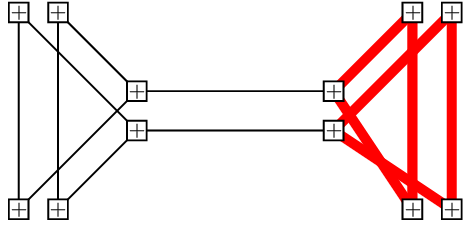, scale=0.50} \\[0.15cm]
            \epsfig{file=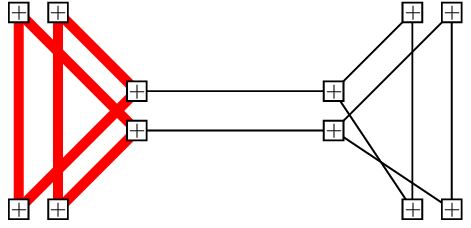, scale=0.50}
            &
            \epsfig{file=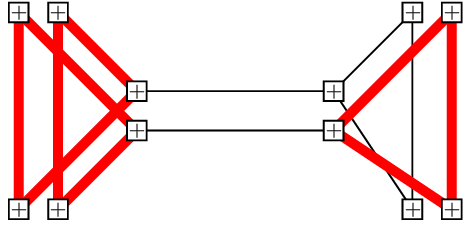, scale=0.50}
            &
            \epsfig{file=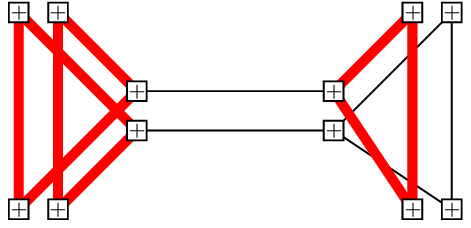, scale=0.50}
            &
            \epsfig{file=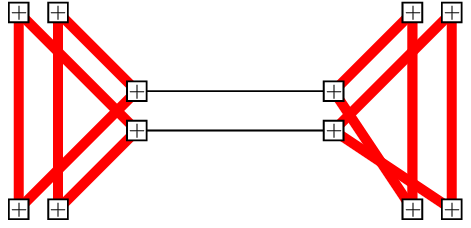, scale=0.50}
          \end{tabular}
        \end{center}
      \end{minipage} \\
      & & \\
      \hline
      & & \\
      \begin{sideways}\hspace{-0.70cm}$2$-cover $\cgraph{N}_{2,2}$\end{sideways} 
      & 
      \begin{minipage}[c]{0.20\linewidth}
        \begin{center}
          \begin{tabular}{c}
            \epsfig{file=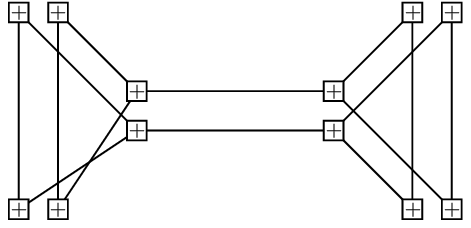, scale=0.50}
          \end{tabular}
        \end{center}
      \end{minipage}
      &
      \begin{minipage}[c]{0.60\linewidth}
        \begin{center}
          \begin{tabular}{cccc}
            \epsfig{file=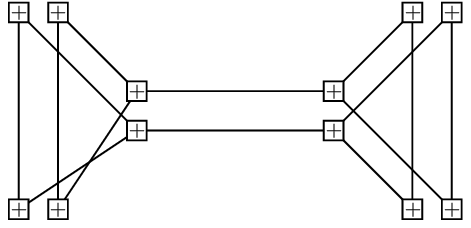, scale=0.50}
            &
            \epsfig{file=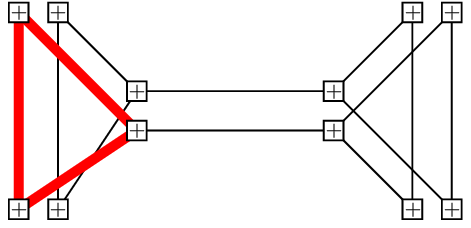, scale=0.50}
            &
            \epsfig{file=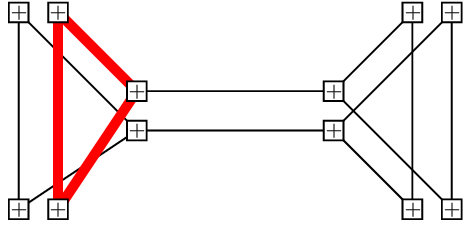, scale=0.50}
            &
            \epsfig{file=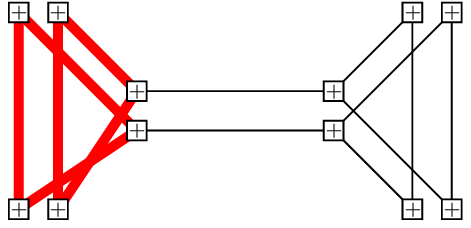, scale=0.50} \\[0.15cm]
            \epsfig{file=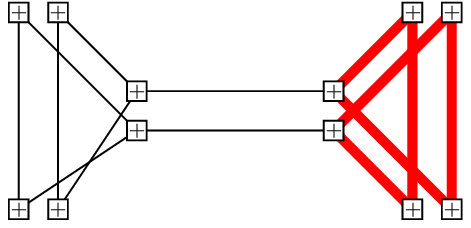, scale=0.50}
            &
            \epsfig{file=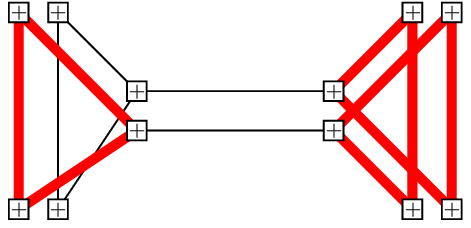, scale=0.50}
            &
            \epsfig{file=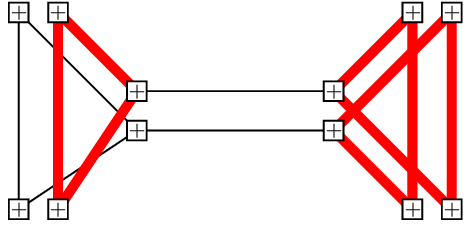, scale=0.50}
            &
            \epsfig{file=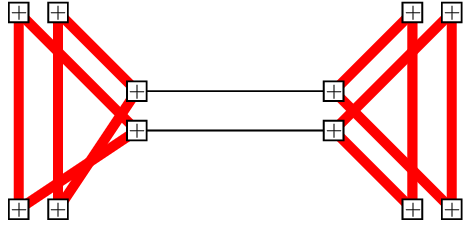, scale=0.50}
          \end{tabular}
        \end{center}
      \end{minipage} \\
      & & \\
      \hline
      & & \\
      \begin{sideways}\hspace{-0.70cm}$2$-cover $\cgraph{N}_{2,1}$\end{sideways} 
      & 
      \begin{minipage}[c]{0.20\linewidth}
        \begin{center}
          \begin{tabular}{c}
            \epsfig{file=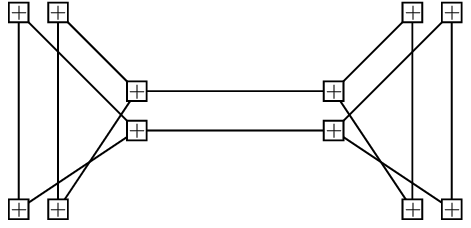, scale=0.50}
          \end{tabular}
        \end{center}
      \end{minipage}
      &
      \begin{minipage}[c]{0.60\linewidth}
        \begin{center}
          \begin{tabular}{cccc}
            \epsfig{file=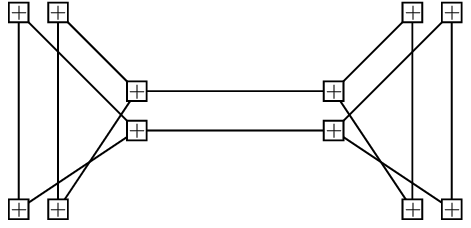, scale=0.50}
            &
            \epsfig{file=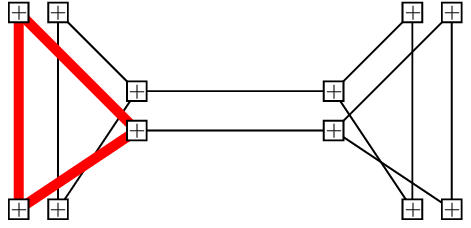, scale=0.50}
            &
            \epsfig{file=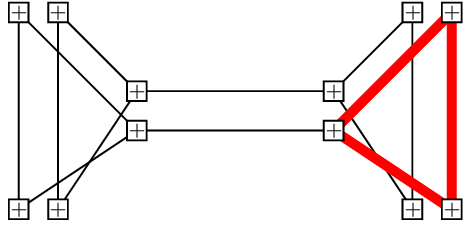, scale=0.50}
            &
            \epsfig{file=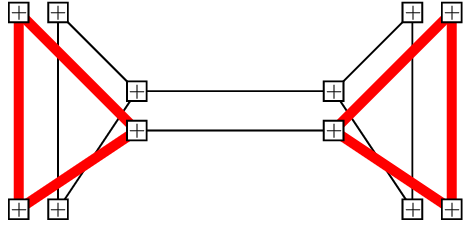, scale=0.50} \\[0.15cm]
            \epsfig{file=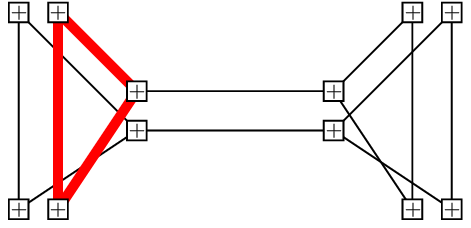, scale=0.50}
            &
            \epsfig{file=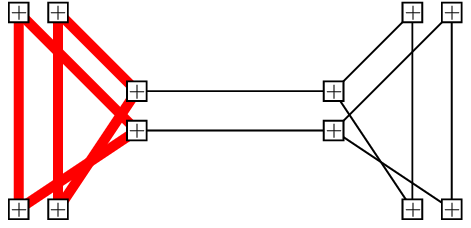, scale=0.50}
            &
            \epsfig{file=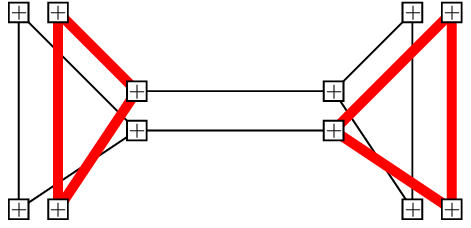, scale=0.50}
            &
            \epsfig{file=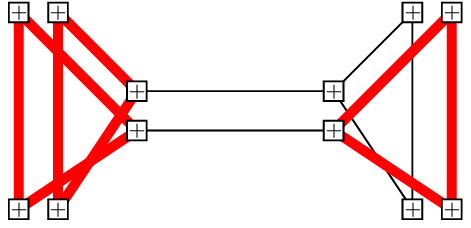, scale=0.50} \\[0.15cm]
            \epsfig{file=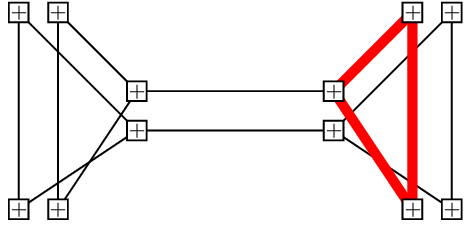, scale=0.50}
            &
            \epsfig{file=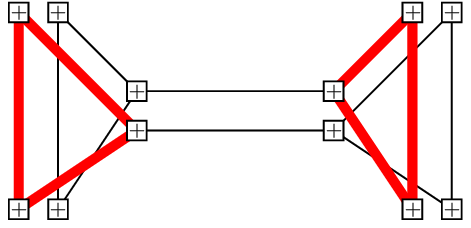, scale=0.50}
            &
            \epsfig{file=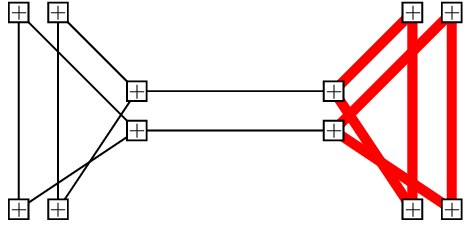, scale=0.50}
            &
            \epsfig{file=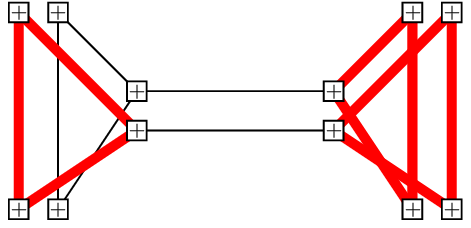, scale=0.50} \\[0.15cm]
            \epsfig{file=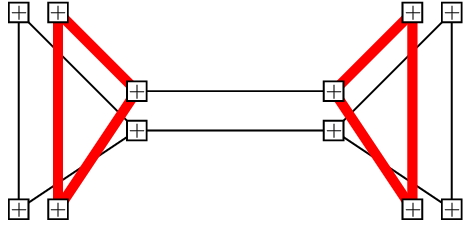, scale=0.50}
            &
            \epsfig{file=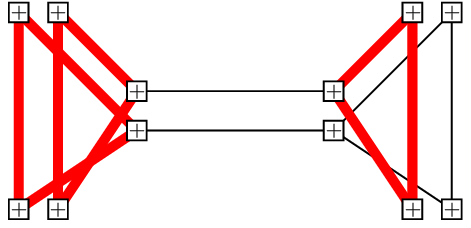, scale=0.50}
            &
            \epsfig{file=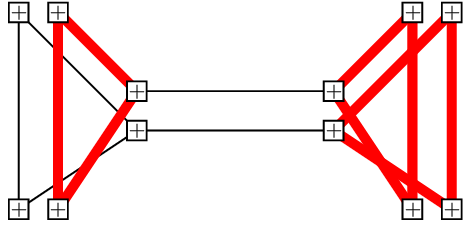, scale=0.50}
            &
            \epsfig{file=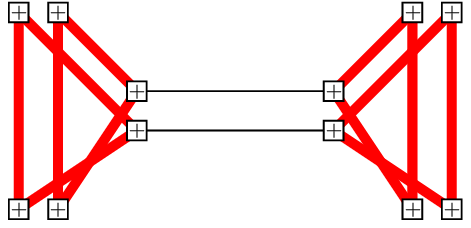, scale=0.50}
          \end{tabular}
        \end{center}
      \end{minipage} \\
      & & \\
      \hline
      \hline
      & & \\
      \begin{sideways}\hspace{-0.60cm}$1$-cover $\graphN$ \end{sideways} 
      & 
      \begin{minipage}[c]{0.20\linewidth}
        \begin{center}
          \begin{tabular}{c}
            \epsfig{file=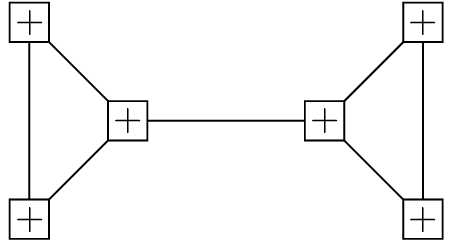, scale=0.50}
          \end{tabular}
        \end{center}
      \end{minipage}
      &
      \begin{minipage}[c]{0.60\linewidth}
        \begin{center}
          \begin{tabular}{cccc}
            \epsfig{file=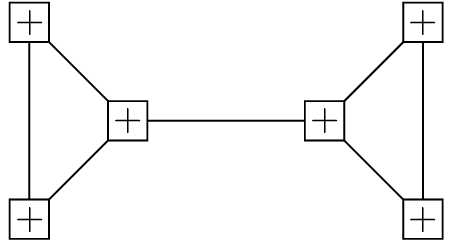, scale=0.50}
            &
            \epsfig{file=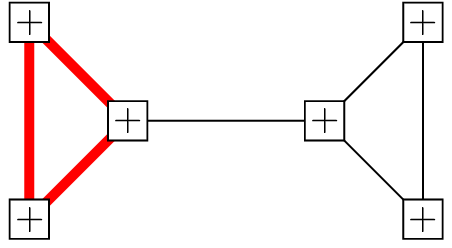, scale=0.50}
            &
            \epsfig{file=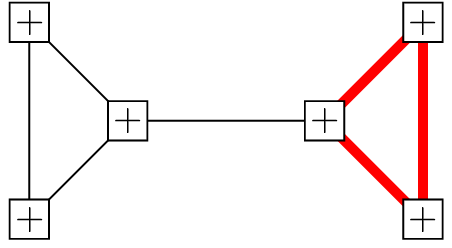, scale=0.50}
            &
            \epsfig{file=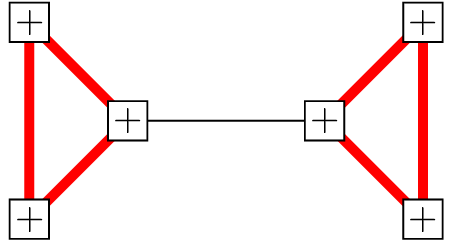, scale=0.50}
          \end{tabular}
        \end{center}
      \end{minipage} \\
      & & \\
      \hline
    \end{tabular}
  \end{center}
  \caption{NFGs that are used in Example~\ref{example:dumbbell:ffg:1}, along
    with their valid configurations. Concerning the NFGs that appear in the
    ``valid configurations'' column: for every $(e,m) \in \setE \times [M]$,
    if $\cover{c}_{e,m} = 0$ then the edge $(e,m)$ is thin and in black,
    whereas if $\cover{c}_{e,m} = 1$ then the edge $(e,m)$ is thick and in
    red.}
  \label{fig:dumbbell:ffg:hierarchy:1}
\end{figure*}

\begin{Example}
  \label{example:dumbbell:ffg:1}

  For improving one's understanding of $\ZBetheM{M}$, it is helpful to
  explictly compute this quantity for small NFGs and small values of $M$. To
  this end, consider the NFG $\graphN$ in the lower left corner of
  Fig.~\ref{fig:dumbbell:ffg:hierarchy:1}. Assume that the variable alphabets
  and local functions are defined analogously to variable alphabets and local
  functions of the NFG in Example~\ref{example:simple:ffg:1:cont:0}.  (The
  same NFG was also discussed in~\cite[Example~29]{Vontobel:Koetter:05:1:subm}
  and in~\cite[Example~2.4]{Koetter:Li:Vontobel:Walker:07:1}.)

  One can easily verify that all valid configurations take on the global
  function value one. Moreover, because $\graphN$ does not have half-edges,
  the set $\{ e \! \in \! \setE \ | \ c_e \! = \! 1 \}$ associated with a
  valid configuration $\vc$ forms a cycle or an edge-disjoint union of cycles
  in $\graphN$. With this, the set $\setC(\graphN)$ of valid configurations
  contains four elements, as shown in the last row of
  Fig.~\ref{fig:dumbbell:ffg:hierarchy:1}, \ie,
  \begin{align*}
    \ZGibbs(\graphN)
      &= 4.
  \end{align*}

  Because $\graphN$ has seven edges, there are $2^7 = 128$ distinct
  $2$-covers: $32$ of them are (when omitting the cover-related parts of the
  vertex and edge labels) isomorphic to $\cgraph{N}_{2,1}$, $32$ of them are
  isomorphic to $\cgraph{N}_{2,2}$, $32$ of them are isomorphic to
  $\cgraph{N}_{2,3}$, and $32$ of them are isomorphic to $\cgraph{N}_{2,4}$
  shown on the left-hand side of Fig.~\ref{fig:dumbbell:ffg:hierarchy:1}. One
  can verify that
  \begin{align*}
    \ZGibbs(\cgraph{N}_{2,1})
      &= 16, \quad
    \ZGibbs(\cgraph{N}_{2,2})
       = \ZGibbs(\cgraph{N}_{2,3})
       = \ZGibbs(\cgraph{N}_{2,4})
       = 8.
  \end{align*}
  Therefore, the degree-$2$ Bethe partition function is
  \begin{align*}
    \ZBetheM{2}(\graphN)
      &= \sqrt[2]{\frac{1}{128}
                  \sum_{h \in [4]}
                    32 \ZGibbs(\cgraph{N}_{2,h})
                 }
       = \sqrt[2]{10}
       = 3.162 \ldots \ .
  \end{align*}
  Some comments:
  \begin{itemize}
 
  \item The $2$-cover $\cgraph{N}_{2,1}$ consists of two copies of $\graphN$
    and consequently we have $\ZGibbs(\cgraph{N}_{2,1}) = \bigl(
    \ZGibbs(\graphN) \bigr)^2 = 4^2 = 16$. If $\ZGibbs(\cgraph{N}) = \bigl(
    \ZGibbs(\graphN) \bigr)^2$ were true for all $2$-covers of $\graphN$, then
    $\ZBetheM{2}(\graphN) = \ZGibbs(\graphN)$.

  \item As we can see from Fig.~\ref{fig:dumbbell:ffg:hierarchy:1}, there are
    $2$-covers $\cgraph{N}$ such that $\ZGibbs(\cgraph{N}) \neq \bigl(
    \ZGibbs(\graphN) \bigr)^2$. Therefore, it is not surprising that
    $\ZBetheM{2}(\graphN) \neq \ZGibbs(\graphN)$.

  \item If all $2$-covers were like $\cgraph{N}_{2,h}$, $h \in [3]$, then the
    set of $2$-cover lift-realizable pseudo-marginal vectors would satisfy
    $\convhull(\lmpBdown{2}) = \convhull(\lmpBdown{1})$. However, the $2$-cover
    $\cgraph{N}_{2,4}$ contains some configurations whose associated
    pseudo-marginal vector does not lie within
    $\convhull(\lmpBdown{1})$. Therefore, $\convhull(\lmpBdown{2}) \supsetneq
    \convhull(\lmpBdown{1})$, and so, because $\lmpB \supseteq
    \convhull(\lmpBdown{2})$, we literally see why the NFG $\graphN$ is an example
    where the local marginal polytope satisfies $\lmpB \supsetneq
    \convhull(\lmpBdown{1})$. (See Remark~\ref{remark:lmpB:1} for a related
    observation.)

  \item As mentioned in Section~\ref{sec:related:work:1}, one can also give a
    combinatorial characterization of the Bethe partition function of an NFG
    $\graphN$ in terms of computation trees and the universal cover
    $\graph{\hat N}$ of $\graphN$. However, in many respects, finite graph
    covers are easier to deal with, and, as this example shows, many effects
    that are responsible for the similarities and differences between
    $\ZBethe(\graphN)$ and $\ZGibbs(\graphN)$ are already visible in finite
    graph covers with small cover degree $M$.  \exampleend

  \end{itemize}
\end{Example}

As the following lemma shows, it is no coincidence that $\ZBetheM{2}(\graphN)$
is a lower bound of $\ZGibbs(\graphN)$ for the NFG $\graphN$ in
Example~\ref{lemma:circuit:NFG:1}. Let $\ncomponents\graphN$ be the number of
connected components of $\graphN$, when $\graphN$ is considered as a graph.

\begin{Lemma}
  \label{lemma:circuit:NFG:1}

  Consider an NFG $\graphN$ as defined in
  Example~\ref{example:dumbbell:ffg:1}, in particular without half-edges. For
  any $M \in \Zpp$ it holds that
  \begin{alignat}{2}
    2^{-((M-1)/M) \cdot \ncomponents\graphN}
      \cdot
      \ZGibbs(\graphN)
      &\ \leq \ 
    \ZBetheM{M}(\graphN)
      &{} \ \leq \ 
         \ZGibbs(\graphN), \nonumber \\
    2^{-\ncomponents\graphN}
      \cdot
      \ZGibbs(\graphN)
      &{} \ \leq \ 
    \ZBethe(\graphN)
      &\ \leq \ 
         \ZGibbs(\graphN).
           \label{eq:lemma:circuit:NFG:ZBethe:1}
  \end{alignat}
  Equivalently,
  \begin{align}
    \ZBetheM{M}(\graphN)
      &\ \leq \ 
         \ZGibbs(\graphN)
       \ \leq \
         2^{((M-1)/M) \cdot \ncomponents\graphN}
           \cdot
           \ZBetheM{M}(\graphN), \nonumber \\
    \ZBethe(\graphN)
      &\ \leq \ 
         \ZGibbs(\graphN)
       \ \leq \ 
         2^{\ncomponents\graphN}
           \cdot
           \ZBethe(\graphN).
             \label{eq:lemma:circuit:NFG:ZBethe:2}
  \end{align}
  
\end{Lemma}

\begin{Proof}
  Because $\ZGibbs(\graphN)$ equals the number of cycles and edge-disjoint
  unions of cycles of $\graphN$, we get
  \begin{align}
    \ZGibbs(\graphN)
      &= 2^{\cirrank(\graphN)},
           \label{eq:ZGibbs:circuit:rank:1}
  \end{align}
  where
  \begin{align}
    \cirrank(\graphN)
      &= |\setE(\graphN)|
         -
         |\setF(\graphN)|
         +
         \ncomponents\graphN
           \label{eq:circuit:rank:1}
  \end{align}
  is the circuit rank of $\graphN$. Similarly, for any $M$-cover $\cgraph{N}$
  of $\graphN$ we obtain
  \begin{align}
    \ZGibbs(\cgraph{N})
      &= 2^{\cirrank(\cgraph{N})},
           \label{eq:ZGibbs:circuit:rank:cover:1}
  \end{align}
  where
  \begin{align}
    \cirrank(\cgraph{N})
      &= |\setE(\cgraph{N})|
         -
         |\setF(\cgraph{N})|
         +
         \ncomponents\cgraph{N}.
           \label{eq:circuit:rank:cover:1}
  \end{align}
  From straightforward graph-theoretic considerations of $M$-covers, in
  particular~\eqref{eq:ncomponents:inequalities:1}, it follows that
  \begin{align}
    |\setE(\cgraph{N})|
      &= M
           \cdot
           |\setE(\graphN)|,
             \label{eq:circuit:rank:setE:cover:1} \\
    |\setF(\cgraph{N})|
      &= M
           \cdot
           |\setF(\graphN)|,
             \label{eq:circuit:rank:setF:cover:1} \\
    \ncomponents\graphN
       \leq
         \ncomponents\cgraph{N}
      &\leq
         M
           \cdot
           \ncomponents\graphN.
             \label{eq:circuit:rank:comp:cover:1}
  \end{align}
  Combining~\eqref{eq:circuit:rank:1}, \eqref{eq:circuit:rank:cover:1},
  and~\eqref{eq:circuit:rank:setE:cover:1}--\eqref{eq:circuit:rank:comp:cover:1},
  we obtain
  \begin{align*}
    M
      \cdot
      \cirrank(\graphN)
    - 
    (M \! - \! 1)
      \cdot
      \ncomponents\graphN
       \, \leq \,
    \cirrank(\cgraph{N})
      &\, \leq \,
         M
           \cdot
           \cirrank(\graphN).
  \end{align*}
  Then, with the help of~\eqref{eq:ZGibbs:circuit:rank:1}
  and~\eqref{eq:ZGibbs:circuit:rank:cover:1}, we get
  \begin{align*}
    2^{- 
       (M-1) \cdot \ncomponents\graphN
      }
      \cdot
      \big(
        \ZGibbs(\graphN)
      \big)^M
       \, \leq \,
         \ZGibbs(\cgraph{N})
      &\, \leq \,
         \big(
           \ZGibbs(\graphN)
         \big)^M.
  \end{align*}
  Plugging these expressions into the definition of $\ZBetheM{M}(\graphN)$,
  see Definition~\ref{def:degree:M:partition:function:1}, yields the result
  that was promised in the lemma statement.
\end{Proof}

We conclude this subsection with a few comments and observations.
\begin{itemize}

\item Of course, the complexity of computing $\ncomponents\graphN$ and
  $\cirrank(\graphN)$ is polynomial in $|\setF(\graphN)|$ and
  $|\setE(\graphN)|$; therefore, the complexity of computing
  $\ZGibbs(\graphN)$ in Lemma~\ref{lemma:circuit:NFG:1} is polynomial in
  $|\setF(\graphN)|$ and $|\setE(\graphN)|$. Moreover, computing the lower and
  upper bounds in~\eqref{eq:lemma:circuit:NFG:ZBethe:2} is equally complex.
  The relevance of Example~\ref{example:dumbbell:ffg:1} and
  Lemma~\ref{lemma:circuit:NFG:1} is therefore not that an intractable
  partition function $\ZGibbs(\graphN)$ is approximated by lower and upper
  bounds based on a tractable Bethe partition function $\ZBethe(\graphN)$, but
  to exhibit an example where explicit computations can easily be done and
  insight be gained into the formalism presented in this section.

\item The inequalities in~\eqref{eq:lemma:circuit:NFG:ZBethe:1}
  and~\eqref{eq:lemma:circuit:NFG:ZBethe:2} can also easily be obtained by
  minimizing the Bethe free energy function and using the expression in
  Definition~\ref{def:Bethe:partition:function:1}. However, using the
  combinatorial characterization of $\ZBethe(\graphN)$ gives us additional
  insights why~\eqref{eq:lemma:circuit:NFG:ZBethe:1} holds. In fact, analyzing
  the proof of Lemma~\ref{lemma:circuit:NFG:1} we see that these inequalities
  are a straightforward consequence of the graph-theoretic inequalities
  \begin{align*}
    \ncomponents\graphN
       \,\leq\,
         \ncomponents\cgraph{N}
      &\,\leq\,
         M
           \!\cdot\!
           \ncomponents\graphN
  \end{align*}
  that hold for any $M$-cover $\cgraph{N}$ of $\graphN$, see
  also~\eqref{eq:ncomponents:inequalities:1}.

\item Note that proving the inequality $\ZGibbs(\cgraph{N}) \leq \big(
  \ZGibbs(\graphN) \big)^M$ for every $M$-cover $\cgraph{N}$ of $\graphN$, as
  was done in the proof of Lemma~\ref{lemma:circuit:NFG:1}, was also at the
  heart of the recent proof by Ruozzi~\cite{Ruozzi:12:1} of the inequality
  $\ZBethe(\graphN) \leq \ZGibbs(\graphN)$ for log-supermodular graphical
  models $\graphN$. (This verified a conjecture by Sudderth, Wainwright, and
  Willsky~\cite{Sudderth:Wainwright:Willsky:07:1}.)

  Moreover, the paper~\cite{Vontobel:11:3:subm} presented setups where the
  inequality $\ZGibbs(\cgraph{N}) \leq \big( \ZGibbs(\graphN) \big)^M$ holds
  for every $M$-cover $\cgraph{N}$ of an NFG $\graphN$ whose partition
  function represents the permanent of a non-negative matrix, and pointed out
  setups where this inequality is conjectured to hold. Further NFGs where this
  inequality is conjectured to hold were listed by
  Watanabe~\cite{Watanabe:11:1}.

\begin{figure}
  \begin{alignat*}{2}
    &\big.
       \ZBetheM{M}(\graphN)
     \big|_{M \to \infty}
        &&= \ZBethe(\graphN) \\
    &\hskip1cm \Big\vert \\
    &\big.
       \ZBetheM{M}(\graphN)
     \big. && \\
    &\hskip1cm \Big\vert \\
    &\big.
       \ZBetheM{M}(\graphN)
     \big|_{M = 1}
       &&= \ZGibbs(\graphN)
  \end{alignat*}
  \caption{The degree-$M$ Bethe partition function of the NFG $\graphN$ for
    different values of $M$.}
  \label{fig:degree:M:Bethe:partition:function:1}
\end{figure}

\item When considering the value of $\ZBetheM{M}(\graphN)$ from $M = 1$ to $M
  = \infty$, one goes from $\ZGibbs(\graphN)$ to $\ZBethe(\graphN)$,
  see~Fig.~\ref{fig:degree:M:Bethe:partition:function:1}. It is worthwhile to
  consider the inequalities that appear in Lemma~\ref{lemma:circuit:NFG:1}
  under this perspective.

\item We can write the ratio $\ZBethe(\graphN) / \ZGibbs(\graphN)$ as the
  following telescoping product
  \begin{align*}
    \frac{\ZBethe(\graphN)}
         {\ZGibbs(\graphN)}
      &= \lim_{M \to \infty}
           \prod_{M' \in [M]}
             \frac{\ZBetheM{M'+1}(\graphN)}
                  {\ZBetheM{M'}(\graphN)}.
  \end{align*}
  Towards a better understanding of the ratio $\ZBethe(\graphN) /
  \ZGibbs(\graphN)$, it might therefore be worthwhile to study the ratios
  $\ZBetheM{M+1}(\graphN) / \ZBetheM{M}(\graphN)$, $M \in \Zpp$. We leave it
  as an open problem to see if general statements can be made about them.

\item Let $\cset{N}^{\parallel}_M$ be the subset of $\cset{N}_M$ that contains
  all $M$-covers $\cgraph{N}$ that consist of $M$ disconnected copies of
  $\graphN$. It holds that $\ZGibbs(\cgraph{N}) = \bigl( \ZGibbs(\graphN)
  \bigr)^M$ for any $\cgraph{N} \in \cset{N}^{\parallel}_M$, and so, trivially,
  \begin{align*}
    \ZGibbs(\graphN)
      &\defeq
         \sqrt[M]{\Big\langle \!
                    \ZGibbs(\cgraph{N})
                  \! \Big\rangle_{\cgraph{N} \in \cset{N}^{\parallel}_{M}}}.
  \end{align*}
  
\item If $\graphN$ does \emph{not} contain any cycles, then
  \begin{align*}
    \cset{N}^{\parallel}_{M}
    &= \cset{N}_{M},
         \quad M \in \Zpp, 
  \end{align*}
  and so $\ZGibbs(\graphN) = \ZBethe(\graphN)$.

\item If $\graphN$ does contain cycles then
  \begin{align*}
    \cset{N}^{\parallel}_{M}
      &\subsetneq
         \cset{N}_{M},
           \quad M \geq 2.
  \end{align*}
  Because usually $\ZGibbs(\cgraph{N}) \neq \bigl( \ZGibbs(\graphN) \bigr)^M$
  for $\cgraph{N} \in \cset{N}_{M} \setminus \cset{N}^{\parallel}_{M}$, it is
  not surprising that usually $\ZGibbs(\graphN) \neq \ZBethe(\graphN)$ for an
  NFG $\graphN$ with cycles.

\end{itemize}

\subsection{Similarities and Differences w.r.t.\ the Replica Method}
\label{sec:replica:method:1}

In this subsection we discuss similarities and differences between, on the one
hand, the concepts and the mathematical expressions that have so far appeared
in this section, and, on the other hand, concepts and mathematical expressions
that appear in the replica theory (see, \eg,
\cite{Mezard:Parisi:Virasoro:87:1}, \cite[Chapter~8]{Mezard:Montanari:09:1},
\cite[Appendix~I]{Tanaka:02:1}).

Let $\graphN_{N,R}$ be an NFGs with ``size parameter'' $N$ whose local
functions depend on a random variable (or random vector) $R$. Assume that
$\graphN_{N,R}$ represents some physical system. Many interesting physical
quantities about this physical system can then be derived from the normalized
log-partition function $\frac{1}{N} \log \bigl( \ZGibbsext{\graphN_{N,R}}
\bigr)$. However, because this expression is usually not tractable, one
studies the ensemble average $\Expec \bigl[ \frac{1}{N} \log \bigl(
\ZGibbsext{\graphN_{N,R}} \bigr) \bigr]$, where the expectation value is
w.r.t.~$R$. If ``measure concentration'' happens for large $N$, then the
normalized log-partition function of the ``typical'' NFG will be close to this
expression for large $N$.

Direct evaluation of $\Expec \bigl[ \frac{1}{N} \log \bigl(
\ZGibbsext{\graphN_{N,R}} \bigr) \bigr]$ is often not possible. Here is the
point where the replica method comes in. Inspired by the equation
\begin{align*}
  \log(z)
    &= \lim_{M \downarrow 0}
         \frac{z^M - 1}
              {M},
           \quad z \in \Rpp,
\end{align*}
where $M$ is considered to be a real number, the replica method proposes the
following reformulation of the above expectation value
\begin{align*}
  \Expec
    \left[
      \frac{1}{N}
      \log
        \big(
          \ZGibbsext{\graphN_{N,R}}
        \big)
    \right]
    &= \lim_{M \downarrow 0} \, 
         \frac{\Expec\Big[ \big( \ZGibbsext{\graphN_{N,R}} \big)^M \Big]
               -
               1}
              {N M}.
\end{align*}
One then notices that for \emph{positive integers} $M$ the term $\bigl(
\ZGibbsext{\graphN_{N,R}} \bigr)^M$, which appears on the right-hand side of
the above expression, corresponds to considering the partition function of $M$
independent copies of $\graphN_{N,R}$ (hence the name ``replica
theory''). After evaluating $\Expec\bigl[ \bigl( \ZGibbsext{\graphN_{N,R}}
\bigr)^M \bigr]$ for positive integers $M$, one then \emph{drops this
  requirement on} $M$, and evaluates the limit $M \downarrow 0$. This is the
gist behind the replica method. Much more can, and needs to be said, for which
we refer to~\cite{Mezard:Parisi:Virasoro:87:1},
\cite[Chapter~8]{Mezard:Montanari:09:1}, \cite[Appendix~I]{Tanaka:02:1} (and
references therein).

Clearly, there are similarities between the replica method and the
developments in this paper. However, there are also some stark differences.
\begin{itemize}

\item The NFG $\graphN_{N,R}$ depends on a random variable
  $R$, whereas the NFG $\graphN$ that is studied in this paper
  is deterministic. (Of course, our setup also allows NFGs
  that depend on a random variable, but that is not necessary.)

\item Because the random variable $R$ is the same in all $M$ copies, there is
  the appearance of some ``coupling effect'' between the $M$ copies of
  $\graphN_{N,R}$ when evaluating $\Expec\bigl[ \bigl(
  \ZGibbsext{\graphN_{N,R}} \bigr)^M \bigr]$. This is in contrast to the
  ``coupling effect'' that appears in an $M$-cover of $\graphN$ as a result of
  the graph-cover construction process where the edges of $M$ independent
  copies of $\graphN$ are permuted.

\item The replica method is based on studying the limit $M \downarrow 0$,
  whereas this paper typically studies the limit $M \to \infty$. Moreover,
  this paper never drops the requirement that $M$ is a positive integer.

\end{itemize}

Note that in coding theory, when studying the growth rate of the average
Hamming weight enumerator of a code ensemble~\cite{Gallager:63}, one usually
evaluates an expression like $\frac{1}{N} \log \bigl( \Expec \bigl[
\ZGibbsext{\graphN_{N,R}} \bigr] \bigr)$. This quantity can be the same as the
above-mentioned $\Expec \bigl[ \frac{1}{N} \log \bigl(
\ZGibbsext{\graphN_{N,R}} \bigr) \bigr]$ but in general one can only state
that
\begin{align*}
     \frac{1}{N}
     \log
       \Big(
         \Expec
           \big[
             \ZGibbsext{\graphN_{N,R}}
           \big]
       \Big)
    &\geq
       \Expec
         \Big[
           \frac{1}{N}
           \log
             \big(
               \ZGibbsext{\graphN_{N,R}}
             \big)
         \Big],
\end{align*}
which is a consequence of Jensen's inequality. For more information on these
types of issues we refer to, \eg, \cite{Di:Montanari:Urbanke:04:1}.

Let us conclude this subsection by mentioning a recent paper by
Mori~\cite{Mori:11:1} that was inspired by an earlier version of the present
paper and that offers an alternative (and simpler) approach to some
computations that are done in the context of the replica method. For more
details we refer to Mori's paper. See also~\cite{Mori:Tanaka:12:1}.

\section{NFGs for Channel Coding}
\label{sec:normal:factor:graphs:for:channel:coding:1}

The main purpose of this section is to introduce some notation and concepts
that will be useful for the next two sections, namely for
Section~\ref{sec:graph:cover:decoding:1} on graph-cover decoding and for
Section~\ref{sec:influence:minimum:Hamming:distance:1} on a connection between
the minimum Hamming distance of a code and the non-concavity of the Bethe
entropy function of some graphical model that represents this code.

For the following definition, we remind the reader of the sets $\codeCedge$
and $\codeCedgehalf$ that were specified in
Definition~\ref{def:normal:factor:graph:2} and the modified Gibbs partition
function that was specified in Lemma~\ref{lemma:F:Gibbs:minimum:2}.

\begin{Definition}
  \label{def:normal:factor:graph:representing:a:code:1}
  
  Let $\setX$ be some finite set and let $\codeCchannel$ be a length-$n$
  channel code over $\setX$, \ie, $\codeCchannel \subseteq \setX^n$. We say
  that an NFG $\graphN(\setF,\setE, \setA, \setG)$ represents the code
  $\codeCchannel$ if the following four conditions are satisfied.
  \begin{itemize}
  
  \item All local functions are indicator functions.

  \item For every $e \in \setEhalf$ we have $\setA_e = \setX$.

  \item The code $\codeCchannel$ is the projection of $\codeCedge$ to
    $\setEhalf$, \ie,
    \begin{align*}
      \codeCchannel
        &= \codeCedgehalf
         \defeq
           \big\{ 
             (c_e)_{e \in \setEhalf}
           \ \big| \ 
             \vc \in \codeCedge
           \big\}.
    \end{align*}
  
  \item There is a $t_{\graphN} \in \Zpp$ such that
    \begin{align*}
      \Big|
        \bigl\{
          \vc \in \codeCedge
          \bigm|
          (c_e)_{e \in \setEhalf} = \vx
        \bigr\}
      \Big|
        &= t_{\graphN}
             \quad \text{(for all $\vx \in \codeCedgehalf$)},
    \end{align*}
    \ie, for every $\vx \in \codeCedgehalf$ there are $t_{\graphN}$ valid
    configurations $\vc \in \codeCedge$ whose restriction to $\setEhalf$
    equals $\vx$. \defend

  \end{itemize}
\end{Definition}

\noindent
Let us comment on this definition.
\begin{itemize}

\item One can verify that the last condition is always satisfied in the
  following important special case: namely the case where all edge alphabets
  are equal to some group and all local functions represent indicator
  functions of subgroups of this group. (The proof of this statement uses the
  fact that all cosets of a subgroup have the same size. We leave the details
  to the reader.)

\item Note that in Example~\ref{example:simple:ffg:1:cont:0}, for every $\vx
  \in \codeCedgehalf$, there are $t_{\graphN} = 4$ valid configurations in
  $\codeC$ whose restriction to $\setEhalf$ equals $\vx$.

\item Usually, an NFG that represents a code is set up such that $t_{\graphN}
  = 1$. However, sometimes it is more natural to set up $\graphN$ such that
  $t_{\graphN} > 1$. For a more detailed discussion of this and related issues
  we refer the interested reader to, \eg, \cite{Forney:GluesingLuerssen:12:1}.

\end{itemize}

\begin{figure}
  \hskip0.1cm
  \begin{minipage}[c]{0.40\linewidth}
    \epsfig{file=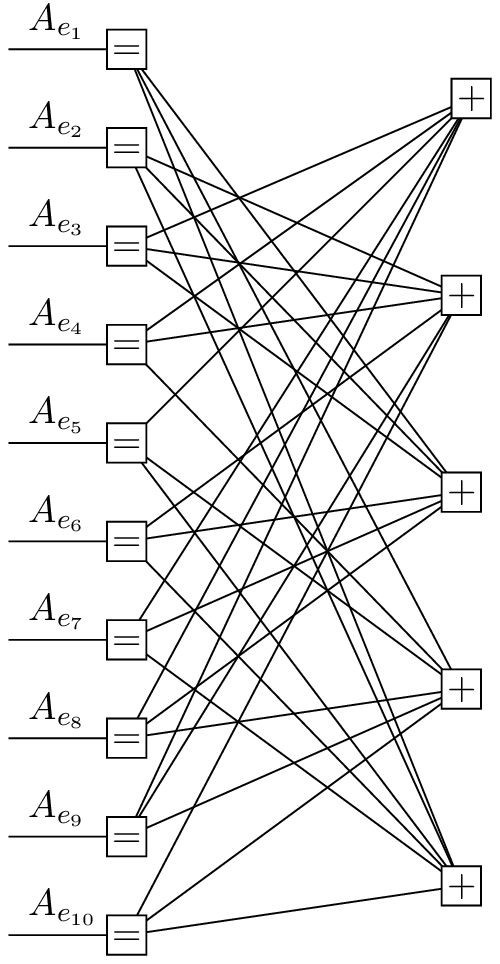, scale=0.70}
  \end{minipage}
  \hskip1cm
  \begin{minipage}[c]{0.40\linewidth}
    \mbox{} \\[0.2cm]
    \epsfig{file=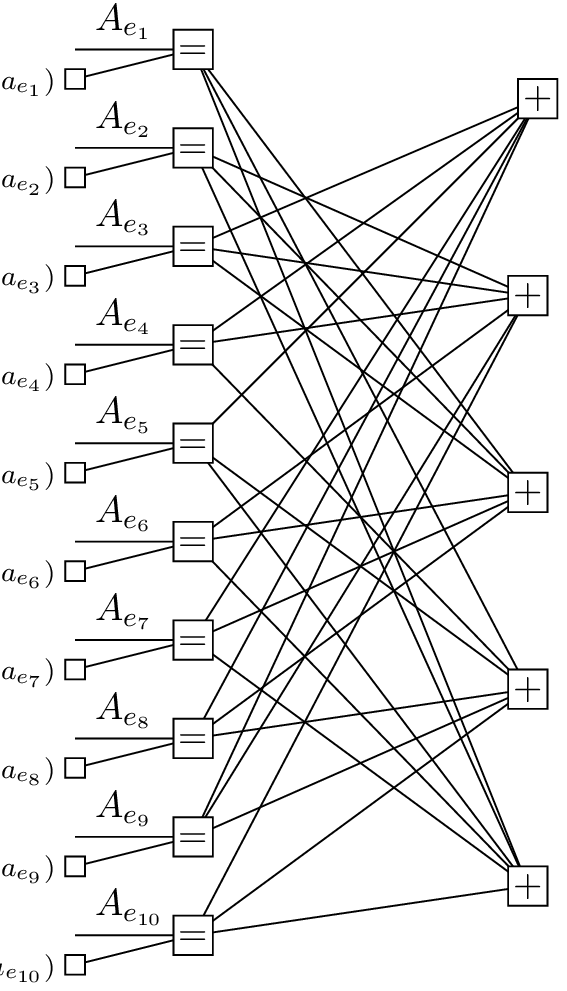, scale=0.70}
  \end{minipage}
  \caption{Left: NFG representing the code $\codeCchannel$ in
    Example~\ref{example:normal:factor:graph:ldpc:code:1}. Right: normal
    factor graph $\graphN(\vy)$ that is discussed in
    Example~\ref{example:normal:factor:graph:ldpc:code:2}.}
  \label{fig:normal:factor:graph:ldpc:code:1}
\end{figure}

\begin{Example}
  \label{example:normal:factor:graph:ldpc:code:1}  

  Consider the length-$10$ code $\codeCchannel$ over $\GF{2}$ defined by the
  parity-check matrix
  \begin{align*}
    \matrH
      \defeq
        \begin{bmatrix}
          0 & 0 & 1 & 1 & 1 & 0 & 1 & 1 & 1 & 0 \\
          0 & 1 & 1 & 1 & 0 & 1 & 0 & 0 & 1 & 1 \\
          1 & 1 & 1 & 0 & 0 & 1 & 1 & 1 & 0 & 0 \\
          1 & 0 & 0 & 1 & 1 & 0 & 0 & 1 & 1 & 1 \\
          1 & 1 & 0 & 0 & 1 & 1 & 1 & 0 & 0 & 1
        \end{bmatrix},
  \end{align*}
  \ie, $\codeCchannel \defeq \bigl\{ \vx \in \GF{2}^n \bigm| \matrH
  \cdot \vx^\tr = \vect{0}^\tr \, (\text{in $\GF{2}$}) \bigr\}$, where vectors
  are row vectors and where $(\,\cdot\,)^\tr$ denotes vector
  transposition. This code can be represented by the NFG shown in
  Fig.~\ref{fig:normal:factor:graph:ldpc:code:1}~(left). Here, all edge
  alphabets are equal to $\GF{2}$, all function nodes on the left-hand side
  represent indicator function nodes of repetition codes, and all function
  nodes on the right-hand side represent indicator function nodes of single
  parity-check codes. It can easily be verified that for this NFG we have
  $t_{\graphN} = 1$.  \exampleend
\end{Example}

This example is formalized in the following definition.

\begin{Definition}
  \label{def:normal:factor:graph:ldpc:code:1}  

  Consider a code $\codeCchannel$ over $\GF{2}$ defined by some parity-check
  matrix $\matrH = [h_{j,i}]_{j \in \setJ, \, i \in \setI}$, where $\setJ$ and
  $\setI$ are the set of row and column indices of $\matrH$, respectively. The
  code $\codeCchannel$ can be represented by an NFG $\graphN(\matrH) \defeq
  \graphN(\setF,\setE, \setA, \setG)$ as follows.
  \begin{itemize}
    
  \item The set of local function nodes is $\setF \defeq \setI \cup \setJ$.

  \item The set of edges is $\setE = \setEhalf \cup \setEfull$, where
    $\setEhalf = \setI$ and where $\setEfull = \bigl\{ (i,j) \in \setI \times
    \setJ \bigm| h_{j,i} = 1 \bigr\}$.

  \item For every $e \in \setE$, the edge alphabet is $\setA_e = \GF{2}$.

  \item For every $i \in \setI$, the local function $g_i$ equals the indicator
    function of a length-$(|\setE_i| \! + \! 1)$ repetition code.

  \item For every $j \in \setJ$, the local function $g_j$ equals the indicator
    function of a length-$|\setE_j|$ single parity-check code.

  \end{itemize}
  If the parity-check matrix $\matrH$ is such that all columns of $\matrH$
  have Hamming weight $\dleft$ and all rows of $\matrH$ have Hamming weight
  $\dright$, then $\matrH$ is called a $(\dleft, \dright)$-regular
  parity-check matrix. (For example, the parity-check matrix $\matrH$ in
  Example~\ref{example:normal:factor:graph:ldpc:code:1} is $(3,6)$-regular.)
  If the parity-check matrix $\matrH$ is sparsely populated then the code
  $\codeCchannel$ is called a low-density parity-check (LDPC)
  code. Consequently, if the parity-check matrix $\matrH$ of an LDPC code is
  $(\dleft, \dright)$-regular then $\codeCchannel$ is called a $(\dleft,
  \dright)$-regular LDPC code, otherwise $\codeCchannel$ is called an
  irregular LDPC code.  \defend
\end{Definition}

The following definition is a generalization of the definition of the
fundamental polytope and the fundamental cone in~\cite{Koetter:Vontobel:03:1,
  Vontobel:Koetter:05:1:subm}.

\begin{Definition}
  \label{def:fundamental:polytope:1}

  Let $\codeCchannel$ be a code over $\GF{2}$, let $\graphN(\setF,\setE, \setA,
  \setG)$ be an NFG that represents $\codeCchannel$, and let $\lmpB$
  be the local marginal polytope of $\graphN$. We define the fundamental
  polytope $\fp{P}$ and the fundamental cone $\fc{K}$ to be, respectively,
  \begin{align*}
    \fp{P}
      &\defeq
         \big\{
           (\bel_{e,1})_{e \in \setEhalf}
         \bigm|
           \vbel \in \lmpB
         \big\}, \\
    \fc{K}
      &\defeq
         \conichull(\fp{P}).
  \end{align*}
  Elements of $\fp{P}$ and $\fc{K}$ are called pseudo-codewords.
  \defend
\end{Definition}

It can easily be verified that $\convhull(\codeCchannel) \subseteq \fp{P}$,
\ie, that the fundamental polytope is a relaxation of the convex hull of the
set of codewords. (Here the codewords are assumed to be embedded in $\R^n$,
where $n$ is the length of the code.)

The following definition is taken from~\cite{Vontobel:10:2}.

\begin{Definition}
  \label{def:induced:bethe:entropy:function:1}

  Let $\codeCchannel$ be a code over $\GF{2}$, let $\graphN(\setF,\setE, \setA,
  \setG)$ be an NFG that represents $\codeCchannel$, and let $\lmpB$
  be the local marginal polytope of $\graphN$.
  \begin{itemize}

  \item Let $\vpsi$ be the surjective mapping
    \begin{align*}
      \vpsi: \ 
        &\lmpB \to     \fp{P}, \quad
         \vbel \mapsto (\bel_{e,1})_{e \in \setEhalf}.
    \end{align*}
    (Clearly, in general there are many $\vbel \in \lmpB$ that map to the same
    pseudo-codeword in $\fp{P}$.)

  \item Let $\vPsiBME$ be the mapping
    \begin{align}
      \vPsiBME: \
        &\fp{P}  \to     \lmpB, \quad
         \vomega \mapsto \argmax_{\vbel \in \lmpB: \, 
                                  \vpsi(\vbel) = \vomega}
                           \HBethe(\vbel),
                                     \label{eq:vPsiBME:def:1}
    \end{align}
    where ``BME'' stands for ``Bethe Max-Entropy.'' This mapping gives for
    each $\vomega \in \fp{P}$ the $\vbel$ among all the $\vpsi$-pre-images of
    $\vomega$ that has the maximal Bethe entropy function value.

  \item The induced Bethe entropy function is defined to be
    \begin{align*}
      \HBethe: \ 
        &\fp{P}  \to     \R, \ 
         \vomega \mapsto \HBethe
             \big(
               \vPsiBME(\vomega)
             \big).
    \end{align*}
    (Note that the argument of $\HBethe$ determines if $\HBethe$ denotes the
    Bethe entropy function or the induced Bethe entropy function.) \defend

  \end{itemize}
\end{Definition}

\section{Graph-Cover Decoding}
\label{sec:graph:cover:decoding:1}

\begin{table*}
  \caption{
    \mbox{Expressions for the pseudo-marginal vectors that appear in a unified
    formulation of BMAPD, BGCD, SMAPD, and SGCD}
    \mbox{in terms of the global
    function of the NFG $\graphN \defeq \graphN(\vy)$ and its
    finite graph covers.}
    \mbox{(For
    every $M \in \Zpp$, the scalar $Z'_M(\vy) \in \Rpp$ is some suitably defined
    constant.)}}
  \begin{center}
    \begin{tabular}{|l|c|c|}
      \hline
      &
      maximum a-posteriori decoding
      &
      graph-cover decoding \\
      \hline
      \hline
      blockwise
      &
      \begin{minipage}{0.42\linewidth}
        \begin{align*}
          \hskip-0.15cm
          \hvbel^{\BMAPD}(\vy)
             \defeq
               \left.
                 \varphiM
                   \left(
                     \argmax_{\cgraph{N} \in \cset{N}_M, \, 
                                \cvc \in \codeC(\cgraph{N})} \,
                       \cover{g}_{\cgraph{N}}(\cvc)
                   \right)
               \right|_{M = 1}
          \hskip-0.15cm
        \end{align*}
      \end{minipage} &
      \begin{minipage}{0.41\linewidth}
        \begin{align*}
          \hskip-0.05cm
          \hvbel^{\BGCD}(\vy)
             \defeq
               \lim_{M \to \infty}
                 \varphiM
                   \left(
                     \argmax_{\cgraph{N} \in \cset{N}_M, \, 
                                \cvc \in \codeC(\cgraph{N})} \,
                       \cover{g}_{\cgraph{N}}(\cvc)
                   \right)
          \hskip-0.05cm
        \end{align*}
      \end{minipage} \\[0.5cm]
      \hline
      symbolwise &
      \begin{minipage}{0.42\linewidth}
        \begin{align*}
          \hskip-0.15cm
          \hvbel^{\SMAPD}(\vy)
             \defeq
               \left.
                 \frac{1}{Z'_M(\vy)}
                   \sum_{\cgraph{N} \in \cset{N}_M} \ 
                     \sum_{\cvc \in \codeC(\cgraph{N})}
                       \cover{g}_{\cgraph{N}}(\cvc)
                       \cdot
                       \varphiM\big( \cgraph{N},\cvc \big)
               \right|_{M = 1}
               \hskip-0.15cm
        \end{align*}
      \end{minipage} &
      \begin{minipage}{0.41\linewidth}
        \begin{align*}
          \hskip-0.05cm
          \hvbel^{\SGCD}(\vy)
             \defeq
               \lim_{M \to \infty}
                 \frac{1}{Z'_M(\vy)}
                   \sum_{\cgraph{N} \in \cset{N}_M} \ 
                     \sum_{\cvc \in \codeC(\cgraph{N})}
                       \cover{g}_{\cgraph{N}}(\cvc)
                       \cdot
                       \varphiM\big( \cgraph{N},\cvc \big)
          \hskip-0.05cm
        \end{align*}
      \end{minipage} \\[0.75cm]
      \hline
    \end{tabular}
  \end{center}
  \label{table:decoder:comparision:1}
\end{table*}

\begin{table*}
  \caption{
    \mbox{Expressions for the pseudo-marginal vectors that appear in a
      unified formulation of BMAPD, BGCD, SMAPD, and SGCD}
    \mbox{in terms of Gibbs and Bethe free energy functions.}
    Here, if $\vbel \in \convhull(\lmpBdown{1})$, \ie, 
    $\vbel$ is a globally realizable pseudo-marginal vector corresponding to 
    $\vp \in \setPi_{\codeC}$, we define $\FGibbs(\vbel) \defeq \FGibbs(\vp)$.
    (See also Remark~\ref{remark:lmpB:1}.)}
  \begin{center}
    \begin{tabular}{|l|c|c|}
      \hline
      &
      maximum a-posteriori decoding
      &
      graph-cover decoding \\
      \hline
      \hline
      & & \\[-0.25cm]
      blockwise
      &
      \begin{minipage}{0.43\linewidth}
        \begin{align*}
          \hvbel^{\BMAPD}(\vy)
             = \argmin_{\vbel \in \convhull(\lmpBdown{1})} \ 
                 \Big.
                   \FGibbs\big( \vbel \big)
                 \Big|_{T = 0}
        \end{align*}
      \end{minipage} &
      \begin{minipage}{0.42\linewidth}
        \begin{align*}
          \hvbel^{\BGCD}(\vy)
             = \argmin_{\vbel \in \lmpB} \ 
                 \Big.
                   \FBethe(\vbel)
                 \Big|_{T = 0}
        \end{align*}
      \end{minipage} \\[0.5cm]
      \hline
      symbolwise
      &
      \begin{minipage}{0.43\linewidth}
        \begin{align*}
          \hvbel^{\SMAPD}(\vy)
             = \argmin_{\vbel \in \convhull(\lmpBdown{1})} \ 
                 \Big.
                   \FGibbs(\vbel)
                 \Big|_{T = 1}
        \end{align*}
      \end{minipage} &
      \begin{minipage}{0.42\linewidth}
        \begin{align*}
          \hvbel^{\SGCD}(\vy)
             = \argmin_{\vbel \in \lmpB} \ 
                 \Big.
                   \FBethe(\vbel)
                 \Big|_{T = 1}
        \end{align*}
      \end{minipage} \\[0.5cm]
      \hline
    \end{tabular}
  \end{center}
  \label{table:decoder:comparision:2}
\end{table*}

As discussed in Section~\ref{sec:introduction:1}, and shown in
Figs.~\ref{fig:BGCD:overview:1} and~\ref{fig:SGCD:overview:1}, graph-cover
decoding is a theoretical tool to connect a variety of known decoders. In this
section, we first review blockwise maximum a-posteriori decoding (BMAPD),
which will set the stage for discussing blockwise graph-cover decoding
(BGCD). Afterwards, we review symbolwise maximum a-posteriori decoding
(SMAPD), upon which we introduce symbolwise graph-cover decoding (SGCD). These
decoders are summarized in Tables~\ref{table:decoder:comparision:1}
and~\ref{table:decoder:comparision:2}.

Note that blockwise graph-cover decoding was simply called graph-cover
decoding in~\cite[Sec.~4]{Vontobel:Koetter:05:1:subm} and that the exposition
here is slightly more general than in~\cite{Vontobel:Koetter:05:1:subm}
because we do not restrict ourselves to binary codes.

\begin{Definition}
  \label{def:coding:normal:factor:graph:1}

  The setup in this section is as follows. (See also the upcoming
  Example~\ref{example:normal:factor:graph:ldpc:code:2}.) We consider a
  discrete memoryless channel with an arbitrary input alphabet $\setX$, an
  arbitrary output alphabet $\setY$, and arbitrary channel law $\bigl\{ W(y|x)
  \bigr\}_{y \in \setY, \, x \in \setX}$~\!, \ie, the probability of observing
  the symbol $y \in \setY$ at the channel output given that the symbol $x \in
  \setX$ was sent is $W(y|x)$. Moreover, let $\codeCchannel$ be a block code
  of length $n$ and with alphabet $\setX$ that is used for data transmission
  over this discrete memoryless channel. We let $\vX = (X_1, \ldots, X_n)$ and
  $\vY = (Y_1, \ldots, Y_n)$ be the random vectors corresponding to,
  respectively, the channel input and output symbols of $n$ channel uses. We
  assume that a codeword $\vx \in \codeCchannel$ is selected with probability
  $P_{\vX}(\vx)$. (Of course, $P_{\vX}(\vx) = 0$ for $\vx \notin
  \codeCchannel$.) The joint probability mass function of $\vX$ and $\vY$ is
  then given by
  \begin{align*}
    P_{\vX, \vY}(\vx, \vy)
     &= P_{\vX}(\vx)
        \! \cdot \!
        P_{\vY|\vX}(\vy | \vx)
      = P_{\vX}(\vx)
        \cdot
        \!\! \prod_{i \in [n]} \!\!
          W(y_i | x_i).
  \end{align*}
  For a given channel output vector $\vy = (y_i)_{i \in [n]} \in \setY^n$,
  consider an NFG $\graphN(\vy) \defeq \graphN(\setF,\setE, \setA, \setG)$
  with the following properties.
  \begin{itemize}

  \item For all $e \in \setEhalf$ we have $\setA_e = \setX$.

  \item We identify $\setEhalf$ with $[n]$.

  \item We identify $\{ a_e \}_{e \in \setEhalf}$ with $\{ x_i \}_{i \in
      [n]}$.

  \item In order to take the received vector $\vy$ into account, some function
    nodes are parameterized by $y_i$, $i \in [n]$.

  \item For every codeword $\vx \in \codeCchannel$, there is exactly one valid
    configuration $\vc \in \codeCedge$ such that the restriction of $\vc$ to
    $\setEhalf$ equals $\vx$. This valid configuration will be denoted by
    $\vc(\vx)$. In terms of
    Definition~\ref{def:normal:factor:graph:representing:a:code:1} this means
    that we impose $t_{\graphN} = 1$. (With the necessary care, the results of
    this section can be generalized to NFGs for which there exists a constant
    $t_{\graphN}$ with $t_{\graphN} > 1$.)
 
  \item There is some constant $\gamma \in \Rpp$ such that for every codeword
    $\vx \in \codeCchannel$, the global function value of the valid
    configuration $\vc(\vx)$ is
    \begin{align}
      g\big(
         \vc(\vx)
       \big)
        &= \gamma
           \cdot
           P(\vx, \vy).
             \label{eq:coding:global:function:1}
    \end{align}
  \defend
  
  \end{itemize}
\end{Definition}

\begin{Example}
  \label{example:normal:factor:graph:ldpc:code:2}

  Consider again the code $\codeCchannel$ from
  Example~\ref{example:normal:factor:graph:ldpc:code:1} (which was represented
  by the NFG in Fig.~\ref{fig:normal:factor:graph:ldpc:code:1}~(left)),
  along with some discrete memoryless channel with input alphabet $\setX
  \defeq \GF{2}$, output alphabet $\setY$, and channel law $\bigl\{ W(y|x)
  \bigr\}_{y \in \setY, \, x \in \setX}$~\!. Let $\vy \in \setY^n$ be a given
  channel output vector. It can be verified that
  Fig.~\ref{fig:normal:factor:graph:ldpc:code:1}~(right) shows a possible
  NFG $\graphN(\vy)$ that has the properties as specified in
  Definition~\ref{def:coding:normal:factor:graph:1}. \exampleend
\end{Example}

Potential ties in upcoming ``$\argmax$'' and ``$\argmin$'' expressions are
assumed to be resolved in a systematic or random manner.

\subsection{Blockwise Maximum A-Posteriori Decoding}
\label{sec:BMAPD:1}

With the setup as in Definition~\ref{def:coding:normal:factor:graph:1}, let
$\hvxBMAPD(\vy) \in \setX^n$ be the decision vector obtained by BMAPD based on
the received vector $\vy$. (Recall that BMAPD is the decision rule that
minimizes the block decision error probability $\operatorname{Pr}\bigl(
\hvxBMAPD(\vY) \neq \vX \bigr)$.)

\begin{Definition}
  \label{def:BMAPD:1}

  Given a channel output vector $\vy$, BMAPD yields the decision
  rule
  \begin{align}
    \hvxBMAPD(\vy)
      &\defeq
         \argmax_{\vx} \,
           P_{\vX|\vY}(\vx | \vy) \nonumber \\
      &= \argmax_{\vx} \,
           P_{\vX, \vY}(\vx, \vy).
             \label{eq:BMAPD:1}
  \end{align}
  \defend
\end{Definition}

On the side, we note that if all codewords are selected equally likely, \ie,
$P(\vx) = \frac{1}{\card{\codeCchannel}} \cdot [\vx \in \codeCchannel]$, then
this decision rule equals the blockwise maximum likelihood decoding rule.

\begin{Lemma}
  \label{lemma:BMAPD:reformulation:1}

  Given a channel output vector $\vy$, consider the NFG
  $\graphN \defeq \graphN(\vy)$ from
  Definition~\ref{def:coding:normal:factor:graph:1}. The vector
  $\hvxBMAPD(\vy)$ satisfies
  \begin{align*}
    \hvxBMAPD(\vy)
      &= \argmax_{\vchalf} \,
           \max_{\vcfull}
             g(\vchalf, \vcfull)
  \end{align*}
  and
  \begin{align*}
    \vc
      \left(
        \hvxBMAPD(\vy)
      \right)
      &= \argmax_{\vc} \,
           g(\vc).
  \end{align*}
  In terms of the pseudo-marginal vector $\hvbel^{\BMAPD}(\vy)$ that is
  defined in Table~\ref{table:decoder:comparision:1}, one can therefore write
  \begin{align*}
    \left[
      a_e \! = \! \hxBMAPD_e(\vy)
    \right]
      &= \hbel_{e,a_e}^{\BMAPD}(\vy),
           \quad e \in \setEhalf, \, a_e \in \setA_e,
  \end{align*}
  and so
  \begin{align*}
    \hxBMAPD_e(\vy)
      &= \argmax_{a_e} \,
           \hbel_{e,a_e}^{\BMAPD}(\vy),
             \quad e \in \setEhalf.
  \end{align*}
\end{Lemma}

\begin{Proof}
  Follows from~\eqref{eq:coding:global:function:1}
  and~\eqref{eq:BMAPD:1}. Note that $\cset{N}_1 = \bigl\{ \graphN(\vy)
  \bigr\}$.
\end{Proof}

As shown in the following lemma, the BMAPD rule can also be cast as a Gibbs
free energy function minimization problem (with temperature $T = 0$).

\begin{Lemma}
  \label{lemma:BMAPD:reformulation:2}

  Given a channel output vector $\vy$, consider the NFG
  $\graphN \defeq \graphN(\vy)$ from
  Definition~\ref{def:coding:normal:factor:graph:1} and define
  \begin{align*}
    \hat \vp
      &\defeq
         \argmin_{\vp \in \setPi_\codeC} \ 
           \Big.
             \FGibbs(\vp)
           \Big|_{T = 0}.
  \end{align*}
  The vector $\hat \vp$ is such that
  \begin{align*}
    \hat \vp_{\vc}
      = \begin{cases}
          1 & \text{if $\vc = \vc\bigl( \hvxBMAPD(\vy) \bigr)$} \\
          0 & \text{otherwise}
        \end{cases}.
  \end{align*}
\end{Lemma}

\begin{Proof}
  Follows from~Definition~\ref{def:F:Gibbs:1},
  Eqs.~\eqref{eq:coding:global:function:1} and~\eqref{eq:BMAPD:1}, and the
  fact that $\FGibbs(\vp) = \UGibbs(\vp)$ for temperature $T = 0$.
\end{Proof}

From Lemma~\ref{lemma:BMAPD:reformulation:2}, Remark~\ref{remark:lmpB:1}, and
the fact that $\cset{N}_1 = \bigl\{ \graphN(\vy) \bigr\}$, it follows that
$\hvbel^{\BMAPD}(\vy)$ can also be written as shown in
Table~\ref{table:decoder:comparision:2}.

\subsection{Blockwise Graph-Cover Decoding}
\label{sec:BGCD:1}

We consider the setup as in
Definition~\ref{def:coding:normal:factor:graph:1}. Recall that BMAPD can be
seen as a competition of all codewords to be the best explanation of the
observed channel output vector. In this subsection we revisit blockwise
graph-cover decoding (BGCD), which was originally introduced
in~\cite[Section~4]{Vontobel:Koetter:05:1:subm}. Actually, we will define this
decoder slightly differently than
in~\cite[Section~4]{Vontobel:Koetter:05:1:subm}. Namely, whereas
in~\cite[Section~4]{Vontobel:Koetter:05:1:subm} all codewords in all finite
covers of an NFG were competing to be the best explanation of a channel output
vector, here we restrict the competition to all codewords in all $M$-covers of
an NFG, and then we let $M$ go to infinity.

\begin{Definition}
  \label{def:BGCD:1}
  
  Given a channel output vector $\vy$, consider the NFG
  $\graphN \defeq \graphN(\vy)$ from
  Definition~\ref{def:coding:normal:factor:graph:1}. For any $M \in \Zpp$, we
  define degree-$M$ BGCD to be the decoding rule that gives back the
  pseudo-marginal vector
  \begin{align*}
    \hvbel^{\BGCD(M)}(\vy)
      &\defeq
         \varphiM
           \left(
             \big( \hcN, \hcvc \big)^{\BGCD(M)}(\vy)
           \right),
  \end{align*}
  where
  \begin{align}
    \big( \hcN, \hcvc \big)^{\BGCD(M)}(\vy)
      &= \argmax_{\cgraph{N} \in \cset{N}_M, \, \cvc \in \codeC(\cgraph{N})} \,
           \cover{g}_{\cgraph{N}}(\cvc).
                 \label{eq:BGCD:1}
  \end{align}
  In the limit $M \to \infty$, we define BGCD to be the decoding rule that
  gives back the pseudo-marginal vector
  \begin{align*}
    \hvbel^{\BGCD}(\vy)
      &\defeq
         \lim_{M \to \infty}
           \hvbel^{\BGCD(M)}(\vy).
  \end{align*}
  (This latter expression is also shown in
  Table~\ref{table:decoder:comparision:1}.)
  \defend
\end{Definition}

In the case $\setX = \GF{2}$, one could have defined BGCD to give back the
pseudo-codeword $\vpsi\bigl( \hvbel^{\BGCD}(\vy) \bigr)$ (with suitable
generalizations for other alphabets $\setX$), however, for simplicity of
notation we will not pursue this option here.

\begin{Theorem}
  \label{theorem:BGCD:characterization:1}

  Given a channel output vector $\vy$, consider the NFG
  $\graphN \defeq \graphN(\vy)$ from
  Definition~\ref{def:coding:normal:factor:graph:1}. Then
  \begin{align*}
    \hvbel^{\BGCD}(\vy)
      &= \argmin_{\vbel \in \lmpB} \ 
           \Big.
             \FBethe(\vbel)
           \Big|_{T = 0}.
  \end{align*}
\end{Theorem}

\begin{Proof}
  This follows from Theorems~\ref{theorem:down:projection:1}
  and~\ref{theorem:asymptotic:average:function:value:pre:images:in:graph:covers:1}
  and the fact that $\FBethe(\vbel) = \UBethe(\vbel)$ for temperature $T = 0$.
\end{Proof}

The decoder relationships that are highlighted in
Fig.~\ref{fig:BGCD:overview:1} are a consequence of the following
observations.
\begin{itemize}

\item Finding the minimum of the Bethe free energy function at temperature $T
  = 0$ is equivalent to linear programming decoding.

\item As shown in Theorem~\ref{theorem:BGCD:characterization:1}, blockwise
  graph-cover decoding is equivalent to finding the minimum of the Bethe free
  energy function at temperature $T = 0$.

\item As discussed in~\cite{Koetter:Vontobel:03:1, Vontobel:Koetter:05:1:subm}
  and in Section~\ref{sec:introduction:1}, a locally operating algorithm like
  the max-product (min-sum) algorithm ``cannot distinguish'' if it is
  operating on an NFG $\graphN$ or, implicitly, on any of its covers. (In
  particular, note that the fact that any finite graph cover $\cgraph{N}$ of
  $\graphN$ looks locally the same as $\graphN$ implies that the collection of
  computation trees of $\graphN$ equals the collection of computation trees of
  $\cgraph{N}$.) With this, BGCD can be considered to be a ``model'' for the
  behavior of max-product (min-sum) algorithm decoding. Note that the
  connection between BGCD and max-product (min-sum) algorithm decoding is in
  general only an approximate one. However, in all cases where analytical
  tools are known that exactly characterize the behavior of the max-product
  algorithm decoder, the connection between the BGCD and the max-product
  (min-sum) algorithm decoder is exact.

\end{itemize}

Note that if the NFG $\graphN$ does not contain cycles then max-product
algorithm decoding and linear programming decoding yield the same decision as
BMAPD~\cite{Kschischang:Frey:Loeliger:01}. This is reflected in the
equivalence of $\FGibbs$ and $\FBethe$ for cycle-free NFGs, once the domains
of these two functions have been suitably identified.

\subsection{Symbolwise Maximum A-Posteriori Decoding}
\label{sec:SMAPD:1}

With the setup as in Definition~\ref{def:coding:normal:factor:graph:1}, let
$\hvxSMAPD(\vy) \in \setX^n$ be the decision vector obtained by SMAPD based on
the received vector $\vy$. Recall that SMAPD is the decision rule that
minimizes the symbol decision error probability $\operatorname{Pr}\bigl(
\hxSMAPD_i(\vY) \neq X_i \bigr)$ for every $i \in [n]$ (or, depending on the
definition, for every $i$ that corresponds to an information symbol of
$\codeCchannel$).

\begin{Definition}
  \label{def:SMAPD:1}

  Given a channel output vector $\vy$, SMAPD yields the vector
  $\hvxSMAPD(\vy)$ with components
  \begin{align*}
    \hxSMAPD_i(\vy)
      &\defeq
         \argmax_{x_i} \,
           P_{X_i|\vY}(x_i | \vy)\nonumber \\
      &= \argmax_{x_i} \,
           P_{X_i, \vY}(x_i, \vy),
             \quad i \in [n], 
  \end{align*}
  where
  \begin{align}
    P_{X_i, \vY}(x_i, \vy)
      = \sum_{\vx': \, x'_i = x_i}
          P_{\vX, \vY}(\vx',\vy).
            \label{eq:SMAPD:marginal:1}
  \end{align}
  \mbox{} \\[-0.5cm]
  \defend
\end{Definition}

Note that $\hvxSMAPD(\vy)$, in contrast to $\hvxBMAPD(\vy)$, is not always a
codeword.

\begin{Lemma}
  \label{lemma:SMAPD:reformulation:1}

  Given a channel output vector $\vy$, consider the NFG
  $\graphN \defeq \graphN(\vy)$ from
  Definition~\ref{def:coding:normal:factor:graph:1}. The SMAPD vector
  $\hvxSMAPD(\vy)$ satisfies
  \begin{align}
    \hxSMAPD_e(\vy)
      &= \argmax_{a_e} \,
           \eta_e(a_e),
             \quad e \in \setEhalf.
               \label{eq:SMAPD:decision:2}
  \end{align}
  with
  \begin{align}
    \eta_e(a_e)
      &\defeq
         \frac{1}{\ZGibbs}
         \cdot
         \sum_{\va': \, a'_e = a_e}
           g(\va'),
             \quad e \in \setEhalf, \, a_e \in \setA_e.
             \label{eq:SMAPD:marginal:2}
  \end{align}
  In terms of the pseudo-marginal vector $\hvbel^{\SMAPD}(\vy)$ that is
  defined in Table~\ref{table:decoder:comparision:1}, one can therefore write
  \begin{align*}
    \eta_e(a_e)
      &= \hbel_{e,a_e}^{\SMAPD}(\vy),
           \quad e \in \setEhalf, \, a_e \in \setA_e,
  \end{align*}
  and
  \begin{align*}
    \hxSMAPD_e(\vy)
      &= \argmax_{a_e} \,
           \hbel_{e,a_e}^{\SMAPD}(\vy),
             \quad e \in \setEhalf.
  \end{align*}
\end{Lemma}

\begin{Proof}
  Follows from~\eqref{eq:coding:global:function:1}
  and~\eqref{eq:SMAPD:marginal:1}. Note that $\cset{N}_1 = \bigl\{
  \graphN(\vy) \bigr\}$.
\end{Proof}

From Lemma~\ref{lemma:SMAPD:reformulation:1} it is clear that the main
computational step towards obtaining the SMAPD vector is to compute the
marginals $\bigl\{ \eta_e(a_e) \bigr\}_{e \in \setEhalf, \, a_e \in
  \setA_e}$. Note that the scaling constant in~\eqref{eq:SMAPD:marginal:2} was
chosen such that $\sum_{a_e} \eta_e(a_e) = 1$ for every $e \in \setEhalf$. Of
course, any other positive scaling factor works equally well as it entails the
same decision in~\eqref{eq:SMAPD:decision:2}.

As shown in the following lemma, the SMAPD rule can also be cast as a Gibbs
free energy function minimization problem (with temperature $T = 1$).

\begin{Lemma}
  \label{lemma:SMAPD:reformulation:2}

  Given a channel output vector $\vy$, consider the NFG $\graphN \defeq
  \graphN(\vy)$ from Definition~\ref{def:coding:normal:factor:graph:1} and
  define
  \begin{align*}
    \hat \vp
      &\defeq
         \argmin_{\vp \in \setPi_\codeC} \ 
           \Big.
             \FGibbs(\vp)
           \Big|_{T = 1}.
  \end{align*}
  Using the notation from Lemma~\ref{lemma:SMAPD:reformulation:1}, we have
  \begin{align*}
    \eta_e(a_e)
      &= \sum_{\vc: \, c_e = a_e}
           \hat \vp_{\vc},
           \quad e \in \setEhalf, \, a_e \in \setA_e.
  \end{align*}
\end{Lemma}

\begin{Proof}
  Follows from~Definition~\ref{def:F:Gibbs:1} and
  Lemmas~\ref{lemma:F:Gibbs:minimum:1} and~\ref{lemma:SMAPD:reformulation:1}.
\end{Proof}

From Lemmas~\ref{lemma:SMAPD:reformulation:1}
and~\ref{lemma:SMAPD:reformulation:2}, Remark~\ref{remark:lmpB:1}, and the
fact that $\cset{N}_1 = \bigl\{ \graphN(\vy) \bigr\}$, it follows that
$\hvbel^{\BMAPD}(\vy)$ can also be written as shown in
Table~\ref{table:decoder:comparision:2}.

\subsection{Symbolwise Graph-Cover Decoding}
\label{sec:SGCD:1}

We consider the setup as in Definition~\ref{def:coding:normal:factor:graph:1}.
Recall that SMAPD is based on computing suitable marginals of the global
function represented by the NFG $\graphN(\vy)$. In this subsection we define
symbolwise graph-cover decoding (SGCD), which was outlined
in~\cite{Vontobel:08:3, Vontobel:09:Talk:5}. Similar to the transition from
BMAPD to BGCD, where the competition is extended from all codewords of
$\graphN(\vy)$ to all codewords in all $M$-covers of $\graphN(\vy)$, when
going from SMAPD to SGCD we replace the marginals of the global function of
$\graphN(\vy)$ by a suitable combination of marginals of the global functions
of all $M$-covers of $\graphN(\vy)$.

\begin{Definition}
  \label{def:SGCD:1}
  
  Given a channel output vector $\vy$, consider the NFG
  $\graphN \defeq \graphN(\vy)$ from
  Definition~\ref{def:coding:normal:factor:graph:1}. For any $M \in \Zpp$, we
  define degree-$M$ SGCD to yield the pseudo-marginal vector
  $\hvbel^{\SGCD(M)}(\vy)$ with components
  \begin{alignat*}{2}
    \hbel_{e,a_e}^{\SGCD(M)}(\vy)
      &\defeq
         \eta_{e,M}(a_e),
           &\quad& e \in \setE, \, a_e \in \setA_e, \\
    \hbel_{f,\va_f}^{\SGCD(M)}(\vy)
      &\defeq
         \eta_{f,M}(\va_f),
           &\quad& f \in \setF, \, \va_f \in \setA_f.
  \end{alignat*}
  For every $e \in \setE$, the ``marginal function'' $\eta_{e,M}$ is 
  defined by
  \begin{align*}
    \eta_{e,M}(a_e)
      &\defeq
         \frac{1}{M}
         \cdot
         \sum_{m \in [M]}
           \eta_{e,m,M}(a_e), \\
      &\hskip2.5cm
             e \in \setE, \, a_e \in \setA_e, \\[0.3cm]
    \eta_{e,m,M}(a_e)
      &\defeq
         \frac{1}{Z'_M(\graphN)}
         \cdot
         \sum_{\cgraph{N} \in \cset{N}_M}
           \ZGibbsext{\cgraph{N}}
           \cdot
           \eta_{e,m,\cgraph{N}}(a_e), \\
      &\hskip2.5cm
             e \in \setE, \, m \in [M], \, a_e \in \setA_e, \\[0.3cm]
    \eta_{e,m,\cgraph{N}}(a_e)
      &\defeq
         \frac{1}{\ZGibbsext{\cgraph{N}}}
         \cdot
         \sum_{\cvc \in \codeC(\cgraph{N}): \, \cover{c}_{e,m} = a_e}
           g_{\cgraph{N}}(\cvc), \\
      &\hskip2.5cm
             e \in \setE, \, m \in [M], \, a_e \in \setA_e,
  \end{align*}
  where
  \begin{align}
    Z'_M(\graphN)
      &\defeq
         \sum_{\cgraph{N} \in \cset{N}_M}
           \ZGibbsext{\cgraph{N}}
       = \cardbig{\cset{N}_M} 
         \cdot
         \big( \ZBetheM{M}(\graphN) \big)^M.
           \label{eq:Z:M:def:1}
  \end{align}
  (For a motivation of these expressions, see the paragraph after this
  definition.) For every $f \in \setF$, the ``marginal function'' $\eta_{f,M}$
  is defined analogously. Moreover, taking the limit $M \to \infty$, we define
  SGCD to be the decoder that gives back the pseudo-marginal vector
  \begin{align*}
    \hvbel^{\SGCD}(\vy)
      &\defeq
         \lim_{M \to \infty}
           \hvbel^{\SGCD(M)}(\vy).
  \end{align*}
  \defend
\end{Definition}

Here are the motivations for specifying the ``marginal functions'' as we did
in Definition~\ref{def:SGCD:1}.
\begin{itemize}

\item Fix an arbitrary $M$-cover $\cgraph{N}$ of $\graphN$. If we were to find
  the SMAPD estimate of the code defined by $\cgraph{N}$ then, following
  Lemma~\ref{lemma:SMAPD:reformulation:1}, we would have to compute the
  marginals
  \begin{align*}
    \big\{
      \eta_{e,m,\cgraph{N}}(\cover{a}_{e,m})
    \big\}_{e \in \setEhalf, \, m \in [M], \, \cover{a}_{e,m} \in \setA_e}.
  \end{align*}

\item However, given the fact that no $M$-cover is more special than any other
  $M$-cover, we take the average of these marginals over all $M$-covers and
  obtain the marginals
  \begin{align*}
    \big\{
      \eta_{e,m,M}(a_e)
    \big\}_{e \in \setEhalf, \, m \in [M], \, a_e \in \setA_e}.
  \end{align*}
  Actually, we take a weighted average where the weighting factor for
  $\cgraph{N}$ is chosen to be its Gibbs partition function
  $\ZGibbsext{\cgraph{N}}$.

\item From the symmetries of the setup it is clear that for every $e \in
  \setE$ and every $a_e \in \setA_e$ the quantity $\eta_{e,m,M}(a_e)$ is
  independent of $m \in [M]$. Therefore, the definition of the marginals
  \begin{align*}
    \big\{
      \eta_{e,M}(a_e)
    \big\}_{e \in \setEhalf, \, a_e \in \setA_e}
  \end{align*}
  is somewhat trivial, but notationally and analytically useful.

\item The scaling factors were chosen such that the marginals sum to $1$, when
  summed over their corresponding alphabets. (Note that the reformulation of
  $Z'_M(\graphN)$ on the right-hand side of~\eqref{eq:Z:M:def:1} follows from
  Definition~\ref{def:degree:M:partition:function:1}.)

\end{itemize}

\begin{Theorem}
  \label{theorem:SGCD:characterization:1}

  Given a channel output vector $\vy$, consider the NFG
  $\graphN(\vy)$ from Definition~\ref{def:coding:normal:factor:graph:1} and
  define
  \begin{align*}
    \hvbel
      &= \argmin_{\vbel \in \lmpB} \,
           \Big.
             \FBethe(\vbel)
           \Big|_{T = 1}.
  \end{align*}
  Using the notation from Definition~\ref{def:SGCD:1}, we have
  \begin{align*}
    \lim_{M \to \infty}
      \eta_{e,M}(a_e)
      &= \hbel_{e,a_e},
           \quad e \in \setE, \ a_e \in \setA_e, \\
    \lim_{M \to \infty}
      \eta_{f,M}(\va_f)
      &= \hbel_{f,\va_f},
           \quad f \in \setF, \ \va_f \in \setA_f.
  \end{align*}
\end{Theorem}

\begin{Proof}
  See Appendix~\ref{sec:proof:theorem:SGCD:characterization:1}.
\end{Proof}

In the rather special case where $\FBethe$ has multiple global minima
(necessarily of equal value), Theorem~\ref{theorem:SGCD:characterization:1}
has to be stated somewhat more carefully. Namely, the pseudo-marginal vector
$\hvbel$ has to be replaced by a suitable pseudo-marginal vector in the convex
hull of all pseudo-marginal vectors that minimize $\FBethe$.

The decoder relationships that are highlighted in
Fig.~\ref{fig:SGCD:overview:1} are a consequence of the following
observations.
\begin{itemize}

\item As shown in Theorem~\ref{theorem:SGCD:characterization:1}, symbolwise
  graph-cover decoding is equivalent to finding the minimum of the Bethe free
  energy function at temperature $T = 1$.

\item As discussed in~\cite{Koetter:Vontobel:03:1, Vontobel:Koetter:05:1:subm}
  and in Section~\ref{sec:introduction:1}, a locally operating algorithm like
  the sum-product algorithm ``cannot distinguish'' if is operating on an NFG
  $\graphN$ or, implicitly, on any of its covers. (Again, note that the fact
  that any finite graph cover $\cgraph{N}$ of $\graphN$ looks locally the same
  as $\graphN$ implies that the collection of computation trees of $\graphN$
  equals the collection of computation trees of $\cgraph{N}$.) Therefore, SGCD
  can be considered to be a ``model'' for the behavior of sum-product
  algorithm decoding. Note that the connection between SGCD and SPA decoding
  is in general only an approximate one. However, in many cases where
  analytical tools are known that exactly characterize the behavior of the SPA
  decoder, the connection between SGCD and SPA decoding is exact.

\end{itemize}

Note that if the NFG $\graphN(\vy)$ does not contain cycles
then sum-product algorithm decoding yields the same (pseudo-)marginal vector
as SMAPD. This is reflected in the equivalence of $\FGibbs$ and $\FBethe$ for
cycle-free NFGs, once the domains of these two functions have
been suitably identified.

For an NFG \emph{without} cycles, the meaning of the pseudo-marginal functions
that are computed by the sum-product algorithm is clear (\confer~the
discussion at the beginning of Section~\ref{sec:introduction:1}), but for an
NFG \emph{with} cycles, the meaning of these pseudo-marginal functions is a
priori less clear. However, combining
Theorems~\ref{theorem:asymptotic:average:number:pre:images:in:graph:covers:1}
and~\ref{theorem:SGCD:characterization:1} with the theorem by Yedidia,
Freeman, and Weiss~\cite{Yedidia:Freeman:Weiss:05:1} on the characterization
of fixed points of the sum-product algorithm, one obtains the following
statement. Namely, a fixed point of the sum-product algorithm corresponds to a
certain pseudo-marginal vector of the factor graph under consideration: it is,
after taking a biasing channel-output-dependent term properly into account,
the pseudo-marginal vector that has (locally) an extremal number of pre-images
in all $M$-covers, when $M$ goes to infinity.\footnote{In this statement we
  included the word ``locally'' because the sum-product algorithm can get
  stuck at a local extremum of the Bethe free energy function. Note that here
  the use of the word ``local'' is different than the use of the word
  ``local'' when comparing the global perspective of maximum a-posteriori
  decoding with the local perspective of message-passing iterative
  decoding. However, ultimately, this ``local'' is also a consequence of the
  suboptimal behavior of message-passing iterative decoding stemming from its
  local perspective.}

\section{The Influence of the Minimum Hamming Distance
               of a Code upon the Bethe Entropy Function of 
                its NFG}
\label{sec:influence:minimum:Hamming:distance:1}

It can easily be verified that the Gibbs entropy function is a concave
function of its arguments. However, the Bethe entropy function is in general
\emph{not} a concave function of its arguments. This has important
consequences when trying to minimize the Bethe free energy function because
the curvature of this function is determined by the curvature of the Bethe
entropy function.

In this section we show that choosing a code from an ensemble of regular LDPC
codes with minimum Hamming distance growing (with high probability) linearly
with the block length comes at the price of having to deal with an NFG whose
Bethe entropy function is concave and convex and whose Bethe free energy
function is, therefore, convex and concave. (By a multidimensional function
being ``concave and convex'' we mean that there are points and directions
where the function is locally concave, and points and directions where the
function is locally convex.) Moreover, we show that the choice of a code from
such an ensemble has implications for the accuracy of the pseudo-marginals
that are computed by the sum-product algorithm.

We conjecture that the above results are also valid for ensembles of
\emph{irregular} LDPC codes whose minimum Hamming distance grows (with high
probability) linearly with the block length, however, we prove the above
statements only for the case of regular LDPC codes.

This section is structured as follows. In
Section~\ref{sec:induced:Bethe:entropy:function:along:diagonal:1} we make a
simple observation about the induced Bethe entropy function for regular LDPC
codes. Afterwards, in
Section~\ref{sec:implications:minimum:Hamming:distance:1} we discuss how this
observation implies the above-mentioned results.

\subsection{An Observation about the Induced Bethe Entropy Function}
\label{sec:induced:Bethe:entropy:function:along:diagonal:1}

In this subsection we consider a setup where the expression for the induced
Bethe entropy (see Definition~\ref{def:induced:bethe:entropy:function:1}) can
be simplified significantly. Afterwards we will recognize that the obtained
expression appears also in some other context. This will then lead to the
promised conclusions, which are discussed in the next subsection.

Recall the definition of a $(\dleft, \dright)$-regular LDPC code from
Definition~\ref{def:normal:factor:graph:ldpc:code:1}. Note that the rate of
such a code is lower bounded by $1 - \dleft / \dright$.

\begin{Lemma}
  \label{lemma:hbethe:diagonal:1}

  Consider a $(\dleft,\dright)$-regular length-$n$ LDPC code over $\GF{2}$
  described by some parity-check matrix $\matrH$, and let $\graphN(\matrH)$ be
  the NFG associated with $\matrH$ as in
  Definition~\ref{def:normal:factor:graph:ldpc:code:1}. Then the induced Bethe
  entropy function along the straight line
  \begin{align*}
    \vomega(s)
      &\defeq
         \omega(s)
         \cdot
         \big(
           1, \ldots, 1
         \big),
           \quad s \in \R,
  \end{align*}
  evaluates to
  \begin{align*}
    \HBethe\big( \vomega(s) \big)
      &= n
         \cdot
         h_{\dleft,\dright}(s),
           \quad s \in \R.
  \end{align*}
  Here we have used the functions
  \begin{alignat*}{2}
    h_{\dleft,\dright}: \
     &\ \R \ &\to     \ &\R \\
     &\ s  \ &\mapsto \
           &\! 
            -
            (\dleft \! - \! 1)
              \!
              \cdot
              \!
              h\big( \omega(s) \big)
            -
            \dleft
              \!
              \cdot
              \!
              s
              \!
              \cdot
              \!
              \omega(s)
            +
            \frac{\dleft}{\dright}
              \!
              \cdot
              \!
              \theta(s),
  \end{alignat*}
  \begin{alignat*}{3}
    \omega: \ 
      &\ \R \ &\to \     &\R, \ 
       \ s  \ &\mapsto \ &
               \frac{1}{\dright}
               \cdot
               \rdiff{s} \theta(s), \\
    \theta: \ 
      &\ \R \ &\to \     &\R, \ 
       \ s  \ &\mapsto \ &
               \log
                 \left(
                   \sum_{w = 0 \atop w \ \mathrm{even}}^{\dright}
                     {\dright \choose w}
                     \exp(s \cdot w)
                 \right) \! , \\
    h: \ &\ [0,1] \ &\to     \ &\R, \ 
         \ \xi   \ &\mapsto \ & - \xi \log(\xi) - (1-\xi) \log(1-\xi).
  \end{alignat*}
\end{Lemma}

\begin{Proof}
  See Appendix~\ref{sec:proof:lemma:hbethe:diagonal:1}.
\end{Proof}

\begin{figure}
  \begin{center}
    \psfrag{xlabel}{\parbox[c]{5cm}{\vspace{0.25cm}\hspace{0.15cm}$\omega(s)$}}
    \psfrag{ylabel}{\parbox[c]{5cm}{\vspace{-0.25cm}\hspace{-0.75cm}
                    $h_{\dleft,\dright}(s)$ [bits]}}
    \epsfig{file=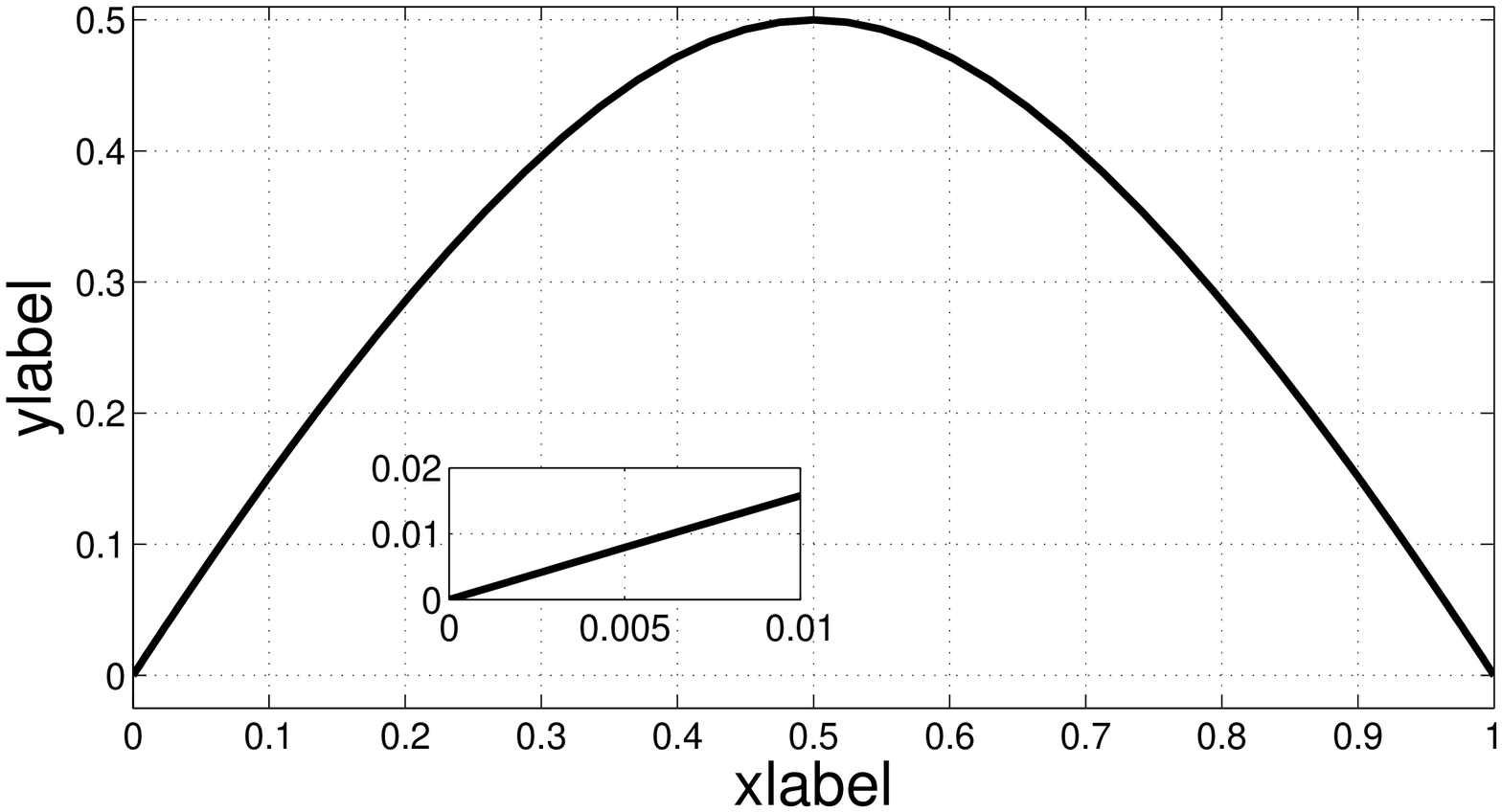, width=0.95\linewidth}
  \end{center}
  \caption{The graph of $s \mapsto \bigl( \omega(s), h_{\dleft,\dright}(s)
    \bigr)$ for $(\dleft, \dright) = (2,4)$. The inset zooms into
    the area near the origin.}
  \label{fig:Bethe:entropy:function:aspect:2:4:1}
\end{figure}

\begin{figure}
  \begin{center}
    \psfrag{xlabel}{\parbox[c]{5cm}{\vspace{0.25cm}\hspace{0.15cm}$\omega(s)$}}
    \psfrag{ylabel}{\parbox[c]{5cm}{\vspace{-0.25cm}\hspace{-0.75cm}
                    $h_{\dleft,\dright}(s)$ [bits]}}
    \epsfig{file=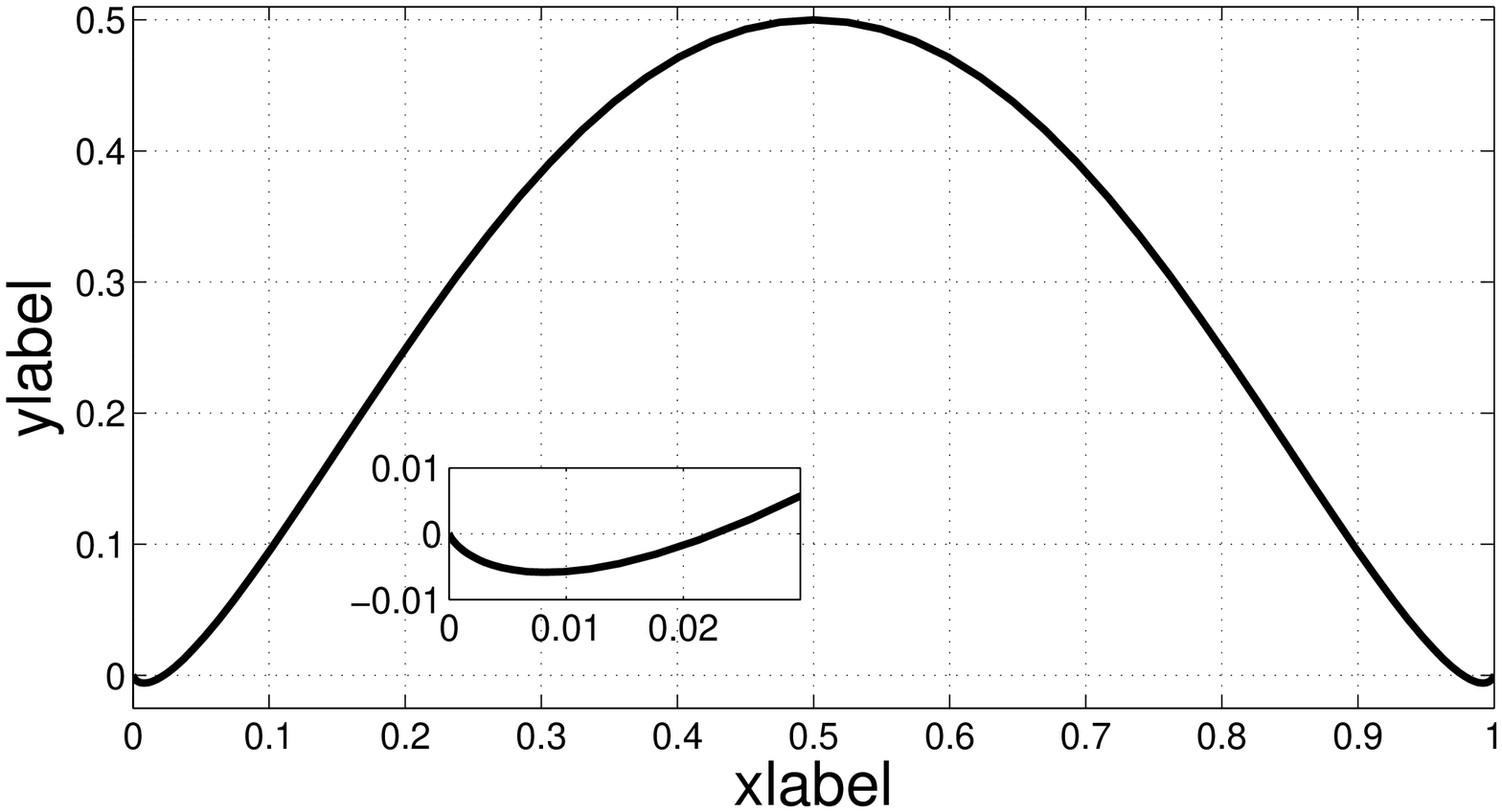, width=0.95\linewidth}
  \end{center}
  \caption{The graph of $s \mapsto \bigl( \omega(s), h_{\dleft,\dright}(s)
    \bigr)$ for $(\dleft, \dright) = (3,6)$. The inset zooms into the area
    near the origin.}
  \label{fig:Bethe:entropy:function:aspect:3:6:1}
\end{figure}

\begin{Example}
  \label{example:induced:Bethe:entropy:plot:1}

  For $(\dleft,\dright) = (2,4)$ and $(\dleft,\dright) = (3,6)$ the graph of
  $s \mapsto \bigl( \omega(s), h_{\dleft,\dright}(s) \bigr)$ is visualized in
  Figs.~\ref{fig:Bethe:entropy:function:aspect:2:4:1}
  and~\ref{fig:Bethe:entropy:function:aspect:3:6:1}, respectively. We make the
  following observations with respect to the shapes of these curves and the
  values that $h_{\dleft,\dright}(s)$ takes on.
  \begin{itemize}
    
  \item In the case $(\dleft,\dright) = (2,4)$, it can be verified that the
    graph $s \mapsto \bigl( \omega(s), h_{\dleft,\dright}(s) \bigr)$ is
    concave and that $h_{\dleft,\dright}(s)$ is always non-negative.

  \item In the case $(\dleft,\dright) = (3,6)$, it can be verified that the
    graph $s \mapsto \bigl( \omega(s), h_{\dleft,\dright}(s) \bigr)$ is
    concave and convex. Moreover, for small $\omega(s) > 0$ the value of
    $h_{\dleft,\dright}(s)$ is negative. (This behavior is actually typical
    for the behavior of $s \mapsto \bigl( \omega(s), h_{\dleft,\dright}(s)
    \bigr)$ for any $(\dleft,\dright)$-regular LDPC code with $3 \leq \dleft <
    \dright$.) \exampleend

  \end{itemize}
\end{Example}

It is worth emphasizing that the expression for $\HBethe\big( \vomega(s)
\big)$ in Lemma~\ref{lemma:hbethe:diagonal:1} holds for \emph{any}
$(\dleft,\dright)$-regular LDPC code over $\GF{2}$ of length $n$, \ie, it is
neither an ensemble average result, nor an asymptotic (in $n$) result.

\begin{Remark}
  \label{remark:asymptotic:growth:rate:Hamming:weight:spectrum:1}

  Interestingly enough, the functions $\omega$ and $h_{\dleft,\dright}$ from
  Lemma~\ref{lemma:hbethe:diagonal:1} appear also when studying the ensemble
  of $(\dleft,\dright)$-regular LDPC codes with block length going to
  infinity. Namely, the asymptotic growth rate of the average number of
  codewords of relative Hamming weight $\omega(s)$ is given by
  $h_{\dleft,\dright}(s)$, where the average is taken over Gallager's ensemble
  of $(\dleft,\dright)$-regular LDPC codes~\cite[Section~2]{Gallager:63}. (The
  same asymptotic growth rate is also obtained for the ensemble of all
  $(\dleft,\dright)$-regular LDPC codes as defined by Richardson and
  Urbanke~\cite{Richardson:Urbanke:01:2}, see
  also~\cite{Richardson:Urbanke:08:1, Litsyn:Shevelev:02:1,
    Di:Montanari:Urbanke:04:1}.) \remarkend
\end{Remark}

Using the interpretation of the Bethe entropy function that was given in
Section~\ref{sec:counting:in:finite:graph:covers:1}, this equivalence is not
totally surprising considering the following facts. (Here, $\graphN(\matrH)$
refers to the NFG in Lemma~\ref{lemma:hbethe:diagonal:1}.)
\begin{itemize}

\item Because $\graphN(\matrH)$ represents a $(\dleft,\dright)$-regular LDPC
  code, any finite graph cover of $\graphN(\matrH)$ also represents a
  $(\dleft,\dright)$-regular LDPC code.

\item A ``typical'' codeword of relative Hamming weight $\omega(s)$ in a
  finite graph cover of $\graphN(\matrH)$ maps down to a pseudo-codeword that
  is very close to $\omega(s) \cdot (1, \ldots, 1)$.

\item The Bethe entropy function value of some pseudo-marginal vector in the
  local marginal polytope of $\graphN(\matrH)$ ``counts'' the number of valid
  configurations in finite graph covers of $\graphN(\matrH)$ that map down to
  that pseudo-marginal vector. (See
  Section~\ref{sec:counting:in:finite:graph:covers:1} for a more precise
  statement.)

\end{itemize}
We leave it as an open problem to find a suitable generalization of
Lemma~\ref{lemma:hbethe:diagonal:1} to irregular LDPC codes and ensembles of
irregular LDPC codes.

\subsection{Implications of the Above Observation}
\label{sec:implications:minimum:Hamming:distance:1}

In this subsection we explore some of the implications of the observation that
the graph $s \mapsto \bigl( \omega(s), h_{\dleft,\dright}(s) \bigr)$ appears
in two different setups, namely in the setup of
Lemma~\ref{lemma:hbethe:diagonal:1} and in the setup of
Remark~\ref{remark:asymptotic:growth:rate:Hamming:weight:spectrum:1}. For this
discussion, recall that a function with a multi-dimensional domain is called
concave if at every point of its domain the function is concave in every
direction.

Let us first consider Gallager's ensemble of $(\dleft,\dright)$-regular LDPC
codes where $3 \leq \dleft < \dright$. It was already observed by
Gallager~\cite{Gallager:63} that codes from this ensemble have a minimum
Hamming distance that grows (with high probability) linearly with the block
length. A necessary condition for this to happen is that the function
$h_{\dleft,\dright}(s)$ is \emph{negative} for small $\omega(s) > 0$
(see~Fig.~\ref{fig:Bethe:entropy:function:aspect:3:6:1} for the case
$(\dleft,\dright) = (3,6)$). Because $h_{\dleft,\dright}(s) = 0$ for
$\omega(s) = 0$ and because $h_{\dleft,\dright}(s) > 0$ for sufficiently large
$0 < \omega(s) < 1$, the function $h_{\dleft,\dright}(s)$ must be a
\emph{convex} function of $\omega(s)$ for small $\omega(s)$. Combining this
observation with Lemma~\ref{lemma:hbethe:diagonal:1} and
Remark~\ref{remark:asymptotic:growth:rate:Hamming:weight:spectrum:1} yields
the conclusion that the induced Bethe entropy function of an NFG of a code
from this ensemble is concave and convex. (See the beginning of this section
for our definition of ``concave and convex.'') A slightly more involved
analysis then also yields the conclusion that the Bethe entropy function of an
NFG of a code from this ensemble is concave and convex.

Still talking about Gallager's ensemble of $(\dleft,\dright)$-regular LDPC
codes where $3 \leq \dleft < \dright$, these observations have also
consequences for the computation of pseudo-marginal vectors with the help of
fixed points of the sum-product algorithm. Recall that the theorem by Yedidia,
Freeman, and Weiss~\cite{Yedidia:Freeman:Weiss:05:1} showed that fixed points
of the sum-product algorithm correspond to stationary points of the Bethe free
energy function. Now, the fact that the Bethe entropy function is not concave
everywhere implies that the Bethe free energy function is not convex
everywhere, in particular it is not convex in the vicinity of pseudo-marginal
vectors that correspond to codewords. For continuing our argument, assume that
the received vector is such that the true marginal vector is close to the
marginal vector corresponding to some codeword. In order for the sum-product
algorithm to be able to somewhat closely reproduce this true marginal vector,
the sum-product algorithm would have to have a stable fixed point with a
pseudo-marginal vector somewhat close to this marginal vector, \ie, the Bethe
free energy function would need to have a local minimum at a pseudo-marginal
vector somewhat close to this marginal vector. However, the above
non-convexity results of the Bethe free energy function show that this is not
possible for every true marginal vector.\footnote{In fact, the operation of
  the sum-product algorithm on the NFG of a code from this ensembles behaves
  such that once it has ``locked into'' some codeword, the pseudo-marginal
  vector produced by the sum-product algorithm is getting closer and closer to
  the marginal vector corresponding to that codeword.} In conclusion, the
accuracy of the sum-product-algorithm-based estimation of marginal vectors of
NFGs of regular LDPC codes from ensembles with linearly growing minimum
Hamming distance has its limitations. However, if only $0$-\versus-$1$
decisions are important (as it very often is the case in channel coding
theory) then these limitations are usually not that severe.

For completeness, let us also briefly discuss Gallager's ensemble of
$(\dleft,\dright)$-regular LDPC codes where $2 = \dleft < \dright$. As pointed
out by Gallager~\cite{Gallager:63}, codes from this ensemble have a minimum
Hamming distance that grows at most logarithmically with the block
length. This is also reflected by the fact that $h_{\dleft,\dright}(s)$ is
\emph{positive} for small $\omega(s) > 0$
(see~Fig.~\ref{fig:Bethe:entropy:function:aspect:3:6:1} for the case
$(\dleft,\dright) = (2,4)$). (This statement is not strong enough to prove
concavity of $h_{\dleft,\dright}(s)$ in $\omega(s)$. For establishing this, a
detailed analysis of $h_{\dleft,\dright}(s)$ as a function of $\omega(s)$ is
necessary.)

\section{Conclusions}
\label{sec:conclusions:1}

We have shown that it is possible to give a \emph{combinatorial}
characterization of the Bethe entropy function and the Bethe partition
function, two functions that were originally defined \emph{analytically}. The
key was to study finite graph covers of the NFG under
consideration, in particular to count valid configurations in these finite
graph covers. Moreover, we have introduced a theoretical tool called
symbolwise graph-cover decoding that helps to better understand the meaning of
the pseudo-marginal vector at fixed points of the sum-product algorithm. For
all these results, the main mathematical tool that we used was the method of
types.

We finish with a few remarks.
\begin{itemize}

\item It is clear that all the results that were stated in this paper for
  temperature $T = 1$ can be suitably generalized to any temperature $T \in
  \Rpp$.

\item The fractional Bethe approximation (see, \eg,
  \cite{Wiegerinck:Heskes:03:1}) and the Kikuchi approximation (see, \eg,
  ~\cite{Yedidia:Freeman:Weiss:05:1, McEliece:Yildirim:03:1}) are usually
  better approximations than the Bethe approximation. Generalizing the results
  of the present paper, we have outlined a combinatorial characterization of
  the entropy function of these approximations in~\cite{Vontobel:11:1}.

\item Although the main application of symbolwise graph-cover decoding is in
  obtaining a better understanding of fixed points of the sum-product
  algorithm, one wonders if also the transient and the periodic behavior of
  the sum-product algorithm can be characterized in terms of graph covers (or
  variations thereof). Some initial results in that direction were sketched
  in~\cite{Vontobel:09:1}.

\item It might be interesting to study the influence of redundant
  parity-checks of LDPC codes upon the Bethe entropy function.

\end{itemize}

\section*{Acknowledgments}

It is a pleasure to acknowledge discussions with Ralf Koetter and Andi
Loeliger on factor-graph-related topics. Moreover, we greatly appreciate the
reviewers' and the associate editor's constructive comments that lead to an
improved presentation of the results.

\appendices

\section{Proof of Lemma~\ref{lemma:average:number:pre:images:in:graph:covers:1}}
\label{sec:proof:lemma:average:number:pre:images:in:graph:covers:1}

Similar to the proof of Lemma~\ref{lemma:set:of:graph:covers:1}, a possibility
to draw an $M$-cover of $\graphN$ is to first draw $M$ copies of every
function node of $\graphN$, then to draw edges that suitably connect these
function nodes, and finally, where required, to attach half-edges to the
function nodes.

In this proof, we will use this drawing procedure to guide the counting
process. Namely, we start by drawing $M$ copies for each function node, along
with the sockets where later on the edges will be attached to. Let us count in
how many ways we can specify $\{ \cva_{f,m} \}_{f \in \setF, m \in [M]}$ that
are consistent with $\vbel$. (We will call these locally valid
configurations.) It can easily be seen that there are $\prod_{f \in \setF} {M
  \choose M \cdot \vbel_f}$ ways to do this.

Fix such a locally valid configuration that is consistent with $\vbel$. Let us
count how many graph covers $\cgraph{N} \in \cset{N}_M$ specify an edge
connection such that this locally valid configuration induces a valid
configuration in $\cgraph{N}$. By this we mean that there is a unique valid
configuration $\cvc \in \codeCedge(\cgraph{N})$ such that the following holds.
\begin{itemize}

\item For every full-edge $(e,m) \in \setEfull \times [M]$ we have
  \begin{align*}
    \cover{a}_{f',m',e,m}
      &= \cover{c}_{e,m}
       = \cover{a}_{f'',m'',e,m}.
  \end{align*}
  (Here we assumed that the full-edge $(e,m)$ connects the two function nodes
  $(f',m')$ and $(f'',m'')$.)
  
\item For every half-edge $(e,m) \in \setEhalf \times [M]$ we have
  \begin{align*}
    \cover{c}_{e,m}
      &= \cover{a}_{f,m,e,m}.
  \end{align*}
  (Here we assumed that the half-edge $(e,m)$ is connected to the function node
  $(f,m)$.)
\end{itemize}
There are precisely $\prod_{e \in \setEfull} \prod_{a_e} (M \bel_{e,a_e})!$
such $M$-covers. This can be seen as follows. Namely, consider some full-edge
$e \in \setEfull$ that connects the two function nodes $f'$ and $f''$ and fix
some $a_e \in \setA_e$. It follows from the edge consistency constraints of
the local marginal polytope that $\vbel$ is such that
\begin{align*}
  \sum_{\va'_{f'} \in \setA_{f'}: \, a'_{f',e} = a_e} \!\!\!
    \bel_{f', \va'_{f'}} = \bel_{e,a_e}
    &= \sum_{\va''_{f''} \in \setA_{f''}: \, a''_{f'',e} = a_e} \!\!\!
         \bel_{f'', \va''_{f''}}.
\end{align*}
Therefore, the number of $e$-sockets among the $M$ copies of $f'$ that take on
the value $a_e$ is $M \cdot \bel_{e,a_e}$, and the number of $e$-sockets among
the $M$ copies of $f''$ that take on the value $a_e$ is also $M \cdot
\bel_{e,a_e}$.  These sockets can be connected by $M \cdot \bel_{e,a_e}$ edges
in exactly $(M \bel_{e,a_e})!$ ways.

Note that the number $\prod_{e \in \setEfull} \prod_{a_e} (M \bel_{e,a_e})!$
is independent of the chosen locally valid configuration, and so, among all
$M$-covers, the total number of valid configurations that map down to $\vbel$
equals $\prod_{f \in \setF} {M \choose M \cdot \vbel_f} \cdot \prod_{e \in
  \setEfull} \prod_{a_e} (M \bel_{e,a_e})!$\,. The lemma statement is then
obtained by dividing this number by $\cardbig{\cset{N}_M}$, by using the
result in~\eqref{eq:number:M:covers:1}, and by applying the abbreviations that
are defined in~\eqref{eq:M:choose:bel:1}--\eqref{eq:M:choose:bel:2}.

\section{Proof of Theorem~\ref{theorem:degree:M:partition:function:1}}
\label{sec:proof:theorem:degree:M:partition:function:1}

We start by reformulating the $M$th power of $\ZBetheM{M}(\graphN)$. Namely,
we have
\begin{align}
  &
  \hskip-0.25cm
  \big(
    \ZBetheM{M}(\graphN)
  \big)^M \nonumber \\
    &\onestareq
       \Big\langle \!
         \ZGibbs(\cgraph{N})
       \! \Big\rangle_{\cgraph{N} \in \cset{N}_{M}} \nonumber \\
    &\twostarseq
       \frac{1}{\cardbig{\cset{N}_M}}
         \sum_{\cgraph{N} \in \cset{N}_M}
           \sum_{\cvc \in \codeCedge(\cgraph{N})}
             g_{\cgraph{N}}(\cvc)^{1/T} \nonumber \\
    &\threestarseq
       \sum_{\vbel \in \lmpBdown{M}}
         \frac{1}{\cardbig{\cset{N}_M}}
         \sum_{\cgraph{N} \in \cset{N}_M}
           \sum_{\cvc \in \codeCedge(\cgraph{N})}
             \!
             \left[ \varphiM\big( \cgraph{N},\cvc \big) = \vbel \right]
             \cdot
             g_{\cgraph{N}}(\cvc)^{1/T} \nonumber \\
    &\fourstarseq
       \sum_{\vbel \in \lmpBdown{M}}
         \exp\big( - (M/T) \cdot \UBethe(\vbel) \big)
           \nonumber \\[-0.25cm]
    &\quad\quad\quad\quad\quad\quad
         \cdot
         \frac{1}{\cardbig{\cset{N}_M}}
         \sum_{\cgraph{N} \in \cset{N}_M}
           \sum_{\cvc \in \codeCedge(\cgraph{N})}
             \left[ \varphiM\big( \cgraph{N},\cvc \big) = \vbel \right]
               \nonumber
  \end{align}
  \begin{align}
    &\fivestarseq
       \sum_{\vbel \in \lmpBdown{M}}
         \exp\big( - (M/T) \cdot \UBethe(\vbel) \big)
         \cdot
         \bar C_M(\vbel) \nonumber \\
    &\sixstarseq
       \sum_{\vbel \in \lmpBdown{M}}
         \exp\big( - (M/T) \cdot \FBethe(\vbel) + o(M) \big),
           \label{eq:ZBethe:power:1}
\end{align}
where at step~$\onestar$ we have used
Definition~\ref{def:degree:M:partition:function:1}, where at step~$\twostars$
we have used~\eqref{eq:Z:Gibbs:1} and
Definition~\ref{def:average:over:graph:covers:1}, where at step~$\threestars$
we have used Definitions~\ref{def:down:projection:1}
and~\ref{def:down:projection:2}, where at step~$\fourstars$ we have
used~Theorem~\ref{theorem:asymptotic:average:function:value:pre:images:in:graph:covers:1},
where at step~$\fivestars$ we have used Eq.~\eqref{eq:def:C:bar:1} and
Definitions~\ref{def:average:over:graph:covers:1}
and~\ref{def:counting:valid:configurations:compatible:with:beliefs:1}, and
where at step~$\sixstars$ we have
used~\eqref{eq:asymptotic:average:number:pre:images:in:graph:covers:1}.
Consequently we obtain
\begin{align*}
  &
  \hskip-0.5cm
  \limsup_{M \to \infty} \
    \ZBetheM{M}(\graphN) \\
    &\onestareq
       \limsup_{M \to \infty}
         \sqrt[M]{\sum_{\vbel \in \lmpBdown{M}}
                    \exp
                      \big(
                        - (M/T) \cdot \FBethe(\vbel) + o(M)
                      \big)
                 } \\
    &\twostarseq
       \limsup_{M \to \infty}
         \sqrt[M]{\max_{\vbel \in \lmpBdown{M}} 
                   \exp\big( - (M/T) \cdot \FBethe(\vbel) + o(M) \big)
                 } \\
    &\threestarseq
       \limsup_{M \to \infty} \ 
         \max_{\vbel \in \lmpBdown{M}} \ 
           \exp\big( - (1/T) \cdot \FBethe(\vbel) + o(1) \big) \\
    &\fourstarseq
       \sup_{\vbel \in \lmpBdown{}} \ 
       \exp
         \big(
           -
           (1/T)
           \cdot
           \FBethe(\vbel)
         \big) \\
    &= \exp
         \left(
           -
           \frac{1}{T}
           \cdot
           \inf_{\vbel \in \lmpBdown{}}
             \FBethe(\vbel)
         \right) \\
    &\fivestarseq
       \exp
         \left(
           -
           \frac{1}{T}
           \cdot
           \min_{\vbel \in \lmpB}
             \FBethe(\vbel)
         \right) \\
    &\sixstarseq
       \ZBethe(\graphN),
\end{align*}
where at step~$\onestar$ we have used~\eqref{eq:ZBethe:power:1}, where in
step~$\twostars$ the replacement of the sum by the maximum operator is
justified by the fact that the size of the set $\lmpBdown{M}$ grows polynomially
in $M$ (see~Remark~\ref{remark:lmpB:1}), where step~$\threestars$ follows from
taking the maximization operator out of the $M$th root, where in
step~$\fourstars$ we have used the definition of $\lmpBdown{}$, the set of
lift-realizable pseudo-marginal vectors, where at step~$\fivestars$ we have
used the fact that the closure of $\lmpBdown{}$ equals $\lmpB$, which is a
consequence of Theorem~\ref{theorem:down:projection:1}, and the fact that
$\lmpB$ is compact so that the infimum operator can be replaced by a
minimization operator, and where at step~$\sixstars$ we have used
Definition~\ref{def:Bethe:partition:function:1}. This is the result that was
promised in the theorem statement.

\section{Proof of Theorem~\ref{theorem:SGCD:characterization:1}}
\label{sec:proof:theorem:SGCD:characterization:1}

From the derivations in this appendix it will be apparent that there are close
connections to the proof of
Theorem~\ref{theorem:degree:M:partition:function:1} in
Appendix~\ref{sec:proof:theorem:degree:M:partition:function:1}.

For any $e \in \setE$, any $M \in \Zpp$, and any $a_e \in \setA_e$ we have
\begin{align*}
  \eta_{e,M}&(a_e) \\
    &\onestareq
       \frac{1}{M}
       \sum_{m \in [M]}
         \eta_{e,m,M}(a_e) \\
    &\twostarseq
       \frac{1}{M}
       \sum_{m \in [M]}
         \frac{1}{Z'_M(\graphN)}
         \sum_{\cgraph{N} \in \cset{N}_M}
           \ZGibbsext{\cgraph{N}}
           \cdot
           \eta_{e,m,\cgraph{N}}(a_e) \\
  \end{align*}
  \begin{align*}
    &\threestarseq
       \frac{1}{M}
       \sum_{m \in [M]}
         \frac{1}{Z'_M(\graphN)}
         \sum_{\cgraph{N} \in \cset{N}_M} \ 
           \sum_{\cvc \in \codeC(\cgraph{N}): \, \cover{c}_{e,m} = a_e}
             g_{\cgraph{N}}(\cvc) \\
    &= \frac{1}{M}
       \sum_{m \in [M]}
         \frac{1}{Z'_M(\graphN)}
         \sum_{\cgraph{N} \in \cset{N}_M}
           \sum_{\cvc \in \codeC(\cgraph{N})}
             \big[
               \cover{c}_{e,m} \! = \! a_e
             \big]
             \cdot
             g_{\cgraph{N}}(\cvc) \\
    &\fourstarseq
       \sum_{\vbel \in \lmpBdown{M}}
         \frac{1}{M}
         \sum_{m \in [M]}
           \frac{1}{Z'_M(\graphN)}
           \sum_{\cgraph{N} \in \cset{N}_M}
             \sum_{\cvc \in \codeC(\cgraph{N})} \\
    &\quad\quad\quad\quad
               \left[ \varphiM\big( \cgraph{N}, \cvc \big) = \vbel \right]
               \cdot
               \big[
                 \cover{c}_{e,m} \! = \! a_e
               \big]
               \cdot
               g_{\cgraph{N}}(\cvc) \\
    &\fivestarseq
       \sum_{\vbel \in \lmpBdown{M}}
         \exp(- M \cdot \UBethe(\vbel))
         \cdot
         \frac{1}{Z'_M(\graphN)}
         \sum_{\cgraph{N} \in \cset{N}_M}
           \sum_{\cvc \in \codeC(\cgraph{N})} \\
    &\quad\quad\quad\quad
             \left[ \varphiM\big( \cgraph{N}, \cvc \big) = \vbel \right]
             \cdot
             \underbrace{
               \frac{1}{M}
               \cdot
               \sum_{m \in [M]}
                 \big[
                   \cover{c}_{e,m} \! = \! a_e
                 \big]
             }_{= \ \bel_{e,a_e}} \\
    &= \sum_{\vbel \in \lmpBdown{M}}
         \bel_{e,a_e}
         \cdot
         \exp
           \big(
             - M \cdot \UBethe(\vbel)
           \big) \\
    &\quad\quad\quad\quad
         \cdot
         \frac{1}{Z'_M(\graphN)}
         \sum_{\cgraph{N} \in \cset{N}_M}
           \sum_{\cvc \in \codeC(\cgraph{N})}
             \left[ \varphiM\big( \cgraph{N}, \cvc \big) = \vbel \right]
              \\
    &\sixstarseq
       \frac{\cardbig{\cset{N}_M}}{Z'_M(\graphN)}
       \cdot
       \sum_{\vbel \in \lmpBdown{M}}
         \bel_{e,a_e}
         \cdot
         \exp
           \big(
             - M \cdot \UBethe(\vbel)
           \big)
         \cdot
         \bar C_M(\vbel) \\
    &\sevenstarseq
       \frac{\cardbig{\cset{N}_M}}{Z'_M(\graphN)}
       \cdot
       \sum_{\vbel \in \lmpBdown{M}}
         \bel_{e,a_e}
         \cdot
         \exp
           \big(
             - M \cdot \FBethe(\vbel) + o(M)
           \big)
         \cdot
\end{align*}
where at steps~$\onestar$, $\twostars$, and $\threestars$ we have used
Definition~\ref{def:SGCD:1}, where at step~$\fourstars$ we have used
Definitions~\ref{def:down:projection:1} and~\ref{def:down:projection:2}, where
at step $\fivestars$ we have used
Theorem~\ref{theorem:asymptotic:average:function:value:pre:images:in:graph:covers:1},
where at step~$\sixstars$ we have used Eq.~\eqref{eq:def:C:bar:1} and
Definitions~\ref{def:average:over:graph:covers:1}
and~\ref{def:counting:valid:configurations:compatible:with:beliefs:1}, and
where at step~$\sevenstars$ we have
used~\eqref{eq:asymptotic:average:number:pre:images:in:graph:covers:1}.

The next step is to evaluate $\eta_{e,M}(a_e)$ in the limit $M \to
\infty$. Because the size of the set $\lmpBdown{M}$ grows polynomially in $M$
(see~Remark~\ref{remark:lmpB:1}), we can use an approach similar to the one
that was used in Section~\ref{sec:Gibbs:free:energy:arises:naturally:1} to
simplify~\eqref{eq:veta:reformulation:2} in the limit $M \to \infty$.  We
obtain
\begin{align*}
  \lim_{M \to \infty}
    \eta_{e,M}(a_e)
    &= \gamma_e
       \cdot
       \bels_{e,a_e},
         \quad e \in \setE, \, a_e \in \setA_e,
\end{align*}
where $\gamma_e \in \Rpp$ is some suitable constant, and where
\begin{align*}
  \hvbel
    &= \argmin_{\vbel \in \lmpB} \,
         \Big.
           \FBethe(\vbel)
         \Big|_{T = 1}.
\end{align*}
Actually, $\gamma_e = 1$ because $\eta_{e,\infty}(a_e)$ was defined such that
$\sum_{a_e} \eta_{e,\infty}(a_e) = 1$ for all $e \in \setE$.

For any $f \in \setF$ and any $\va_f \in \setA_f$, the proof of the second
statement in Theorem~\ref{theorem:SGCD:characterization:1} is nearly identical
to the above proof. We omit the details.

\section{Proof of Lemma~\ref{lemma:hbethe:diagonal:1}}
\label{sec:proof:lemma:hbethe:diagonal:1}

Recall the definition of the Bethe entropy function from
Definition~\ref{def:Bethe:free:energy:1} and the induced Bethe entropy
function from Definition~\ref{def:induced:bethe:entropy:function:1}. Fix some
pseudo-codeword $\vomega = \omega \cdot (1, \ldots, 1) \in \fp{P}$, $0 \leq
\omega \leq 1$, and let $\vbels \defeq \vPsiBME(\vomega)$. We have to evaluate
\begin{align*}
  \HBethe(\vomega)
    &= \HBethe(\vbels) \\
    &= \sum_f
         \HBethesub{f}(\vbels_f)
       -
       \sum_{e \in \setEfull}
         \HBethesub{e}(\vbels_e) \\
    &= \sum_i
         \HBethesub{i}(\vbels_i)
       +
       \sum_j
         \HBethesub{j}(\vbels_j)
       -
       \sum_{e \in \setEfull}
         \HBethesub{e}(\vbels_e).
\end{align*}
Clearly, for every $i \in \setI$ we have $\bels_{i,(0,\ldots,0)} = 1 - \omega$ and
$\bels_{i,(1,\ldots,1)} = \omega$, and so
\begin{align}
  \HBethesub{i}(\vbels_i)
    &= h(\omega),
         \quad i \in \setI.
           \label{eq:Bethe:entropy:value:i:1}
\end{align}
Moreover, the edge consistency constraints of $\lmpB$ imply that for every $e
\in \setEfull$ it holds that $\bels_{e,0} = 1 - \omega$ and $\bels_{e,1} =
\omega$, and so
\begin{align}
  \HBethesub{e}(\vbels_e)
    &= h(\omega),
         \quad e \in \setEfull.
           \label{eq:Bethe:entropy:value:e:1}
\end{align}
The computations for $\HBethesub{j}(\vbels_j)$ are more involved because we
need to find the maximizing $\vbel = \vbels$ in~\eqref{eq:vPsiBME:def:1}.

These computations are simplified by the observation that
$\HBethesub{j}(\vbels_j)$ can be maximized for every $j \in \setJ$
separately. Therefore, let us fix some $j \in \setJ$. We have to maximize
\begin{align}
  \HBethesub{j}(\vbel_j)
    = -
      \sum_{\va_j}
        \bel_{j,\va_j} \log(\bel_{j,\va_j})
          \label{eq:check:entropy:1}
\end{align}
under the constraints 
\begin{align}
  \sum_{\va_j: \, a_{j,e} = 1}
    \bel_{j,\va_j}
    &= \omega,
       \quad e \in \set{E}_j,
         \label{eq:check:entropy:constraints:1} \\
  \sum_{\va_j} \bel_{j,\va_j}
    &= 1,
         \label{eq:check:entropy:constraints:2}
\end{align}
where the constraints in~\eqref{eq:check:entropy:constraints:1} are implied by
the edge consistency constraints of $\lmpB$. (Strictly speaking, we also have
to impose the inequalities $0 \leq \bel_{j,\va_j} \leq 1$ for all $\va_j \in
\setA_j$, however, we will see that the solution satisfies them
automatically.) Introducing Lagrange multipliers $\{ s_{j,e} \}_{e \in
  \setE_j}$ and $\nu_j$, we obtain the Lagrangian
\begin{align*}
  &
  -
  \sum_{\va_j \in \set{B}_j}
    \bel_{j,\va_j} \log(\bel_{j,\va_j}) \\
  &
  +
  \sum_{e \in \set{E}_j}
    s_{j,e}
    \cdot
    \left(
      \sum_{\va_j \atop a_{j,e} = 1}
        \bel_{j,\va_j}
      -
      \omega
    \right)
  +
  \nu_j
    \cdot
    \left(
      \sum_{\va_j} \bel_{j,\va_j}
      -
      1
    \right).
\end{align*}
Because of the concavity of the Lagrangian in $\{ \bel_{j,\va_j} \}_{\va_j}$,
and because of the symmetry of the setup (\ie, the symmetry of the single
parity-check code and the symmetry of the constraints), all Lagrange
multipliers $\{ s_{j,e} \}_{e \in \setE_j}$ must take on the same value, say
$s_j$. Therefore, the new Lagrangian is
\begin{align*}
  &
  -
  \sum_{\va_j}
    \bel_{j,\va_j} \log(\bel_{j,\va_j}) \\
  &
  +
  s_j
  \sum_{e \in \set{E}_j}
    \left(
      \sum_{\va_j: \, a_{j,e} = 1}
      \bel_{j,\va_j}
      -
      \omega
    \right)
    +
    \nu_j
    \cdot
    \left(
      \sum_{\va_j} \bel_{j,\va_j}
      -
      1
    \right) \\
    &\onestareq
       - \!
       \sum_{\va_j}
         \bel_{j,\va_j} \! \log(\bel_{j,\va_j}) \\
    &\quad\,
       +
       s_j
         \cdot
         \sum_{\va_j}
           \wH(\va_j)
           \bel_{j,\va_j}
       - 
       s_j \dright \omega
       +
       \nu_j
         \cdot
         \sum_{\va_j}
           \bel_{j,\va_j}
       -
       \nu_j,
\end{align*}
where $\wH(\va_j)$ denotes the Hamming weight of $\va_j$, and where at step
$\onestar$ we have used $|\setE_j| = \dright$. Computing the gradient of the
Lagrangian with respect to $\{ \bel_{j,\va_j} \}_{\va_j}$, and setting it
equal to the zero vector, we obtain
\begin{align*}
  -
  \log(\bels_{j,\va_j})
  -
  1
  +
  s_j
  \cdot
  \wH(\va_j)
  +
  \nu_j
    &\overset{!}{=}
       0,
       \quad \va_j \in \set{B}_j.
\end{align*}
Therefore,
\begin{align}
  \bels_{j,\va_j}
    &= \frac{\exp\big( s_j \cdot \wH(\va_j) \big)}
            {\sum_{\va'_j}
               \exp\big( s_j \cdot \wH(\va'_j) \big)
            },
            \quad \va_j \in \set{B}_j.
              \label{eq:optimal:bel:1}
\end{align}
We define
\begin{align}
  \theta_j(s_j)
    &\defeq
       \log
         \left(
           \sum_{\va_j}
             \exp\big( s_j \cdot \wH(\va_j) \big)
         \right).
           \label{def:theta:function:1}
\end{align}
Then the sum of all the constraints in~\eqref{eq:check:entropy:constraints:1}
implies
\begin{align}
  \dright \omega
    &= \sum_{e \in \set{E}_j}
         \sum_{\va_j: \, a_{j,e} = 1}
           \bels_{j,\va_j}
     = \sum_{\va_j}
         \wH(\va_j) \cdot \bels_{j,\va_j} \nonumber \\
    &\onestareq
       \sum_{\va_j}
         \wH(\va_j)
         \cdot
         \frac{\exp\big( s_j \cdot \wH(\va_j) \big)}
            {\sum_{\va'_j}
               \exp\big( s_j \cdot \wH(\va'_j) \big)
            }
     \twostarseq
       \rdiff{s_j} \theta_j(s_j),
         \label{eq:theta:derivative:1}
\end{align}
where at step~$\onestar$ we have used~\eqref{eq:optimal:bel:1}, and where at
step~$\twostars$ we have used~\eqref{def:theta:function:1}. Solving for
$\omega$ we obtain
\begin{align}
  \omega
    &= \omega^{(j)}(s_j)
     \defeq
       \frac{1}{\dright}
       \cdot
       \rdiff{s_j} \theta_j(s_j).
         \label{eq:omega:theta:def:1}
\end{align}
With this, the entropy expression
in~\eqref{eq:check:entropy:1} can be rewritten to read
\begin{align}
  \HBethesub{j}(\vbels_j)
    &\onestareq
       -
       s_j \rdiff{s_j} \theta_j(s_j)
       +
       \theta_j(s_j) \nonumber \\
    &\twostarseq 
       -
       \dright \cdot s_j \cdot \omega^{(j)}(s_j)
       +
       \theta_j(s_j),
         \label{eq:Bethe:entropy:value:j:1}
\end{align}
where at step~$\onestar$ we have used~\eqref{eq:optimal:bel:1},
\eqref{def:theta:function:1}, and~\eqref{eq:theta:derivative:1}, and where at
step~$\twostars$ we have used~\eqref{eq:omega:theta:def:1}. It can be verified
that this is indeed the maximal value of~\eqref{eq:check:entropy:1} under the
constraints
in~\eqref{eq:check:entropy:constraints:1}--\eqref{eq:check:entropy:constraints:2}.

Note that $\theta_j(s_j)$ is a strictly convex function in $s_j$ (see, \eg,
\cite{Boyd:Vandenberghe:04:1}), and so $\rdiff{s_j} \theta_j(s_j)$ is a
strictly monotonically increasing function in $s_j$. This implies that for
every $j \in \setJ$ there is a unique $s_j$ such that $\omega =
\omega^{(j)}(s_j)$.

Because all function nodes $j \in \setJ$ have the same degree, it is clear
that the functions $\theta_j$ and $\omega^{(j)}$ are independent of $j$. This
implies that there is an $s \in \R$ such that $s_j = s$ for all $j \in
\setJ$. It also implies that $\HBethesub{j}(\vbels_j)$ is independent of $j$.

Finally, adding up all entropy terms, the induced Bethe entropy equals
\begin{align*}
  &
  \HBethe(\vomega) \\
    &= \sum_i
         \HBethesub{i}(\vbels_i)
       +
       \sum_j
         \HBethesub{j}(\vbels_j)
       -
       \sum_{e \in \setEfull}
         \HBethesub{e}(\vbels_e) \\
    &\onestareq
       \sum_i
         h(\omega)
       -
       \sum_j
         \dright \cdot s \cdot \omega(s)
       +
       \sum_j
         \theta(s)
       -
       \sum_{e \in \setEfull}
         h(\omega) \\
    &\twostarseq
       -
       n
       \cdot
       (\dleft \! - \! 1) \cdot h\big( \omega(s) \big)
       -
       n
       \cdot
       \dleft \cdot s \cdot \omega(s)
       +
       n
       \cdot
       \frac{\dleft}{\dright}
       \cdot
       \theta(s),
\end{align*}
where at step~$\onestar$ we have used~\eqref{eq:Bethe:entropy:value:i:1},
\eqref{eq:Bethe:entropy:value:e:1}, and~\eqref{eq:Bethe:entropy:value:j:1},
and where at step~$\twostars$ we have used $|\setI| = n$, $|\setEfull| = n
\cdot \dleft$, and $|\setJ| = |\setEfull| / \dright = n \cdot \dleft /
\dright$.

The proof of this lemma is then concluded by observing that $\theta_j(s_j)$
in~\eqref{def:theta:function:1} can also be written as
\begin{align}
  \theta_j(s_j)
    &= \log
         \left(
           \sum_{w = 0 \atop w \ \mathrm{even}}^{\dright}
             {\dright \choose w}
             \exp(s_j \cdot w)
         \right),
\end{align}
where we have used the fact that the local constraint code $\setA_j$ contains
${\dright \choose w}$ codewords of weight $w$ if $w \in \{ 0, 1, \ldots,
\dright \}$ is even, and $0$ codewords of weight $w$ if $w \in \{ 0, 1,
\ldots, \dright \}$ is odd.

\bibliographystyle{IEEEtran}
\bibliography{/home/vontobel/references/references}

\begin{thebibliography}{10}
\providecommand{\url}[1]{#1}
\csname url@samestyle\endcsname
\providecommand{\newblock}{\relax}
\providecommand{\bibinfo}[2]{#2}
\providecommand{\BIBentrySTDinterwordspacing}{\spaceskip=0pt\relax}
\providecommand{\BIBentryALTinterwordstretchfactor}{4}
\providecommand{\BIBentryALTinterwordspacing}{\spaceskip=\fontdimen2\font plus
\BIBentryALTinterwordstretchfactor\fontdimen3\font minus
  \fontdimen4\font\relax}
\providecommand{\BIBforeignlanguage}[2]{{%
\expandafter\ifx\csname l@#1\endcsname\relax
\typeout{** WARNING: IEEEtran.bst: No hyphenation pattern has been}%
\typeout{** loaded for the language `#1'. Using the pattern for}%
\typeout{** the default language instead.}%
\else
\language=\csname l@#1\endcsname
\fi
#2}}
\providecommand{\BIBdecl}{\relax}
\BIBdecl

\bibitem{Wiberg:Loeliger:Koetter:95}
N.~Wiberg, H.-A. Loeliger, and R.~K{\"o}tter, ``Codes and iterative decoding on
  general graphs,'' \emph{Europ.\ Trans.\ on Telecomm.}, vol.~6, pp. 513--525,
  Sep./Oct. 1995.

\bibitem{Wiberg:96}
N.~Wiberg, ``Codes and decoding on general graphs,'' Ph.D. dissertation,
  Department of Electrical Engineering, Link\"oping University, Sweden, 1996.

\bibitem{Yedidia:Freeman:Weiss:05:1}
J.~S. Yedidia, W.~T. Freeman, and Y.~Weiss, ``Constructing free-energy
  approximations and generalized belief propagation algorithms,'' \emph{IEEE
  Trans.\ Inf.\ Theory}, vol.~51, no.~7, pp. 2282--2312, Jul. 2005.

\bibitem{Koetter:Vontobel:03:1}
R.~Koetter and P.~O. Vontobel, ``Graph covers and iterative decoding of
  finite-length codes,'' in \emph{Proc.\ 3rd Intern.\ Symp.\ on Turbo Codes and
  Related Topics}, Brest, France, Sep.~1--5 2003, pp. 75--82.

\bibitem{Vontobel:Koetter:05:1:subm}
P.~O. Vontobel and R.~Koetter, ``Graph-cover decoding and finite-length
  analysis of message-passing iterative decoding of {LDPC} codes,'' \emph{CoRR,
  \emph{\texttt{http://www.arxiv.org/abs/cs.IT/0512078}}}, Dec. 2005.

\bibitem{Vontobel:Koetter:04:2}
------, ``On the relationship between linear programming decoding and min-sum
  algorithm decoding,'' in \emph{Proc.\ Intern.\ Symp.\ on Inf.\ Theory and its
  Applications (ISITA)}, Parma, Italy, Oct.~10--13 2004, pp. 991--996.

\bibitem{Feldman:03:1}
J.~Feldman, ``Decoding error-correcting codes via linear programming,'' Ph.D.
  dissertation, Dept.~of Electrical Engineering and Computer Science,
  Massachusetts Institute of Technology, Cambridge, MA, 2003.

\bibitem{Feldman:Wainwright:Karger:05:1}
J.~Feldman, M.~J. Wainwright, and D.~R. Karger, ``Using linear programming to
  decode binary linear codes,'' \emph{IEEE Trans.\ Inf.\ Theory}, vol.~51,
  no.~3, pp. 954--972, Mar. 2005.

\bibitem{Bethe:35:1}
H.~Bethe, ``Statistical theory of superlattices,'' \emph{Proc.\ Royal Society
  London A}, vol. 150, no. 871, pp. 552--575, Jul. 1935.

\bibitem{Peierls:36:1}
R.~Peierls, ``On {Ising's} model of ferromagnetism,'' \emph{Math.\ Proc.
  Cambr.\ Phil.\ Soc.}, vol.~32, no.~3, pp. 477--481, Oct. 1936.

\bibitem{Kurata:Kikuchi:Watari:53:1}
M.~Kurata, R.~Kikuchi, and T.~Watari, ``A theory of cooperative phenomena.
  {III}. {D}etailed discussions of the cluster variation method,'' \emph{J.\
  Chem.\ Phys.}, vol.~21, no.~3, pp. 434--448, Mar. 1953.

\bibitem{Eggarter:74:1}
T.~P. Eggarter, ``{C}ayley trees, the {I}sing problem, and the thermodynamic
  limit,'' \emph{Phys.\ Rev.\ B}, vol.~9, no.~7, pp. 2989--2992, Apr. 1974.

\bibitem{Thorpe:82:1}
M.~F. Thorpe, ``Bethe lattices,'' in \emph{Excitations in Disordered Systems
  (NATO Advanced Study Institute Series B78)}, M.~F. Thorpe, Ed.\hskip 1em plus
  0.5em minus 0.4em\relax Plenum, New York, 1982, pp. 85--107.

\bibitem{Baxter:82:1}
R.~J. Baxter, \emph{Exactly Solved Models in Statistical Mechanics}.\hskip 1em
  plus 0.5em minus 0.4em\relax London, UK: Academic Press, 1982.

\bibitem{Frey:Koetter:Vardy:01:1}
B.~J. Frey, R.~Koetter, and A.~Vardy, ``Signal-space characterization of
  iterative decoding,'' \emph{IEEE Trans.\ Inf.\ Theory}, vol.~47, no.~2, pp.
  766--781, Feb. 2001.

\bibitem{Fogal:McEliece:Thorpe:05:1}
S.~L. Fogal, R.~McEliece, and J.~Thorpe, ``Enumerators for protograph ensembles
  of {LDPC} codes,'' in \emph{Proc.\ IEEE Int.\ Symp.\ Inf.\ Theory}, Adelaide,
  Australia, Sep.~4--9 2005, pp. 2156--2160.

\bibitem{Divsalar:Jones:Dolinar:Thorpe:05:1}
D.~Divsalar, C.~Jones, S.~Dolinar, and J.~Thorpe, ``Protograph based {LDPC}
  codes with minimum distance linearly growing with block size,'' in
  \emph{Proc.\ IEEE Global Communications Conference}, St. Louis, MO, USA,
  Nov.~28--Dec.~2 2005.

\bibitem{Fu:Anastasopoulos:07:1}
K.~Fu and A.~Anastasopoulos, ``Stopping-set enumerator approximations for
  finite-length protograph {LDPC} codes,'' in \emph{Proc.\ IEEE Int.\ Symp.\
  Inf.\ Theory}, Nice, France, June 24--29 2007, pp. 2946--2950.

\bibitem{Ravazzi:Fagnani:09:1}
C.~Ravazzi and F.~Fagnani, ``Spectra and minimum distances of repeat
  multiple-accumulate codes,'' \emph{IEEE Trans.\ Inf.\ Theory}, vol.~55,
  no.~11, pp. 4905--4924, Nov. 2009.

\bibitem{AbuSurra:Divsalar:Ryan:11:1}
S.~Abu-Surra, D.~Divsalar, and W.~E. Ryan, ``Enumerators for protograph-based
  ensembles of {LDPC} and generalized {LDPC} codes,'' \emph{IEEE Trans.\ Inf.\
  Theory}, vol.~57, no.~2, pp. 858--886, Feb. 2011.

\bibitem{Flanagan:Paolini:Chiani:Fossorier:11:1}
M.~F. Flanagan, E.~Paolini, M.~Chiani, and M.~P.~C. Fossorier, ``On the growth
  rate of the weight distribution of irregular doubly generalized {LDPC}
  codes,'' \emph{IEEE Trans.\ Inf.\ Theory}, vol.~57, no.~6, pp. 3721--3737,
  Jun. 2011.

\bibitem{Divsalar:Dolecek:12:1}
D.~Divsalar and L.~Dolecek, ``Graph cover ensembles of non-binary protograph
  {LDPC} codes,'' in \emph{Proc.\ IEEE Int.\ Symp.\ Inf.\ Theory}, Cambridge,
  MA, USA, Jul.~1--6 2012.

\bibitem{Benedetto:Montorsi:96:1}
S.~Benedetto and G.~Montorsi, ``Unveiling turbo codes: some results on parallel
  concatenated coding schemes,'' \emph{IEEE Trans.\ Inf.\ Theory}, vol.~42,
  no.~2, pp. 409--428, Mar. 1996.

\bibitem{Benedetto:Divsalar:Montorsi:Pollara:98:1}
S.~Benedetto, D.~Divsalar, G.~Montorsi, and F.~Pollara, ``Serial concatenation
  of interleaved codes: performance analysis, design, and iterative decoding,''
  \emph{IEEE Trans.\ Inf.\ Theory}, vol.~44, no.~3, pp. 909--926, May 1998.

\bibitem{Angluin:80:1}
D.~Angluin, ``Local and global properties in networks of processors,'' in
  \emph{Proc.\ 12th Annual ACM Symp.\ on Theory of Computing}, Los Angeles, CA,
  1980, pp. 82--93.

\bibitem{Ruozzi:Thaler:Tatikonda:09:1}
N.~Ruozzi, J.~Thaler, and S.~Tatikonda, ``Graph covers and quadratic
  minimization,'' in \emph{Proc.\ 47th Allerton Conf.\ on Communications,
  Control, and Computing}, Allerton House, Monticello, IL, USA, Sep.~30--Oct.~2
  2009, pp. 1590--1596.

\bibitem{Malioutov:Johnson:Willsky:06:1}
D.~M. Malioutov, J.~K. Johnson, and A.~S. Willsky, ``Walk-sums and belief
  propagation in {G}aussian graphical models,'' \emph{J.\ Mach.\ Learn.\ Res.},
  vol.~7, pp. 2031--2064, Dec. 2006.

\bibitem{Mooij:Kappen:05:1}
J.~M. Mooij and H.~J. Kappen, ``On the properties of the {B}ethe approximation
  and loopy belief propagation on binary networks,'' \emph{J.\ Stat.\ Mech.:
  Theory and Experiment}, p. P11012, Nov. 2005.

\bibitem{Heskes:06:1}
T.~Heskes, ``On the uniqueness of loopy belief propagation fixed points,''
  \emph{Neural Computation}, vol.~16, no.~11, pp. 2379--2413, Nov. 2004.

\bibitem{Chertkov:Chernyak:06:1}
M.~Chertkov and V.~Y. Chernyak, ``Loop series for discrete statistical models
  on graphs,'' \emph{J.\ Stat.\ Mech.: Theory and Experiment}, p. P06009, Jun.
  2006.

\bibitem{Walsh:Regalia:Johnson:06:1}
J.~M. Walsh, P.~A. Regalia, and C.~R. {Johnson, Jr.}, ``Turbo decoding as
  iterative constrained maximum-likelihood sequence detection,'' \emph{IEEE
  Trans.\ Inf.\ Theory}, vol.~52, no.~12, pp. 5426--5437, Dec. 2006.

\bibitem{Regalia:Walsh:07:1}
P.~A. Regalia and J.~M. Walsh, ``Optimality and duality of the turbo decoder,''
  \emph{Proceedings of the IEEE}, vol.~95, no.~6, pp. 1362--1377, Jun. 2007.

\bibitem{Mooij:Kappen:07:1}
J.~M. Mooij and H.~J. Kappen, ``Sufficient conditions for convergence of the
  sum-product algorithm,'' \emph{IEEE Trans.\ Inf.\ Theory}, vol.~53, no.~12,
  pp. 4422--4437, Dec. 2007.

\bibitem{Mezard:Parisi:Virasoro:87:1}
M.~M{\'e}zard, G.~Parisi, and M.~A. Virasoro, \emph{Spin Glass Theory and
  Beyond}.\hskip 1em plus 0.5em minus 0.4em\relax Singapore: World Scientific,
  1987.

\bibitem{Mezard:Montanari:09:1}
M.~M{\'e}zard and A.~Montanari, \emph{Information, Physics, and
  Computation}.\hskip 1em plus 0.5em minus 0.4em\relax New York, NY: Oxford
  University Press, 2009.

\bibitem{Tanaka:02:1}
T.~Tanaka, ``A statistical-mechanics approach to large-system analysis of
  {CDMA} multiuser detectors,'' \emph{IEEE Trans.\ Inf.\ Theory}, vol.~48,
  no.~11, pp. 2888--2910, Nov. 2002.

\bibitem{Mori:11:1}
R.~Mori, ``Connection between annealed free energy and belief propagation on
  random factor graph ensembles,'' in \emph{Proc.\ IEEE Int.\ Symp.\ Inf.\
  Theory}, St.~Petersburg, Russia, Jul.~31--Aug.~5 2011, pp. 2010--2014.

\bibitem{Mori:Tanaka:12:1}
R.~Mori and T.~Tanaka, ``Central approximation in statistical physics and
  information theory,'' in \emph{Proc.\ IEEE Int.\ Symp.\ Inf.\ Theory},
  Cambridge, MA, USA, Jul.~1--6 2012, pp. 1652--1656.

\bibitem{Mori:Tanaka:12:2}
------, ``New generalizations of the {Bethe} approximation via asymptotic
  expansion,'' in \emph{Proc.\ 35th Symp.\ Inf.\ Theory and its Appl.}, Beppu,
  Oita, Japan, Dec.~11--14 2012.

\bibitem{Minc:78}
H.~Minc, \emph{Permanents}.\hskip 1em plus 0.5em minus 0.4em\relax Reading, MA:
  Addison-Wesley, 1978.

\bibitem{Vontobel:11:3:subm}
P.~O. Vontobel, ``The {B}ethe permanent of a non-negative matrix,''
  \emph{accepted for IEEE Trans.\ Inf.\ Theory, available online under
  \emph{\texttt{http://arxiv.org/abs/1107.4196}}}, Jan. 2012.

\bibitem{Smarandache:11:1:subm}
R.~Smarandache, ``Pseudocodewords from {B}ethe permanents,'' \emph{submitted to
  IEEE Trans.\ Inf.\ Theory, available online under \emph{\texttt{http://
  arxiv.org/abs/1112.4625}}}, Dec. 2011.

\bibitem{Greenhill:Janson:Rucinski:10:1}
C.~Greenhill, S.~Janson, and A.~Ruci{\'n}ski, ``On the number of perfect
  matchings in random lifts,'' \emph{Comb., Prob., and Comp.}, vol.~19, no.
  5--6, pp. 791--817, Nov. 2010.

\bibitem{Watanabe:11:1}
Y.~Watanabe, ``A conjecture on independent sets and graph covers,'' \emph{CoRR,
  available online under \emph{\texttt{http://arxiv.org/abs/ 1109.2445}}}, Oct.
  2011.

\bibitem{Ruozzi:12:1}
N.~Ruozzi, ``The {B}ethe partition function of log-supermodular graphical
  models,'' in \emph{Proc.\ Neural Inf.\ Proc.\ Sys.\ Conf.}, Lake Tahoe, NV,
  USA, Dec.~3--6 2012.

\bibitem{Sudderth:Wainwright:Willsky:07:1}
E.~B. Sudderth, M.~J. Wainwright, and A.~S. Willsky, ``Loop series and {B}ethe
  variational bounds in attractive graphical models,'' in \emph{Proc.\ Neural
  Information Processing Systems Conference}, Vancouver, Canada, Dec.~3--8
  2007.

\bibitem{Parvaresh:Vontobel:12:1}
F.~Parvaresh and P.~O. Vontobel, ``Approximately counting the number of
  constrained arrays via the sum-product algorithm,'' in \emph{Proc.\ IEEE
  Int.\ Symp.\ Inf.\ Theory}, Cambridge, MA, USA, Jul.~1--6 2012, pp. 279--283.

\bibitem{Vontobel:12:1}
P.~O. Vontobel, ``The {B}ethe approximation of the pattern maximum likelihood
  distribution,'' in \emph{Proc.\ IEEE Int.\ Symp.\ Inf.\ Theory}, Cambridge,
  MA, USA, Jul.~1--6 2012, pp. 2012--2016.

\bibitem{Boyd:Vandenberghe:04:1}
S.~Boyd and L.~Vandenberghe, \emph{Convex Optimization}.\hskip 1em plus 0.5em
  minus 0.4em\relax Cambridge, UK: Cambridge University Press, 2004.

\bibitem{Kschischang:Frey:Loeliger:01}
F.~R. Kschischang, B.~J. Frey, and H.-A. Loeliger, ``Factor graphs and the
  sum-product algorithm,'' \emph{IEEE Trans.\ Inf.\ Theory}, vol.~47, no.~2,
  pp. 498--519, Feb. 2001.

\bibitem{Forney:01:1}
G.~D. {Forney, Jr.}, ``Codes on graphs: normal realizations,'' \emph{IEEE
  Trans.\ Inf.\ Theory}, vol.~47, no.~2, pp. 520--548, Feb. 2001.

\bibitem{Loeliger:04:1}
H.-A. Loeliger, ``An introduction to factor graphs,'' \emph{IEEE Sig.\ Proc.\
  Mag.}, vol.~21, no.~1, pp. 28--41, Jan. 2004.

\bibitem{Cover:Thomas:91}
T.~M. Cover and J.~A. Thomas, \emph{Elements of Information Theory}.\hskip 1em
  plus 0.5em minus 0.4em\relax New York: John Wiley \& Sons Inc., 1991.

\bibitem{Forney:11:1}
G.~D. {Forney, Jr.}, ``Codes on graphs: duality and {MacWilliams} identities,''
  \emph{IEEE Trans.\ Inf.\ Theory}, vol.~57, no.~3, pp. 1382--1397, Mar. 2011.

\bibitem{AlBashabsheh:Mao:11:1}
A.~Al-Bashabsheh and Y.~Mao, ``Normal factor graphs and holographic
  transformations,'' \emph{IEEE Trans.\ Inf.\ Theory}, vol.~57, no.~2, pp.
  752--763, Feb. 2011.

\bibitem{Csiszar:Korner:81}
I.~Csisz{\'a}r and J.~K{\"o}rner, \emph{Information Theory: Coding Theorems for
  Discrete Memoryless Systems}.\hskip 1em plus 0.5em minus 0.4em\relax
  Budapest: Akad\'emiai Kiad\'o (Publishing House of the Hungarian Academy of
  Sciences), 1981.

\bibitem{Yedidia:01:1}
J.~Yedidia, ``An idiosyncratic journey beyond mean field theory,'' in
  \emph{Advanced Mean Field Methods, Theory and Practice}, M.~Opper and
  D.~Saad, Eds.\hskip 1em plus 0.5em minus 0.4em\relax MIT Press, Jan. 2001,
  pp. 21--36.

\bibitem{Wainwright:Jordan:08:1}
M.~J. Wainwright and M.~I. Jordan, ``Graphical models, exponential families,
  and variational inference,'' \emph{Foundations and Trends in Machine
  Learning}, vol.~1, no. 1--2, pp. 1--305, Jan. 2008.

\bibitem{Stark:Terras:96:1}
H.~M. Stark and A.~A. Terras, ``Zeta functions of finite graphs and
  coverings,'' \emph{Adv.\ in Math.}, vol. 121, no.~1, pp. 124--165, Jul. 1996.

\bibitem{Koetter:Li:Vontobel:Walker:07:1}
R.~Koetter, W.-C.~W. Li, P.~O. Vontobel, and J.~L. Walker, ``Characterizations
  of pseudo-codewords of (low-density) parity-check codes,'' \emph{Adv.\ in
  Math.}, vol. 213, no.~1, pp. 205--229, Aug. 2007.

\bibitem{Kelley:Sridhara:07:1}
C.~A. Kelley and D.~Sridhara, ``Pseudocodewords of {T}anner graphs,''
  \emph{IEEE Trans.\ Inf.\ Theory}, vol.~53, no.~11, pp. 4013--4038, Nov. 2007.

\bibitem{Axvig:Dreher:Morrison:Psota:Perez:Walker:09:1}
N.~Axvig, D.~Dreher, K.~Morrison, E.~Psota, L.~C. Perez, and J.~L. Walker,
  ``Analysis of connections between pseudocodewords,'' \emph{IEEE Trans.\ Inf.\
  Theory}, vol.~55, no.~9, pp. 4099--4107, Sep. 2009.

\bibitem{Massey:77:1}
W.~S. Massey, \emph{Algebraic Topology: an Introduction}.\hskip 1em plus 0.5em
  minus 0.4em\relax New York: Springer-Verlag, 1977, reprint of the 1967
  edition, Graduate Texts in Mathematics, Vol. 56.

\bibitem{Vontobel:08:3}
P.~O. Vontobel, ``Symbolwise graph-cover decoding: connecting sum-product
  algorithm decoding and {B}ethe free energy minimization,'' in \emph{Proc.\
  46th Allerton Conf.\ on Communications, Control, and Computing}, Allerton
  House, Monticello, IL, USA, Sep.~23--26 2008, slides available at
  \texttt{http://www.pseudocodewords.info}.

\bibitem{Vontobel:09:Talk:5}
------, ``Counting, counting, counting (or, finite-length analysis of the
  sum-product algorithm),'' Plenary talk at 2009 Information Theory Workshop in
  Taormina, Italy, Oct.~13 2009, slides available at
  \texttt{http://www.pseudocodewords.info}.

\bibitem{Gallager:63}
R.~G. Gallager, \emph{Low-Density Parity-Check Codes}.\hskip 1em plus 0.5em
  minus 0.4em\relax M.I.T. Press, Cambridge, MA, 1963.

\bibitem{Di:Montanari:Urbanke:04:1}
C.~Di, A.~Montanari, and R.~Urbanke, ``Weight distributions of {LDPC} code
  ensembles: combinatorics meets statistical physics,'' in \emph{Proc.\ IEEE
  Int.\ Symp.\ Inf.\ Theory}, Chicago, IL, USA, June 27--July 2 2004, p. 102.

\bibitem{Forney:GluesingLuerssen:12:1}
G.~D. {Forney, Jr.} and H.~{Gluesing-Luerssen}, ``Observability,
  controllability and local reducibility,'' \emph{to appear in IEEE Trans.\
  Inf.\ Theory}, 2012.

\bibitem{Vontobel:10:2}
P.~O. Vontobel, ``Connecting the {B}ethe entropy and the edge zeta function of
  a cycle code,'' in \emph{Proc.\ IEEE Int.\ Symp.\ Inf.\ Theory}, Austin, TX,
  USA, Jun.~13--18 2010, pp. 704--708.

\bibitem{Richardson:Urbanke:01:2}
T.~J. Richardson and R.~L. Urbanke, ``The capacity of low-density parity-check
  codes under message-passing decoding,'' \emph{IEEE Trans.\ Inf.\ Theory},
  vol.~47, no.~2, pp. 599--618, 2001.

\bibitem{Richardson:Urbanke:08:1}
T.~Richardson and R.~Urbanke, \emph{Modern Coding Theory}.\hskip 1em plus 0.5em
  minus 0.4em\relax New York, NY: Cambridge University Press, 2008.

\bibitem{Litsyn:Shevelev:02:1}
S.~Litsyn and V.~Shevelev, ``On ensembles of low-density parity-check codes:
  asymptotic distance distributions,'' \emph{IEEE Trans.\ Inf.\ Theory},
  vol.~48, no.~4, pp. 887--908, Apr. 2002.

\bibitem{Wiegerinck:Heskes:03:1}
W.~Wiegerinck and T.~Heskes, ``Fractional belief propagation,'' in
  \emph{Advances in Neural Information Processing Systems 15}, S.~Becker,
  S.~Thrun, and K.~Obermayer, Eds.\hskip 1em plus 0.5em minus 0.4em\relax
  Cambridge, MA: MIT Press, 2003, pp. 438--445.

\bibitem{McEliece:Yildirim:03:1}
R.~J. McEliece and M.~Yildirim, ``Belief propagation of partially ordered
  sets,'' in \emph{Mathematical Systems Theory in Biology, Communication,
  Computation, and Finance, IMA Volumes in Math.~\& Appl.}, D.~Gilliam and
  J.~Rosenthal, Eds.\hskip 1em plus 0.5em minus 0.4em\relax Springer Verlag,
  2003.

\bibitem{Vontobel:11:1}
P.~O. Vontobel, ``A combinatorial characterization of the {B}ethe and the
  {K}ikuchi partition functions,'' in \emph{Proc.\ Inf.\ Theory Appl.\
  Workshop}, UC San Diego, La Jolla, CA, USA, Feb.~6--11 2011.

\bibitem{Vontobel:09:1}
------, ``A graph-dynamics interpretation of the sum-product algorithm,'' in
  \emph{Proc.\ Inf.\ Theory Appl.\ Workshop}, UC San Diego, La Jolla, CA, USA,
  Feb.~8--13 2009, slides available at
  \texttt{http://www.pseudocodewords.info}.

\end{thebibliography}

\end{document}